\documentclass[fleqn,usenatbib]{mnras}

\usepackage{newtxtext,newtxmath}

\usepackage[T1]{fontenc}

\DeclareRobustCommand{\VAN}[3]{#2}
\let\VANthebibliography\thebibliography
\def\thebibliography{\DeclareRobustCommand{\VAN}[3]{##3}\VANthebibliography}

\usepackage{graphicx}	
\usepackage{amsmath}	
\usepackage{nccmath}
\usepackage{comment}
\usepackage{arydshln}
\hypersetup{breaklinks=true}
\usepackage{multirow}
\usepackage{subcaption}
\usepackage{notoccite}
\usepackage{booktabs}
\usepackage{float}

\usepackage{units}
\defcitealias{Espejo_Salcedo_2022}{ES22}
\usepackage{arydshln}

\usepackage{subfiles}

\usepackage{xcolor}
\definecolor{lightblue}{RGB}{173,216,230}
\definecolor{darkgreen}{RGB}{0,100,0}
\definecolor{darkorange}{RGB}{220,110,0}

\title[The Fall relation at $1.5<z<2.5$]{A Shallow Slope for the Stellar Mass--Angular Momentum Relation of Star-Forming Galaxies at $1.5 < z < 2.5$}

\author[Juan M. Espejo Salcedo et al.]{Juan M. Espejo Salcedo$^{1,4,5}$\thanks{E-mail: jespejosalcedo@swin.edu.au}, Karl Glazebrook$^{1,4}$, Deanne B. Fisher$^{1,4}$, Sarah M. Sweet$^{3,4}$,\and Danail 
Obreschkow$^{2}$, N. M. F\"orster Schreiber$^{5}$
\\
$^{1}$Centre for Astrophysics \& Supercomputing, Swinburne University of Technology, PO Box 218, Hawthorn, VIC 3122, Australia\\
$^{2}$International Centre for Radio Astronomy Research, University of Western Australia, 7 fairway, Crawley, WA 6009, Australia\\
$^{3}$School of Mathematics and Physics, University of Queensland, Brisbane, QLD 4072, Australia\\
$^{4}$ARC Centre of Excellence for All Sky Astrophysics in 3 Dimensions (ASTRO 3D)\\
$^{5}$Max-Planck-Institut für extraterrestische Physik (MPE), Giessenbachstr., 85748 Garching, Germany
}

\date{Accepted 2024 November 25. Received 2024 November 4; in original form 2024 July 3}

\pubyear{2023}

\begin{document}
\label{firstpage}
\pagerange{\pageref{firstpage}--\pageref{lastpage}}
\maketitle

\begin{abstract}
We present measurements of the specific angular momentum $j_\star$ of 41 star-forming galaxies at $1.5<z<2.5$. These measurements are based on radial profiles inferred from near-IR \textit{HST} photometry, along with multi-resolution emission-line kinematic modelling using integral field spectroscopy (IFS) data from KMOS, SINFONI, and OSIRIS. We identified 24 disks (disk fraction of $58.6\pm 7.7\%$) and used them to parametrize the $j_\star$ \textit{vs} stellar mass $M_\star$ relation (Fall relation) as $j_\star\propto M_\star^{\beta}$. We measure a power-law slope $\beta=0.25\pm0.15$, which deviates by approximately $3\sigma$ from the commonly adopted local value $\beta = 0.67$, indicating a statistically significant difference. We find that two key systematic effects could drive the steep slopes in previous high-redshift studies: first, including irregular (non-disk) systems due to limitations in spatial resolution and second, using the commonly used approximation $\tilde{j}_\star\approx k_n v_s r_\mathrm{eff}$, which depends on global unresolved quantities. In our sample, both effects lead to steeper slopes of $\beta=0.48\pm0.21$ and $\beta=0.61\pm0.21$, respectively. To understand the shallow slope, we discuss observational effects and systematic uncertainties and analyze the retention of $j_\star$ relative to the angular momentum of the halo $j_h$ (angular momentum retention factor $f_j =j_\star/j_h$). For the $M_\star$ range covered by the sample $9.5 <\log_{10} (M_\star/M_\odot) < 11.5$ (halo mass $11.5 < \log_{10} (M_h/M_\odot) < 14$), we find large $f_j$ values ($>1$ in some cases) in low-mass haloes that decrease with increasing mass, suggesting a significant role of efficient angular momentum transport in these gas-rich systems, aided by the removal of low-$j_\star$ gas via feedback-driven outflows in low-mass galaxies.
\end{abstract}




\section{Introduction}

One of the most fundamental physical quantities driving the formation and evolution of galaxies is their angular momentum, a global quantity that controls their dynamical state, the galaxy-wide instabilities as well as their size and shape (\citealp{Fall_Efstathiou}; \citealp{van_den_Bosch}; \citealp{Obreschkow_2014}). Tidal torque theory links the origin of the halo angular momentum ($J_h$) to local cosmic density fluctuations (\citealp{Hoyle}; \citealp{Peebles}; \citealp{Efstathiou_Jones_1979}; \citealp{Mo_galaxy_formation}; \citealp{Liao}). In this framework, there is a tight relationship between the halo mass ($M_h$) and the specific angular momentum ($j_h=J_h/M_h$) of the form $j_h\propto M_h^{\beta}$ with $\beta=2/3$ (\citealp{Catelan_Theuns}; \citealp{Obreschkow_2014}). As baryons cool and collapse into the centre of the dark halo, they are often assumed to approximately preserve the average specific angular momentum of the host halo. Such conservation is approximately consistent with the observed relation between the stellar mass ($M_\star$) and 
specific angular momentum $j_\star$, first studied by \cite{Fall_1983} and commonly known as the ``Fall relation''. The Fall relation provides benchmarks for the outcomes of simulations and the evolution of angular momentum with cosmic time. Measuring the slope and scatter around this relation is useful in galaxy evolution studies as it puts constraints on the different mechanisms that affect (or are affected by) the angular momentum content of galaxies, such as Hubble type, bulge-to-total $B/T$ ratios, merger histories, substructure or stellar feedback (\citealp{Ubler_2014}).

Spatially resolved kinematic measurements of large galaxy samples at $z\approx 0$ have become possible over the last couple of decades, thanks to the significant improvements of integral field spectroscopy (IFS) instrumentation. These large samples cover a large range in morphology and stellar masses, allowing us to understand the role of angular momentum in shaping nearby galaxies (e.g., \citealp[CALIFA:][]{Sanchez_CALIFA}; \citealp[SAMI:][]{Bryant_SAMI}, \citealp[MaNGA:][]{Bundy_MaNGA}). These studies, mainly from optical and $\textrm{HI}$ surveys, find that disk-dominated galaxies follow a relation of the form $j_\star \propto M_\star^{2/3}$ (\citealp{Posti_2018}; \citealp{Hardwick}; \citealp{Du_2022}), while bulge-dominated galaxies exhibit a similar scaling (albeit less well-established) with a negative vertical offset (\citealp{Romanowsky}; \citealp{Obreschkow_2014}; \citealp{Pulsoni2023}), consistent with scenarios of galaxy formation involving merger events and rapid inward gas transport induced by Toomre-scale gravitational instabilities (e.g., \citealp{Bournaud}; \citealp{Fakhouri_2008}; \citealp{Dekel_Burkert};  \citealp{Rodriguez-Gomez_2015}; \citealp{Genzel_2011, Genzel_2020}).

The exploration of kinematics extends to $1\leq z \leq3$, a period commonly known as ``cosmic noon'' (see review by \citealp{Forster_review_2020}) is now possible due to the development of near-infrared integral field units (IFU), capable of tracing prominent emission lines in these systems. This period is interesting since galaxies exhibit high levels of star formation, show complex morphologies, contain numerous bright star-forming clumps, and experience higher accretion rates. Moreover, the transition of star-forming galaxies from clumpy irregular systems at $z\sim2$ to rapidly rotating spirals at $z\sim0$, set by an increase in rotation velocities and a decrease in integrated disordered motions over time (disk settling; \citealp{Kassin_disk_settling}; \citealp{Mortlock_2013}) aligns with the accumulation of specific angular momentum over cosmic time (e.g., \citealp{Obreschkow_2015}; \citealp{Swinbank}; \citealp{Naab}), as well as the average growth in galaxy sizes (e.g., \citealp{Trujillo_2006}; \citealp{Allen_2016}; \citealp{Yang_2021}). Recent studies of $j_\star$ within the redshift range $1\leq z \leq3$ found that the Fall relation is consistent with the scaling $j_\star \propto M_\star^{2/3}$ (e.g., \citealp{Contini_2016}; \citealp{Burkert}; \citealp{Swinbank}; \citealp{Harrison_2017}; \citealp{Alcorn_2018}; \citealp{Gillman_Hizels}; \citealp{Tiley}; \citealp{Gillman}). 

However, given the difficulties in obtaining reliable estimates of $j_\star$, some of these studies centred their discussions on the normalisation and scatter of the Fall relation, assuming a fixed power-law slope of $\beta=2/3$. This assumption is based on the expectation that the slope $\beta$ remains the same as for $z=0$ galaxies (e.g., \citealp{Burkert}; \citealp{Gillman}). This naturally raises the question of whether there is any evolution in the slope with redshift from the cosmic noon epoch and if the clumpy nature of galaxies at cosmic noon could affect it. A related question is to what extent any systematic bias affects the measurements of these complex systems and the fitted slope if it is not fixed.

The vast majority of the current measurements of spatially-resolved ionized gas kinematics at high redshift come from IFUs in ground-based telescopes, which are limited by the seeing of the turbulent atmosphere. These types of observations are commonly referred to as ``seeing limited'' or said to be taken under ``natural seeing'' (NS) conditions. The most notable example of large seeing-limited IFS surveys at $z>1$ are those taken with the $K$-band multi-object spectrograph \hbox{\citep[KMOS;][]{KMOS}} at VLT which has allowed us to address the kinematic state of hundreds of galaxies (e.g., \citealp[KMOS$^\mathrm{3D}$;][]{KMOS3D, Wisnioski_2019}, \citealp[KROSS;][]{Swinbank}; \citealp{Harrison_2017}, \citealp[KGES;][]{Gillman}).

However, the low resolution in the NS-based observations makes the distinction between mergers and disks difficult (\citealp{Rodrigues}; \citealp{Sweet}; \citealp{Simons2019}) and only allows one to resolve spatial scales of $\sim 5$ kpc at $z>1$ which is comparable to or larger than the typical effective radii of galaxies at the same redshift (\citealp{Szomoru_half_mass_radii}). Hence, many studies that measure $j_\star$ rely on a simple approximation based on a small number of global quantities (e.g., \citealp{Burkert}; \citealp{Alcorn_2018}; \citealp{Gillman}; \citealp{Tiley_2021}). The commonly used approximation is of the form $\tilde{j}_\star \approx k_n v_s r_\mathrm{eff}$ (\citealp{Romanowsky}) and hereafter referred to as R\&F approximation, where $v_s$ is the rotational velocity at a specific radius, and $k_n$ and $r_\mathrm{eff}$ are measures of the galaxy shape and size, often limited to the fit of a single component S\'ersic profile (\citealp{Sersic}). Due to the complexity of the morphology and kinematics of these systems, this approximation is likely to introduce systematic effects in the measurement of $j_\star$ and the scaling of the Fall relation. Angular momentum measurements require spatially resolved observations down to kpc scales in both the stellar mass distributions and the kinematics, specifically to trace the rotational velocity and mass distribution across different regions of the galaxy. This ensures that variations in internal structure, such as bulges, disks, and asymmetries, are properly accounted for, reducing the uncertainties introduced by beam smearing and the assumption of a single-component model.

In the case of the stellar mass content, deep near-infrared (IR) imaging is necessary to measure the light profile, which can be used as a proxy for the mass profile $\Sigma(r)$ under the assumption of a constant mass-to-light $(M/L)_\star$ ratio. A simple measurement of the effective radius from a S\'ersic fit may not describe the shape of the light distribution and does not take into account the complex morphological features within the disks, such as the presence of clumps. Moreover, the S\'ersic index $n$ can be greatly influenced by non-symmetrical features, which are common in high-redshift galaxies. The choice of the near-IR band is essential in quantifying different galaxy features, with long wavelengths (redder) tracing old stellar populations that dominate the stellar mass distribution (e.g., \citealp{Lang_2014}), while bluer wavelengths are better suited for measuring clumps, the youngest star-forming sites.

In the case of kinematics, kpc-scale measurements are only effectively possible with IFUs assisted with adaptive optics (AO) in ground-based telescopes with long integration times and more recently with \textit{JWST} NIRSpec with its IFU (e.g., \citealp{D'Eugenio}; \citealp{Perna2}) or via multiplexed slit-stepping using the NIRSpec micro-shutter array (MSA) (e.g., \citealp{Barisic_2024}). For this reason, very few objects at high redshift have been observed at high-spatial resolution (e.g., \citealp{Mieda_IROCKS}; \citealp{Forster_2018}; \citealp{Molina_shizels}; see Figure 3 in \citealp{Forster_review_2020} for an overview). Due to the gain in spatial resolution, observations assisted by adaptive optics techniques allow us to resolve the inner part of the rotation curves and velocity dispersion profiles and to identify small-scale structures. However, they are typically limited to a smaller field of view and have a lower sensitivity per pixel than their seeing-limited counterparts (\citealp{Burkert}; \citealp{Gillman_Hizels}). A combination of the AO-assisted observations with their natural seeing counterparts can leverage the high spatial resolution of the former and the depth of the latter to better distinguish disks from mergers and measure rotation curves with higher accuracy (e.g., \citealp{Obreschkow_2015}; \citealp{Sweet_2018}, and \citealp{Espejo_Salcedo_2022}).

In this paper, we use this combination method to study the angular momentum of 41 star-forming galaxies at $1.5<z<2.5$. This sample is the largest at $z \gtrsim 1.5$ to feature both AO-assisted and seeing-limited deep IFS data, complemented by deep \textit{HST} photometric imaging. We classify the sample morphologically, measure the slope and scatter in the Fall relation for the disk systems, and discuss the implications and possible interpretations of our findings. The paper is organized as follows: Section \S \ref{section: Observations} describes the sample. Section \S \ref{section: substructure} details the methods for identifying clumps from photometry and measuring central light concentrations. Section \S \ref{section: angular momentum} discusses the methodology for measuring specific angular momentum, including photometric and kinematic analysis.
Section \S \ref{section: Results} presents the results, including the classification of disks and the analysis of the $j_\star$ \textit{vs} $M_\star$ Fall relation, revealing a shallow slope.
Section \S \ref{section: Discussion} discusses potential systematic errors in different $j_\star$ estimation methods and provides a physical interpretation based on angular momentum retention factors. Finally, we present the conclusions of this work in Section \S \ref{section: Conclusions}. Appendix \S \ref{appendix: spatial resolution and PSF} contains a discussion of the spatial resolution and PSF modelling of the datasets, and Appendix \S \ref{appendix: case by case} contains figures with a summary of all the individual disk galaxies. Throughout this work, we adopt a $\Lambda$CDM cosmology with $\Omega_\mathrm{m}$ = 0.3, $\Omega_\Lambda$ = 0.7, and $H_0$ = 70 km s$^{-1}$ Mpc$^{-1}$. In this framework, one arcsecond corresponds to $8.46$ kpc at $z=1.5$ and $8.07$ kpc at $z=2.5$.

\section{Observations and sample description.}
\label{section: Observations}

\begin{table*}
\centering
\caption{Summary of the sample including the galaxy ID, redshift estimated from their H$\alpha$ emission (from the AO datasets), Right Ascension and Declination in J2000 coordinates, instruments used to acquire the IFS data (S=SINFONI; O=OSIRIS; K=KMOS), \textit{HST} photometric band, the point-spread function full width at half-maximum (PSF FWHM) from the acquisition stars (see Appendix \S\ref{appendix: PSF modelling}), effective radius, star formation rate, and stellar mass measurements from \protect\cite{Gillman} at $z\sim1.5$ and \protect\cite{Forster_2018} at $z\sim2.2$. We adopt typical uncertainties in the mass measurements of $\Delta M_\star = 0.2$ dex in $\log(M_\star)$, consistent with estimates from \protect\cite{Mobasher}.}
	\label{tab:sample_description}
\begin{tabular}{ccccccccccc}
\toprule
Name  & $z$   & RA   & DEC  & Instrument & Photometry & \multicolumn{2}{c}{PSF FWHM} & $r_\mathrm{eff}$  & SFR & $M_\star$  \\
\cmidrule(lr){7-8}
(ID) & & (hh:mm:ss) & ($^\circ$:':") & NS + AO & Band(s) & NS (") & AO (") & (kpc)  & ($M_\odot \mathrm{yr}^{-1}$) &  ($10^{10}M_\odot$) \\
\hline
Q1623-BX455    & 2.4078 & 16:25:51.7    & $+$26:46:55     & S + S & \textit{H}        & 0.58     &  0.13    & 2.1   & 15  & 1.03     \\
Q1623-BX543    & 2.5209 & 16:25:57.7    & $+$26:50:09     & S + S & \textit{H}        & 0.78     &  0.22    & 3.3   & 145 & 0.94     \\
Q1623-BX599    & 2.3313 & 16:26:02.6    & $+$26:45:32     & S + S & \textit{H, J}     & 0.57     &  0.29    & 2.4   & 34  & 5.66     \\
Q2343-BX389    & 2.1733 & 23:46:28.9    & $+$12:47:34     & S + S & \textit{H, J}     & 0.60     &  0.24    & 6.2   & 25  & 4.12     \\
Q2343-BX513    & 2.1080  & 23:46:11.1   & $+$12:48:32     & S + S & ...               & 0.57     &  0.21    & 2.6   & 10  & 2.7      \\
Q2343-BX610    & 2.2103 & 23:46:09.4    & $+$12:49:19     & S + S & \textit{H, J}     & 0.60     &  0.31    & 4.5   & 60  & 10       \\
Q2346-BX482    & 2.2571 & 23:48:13.0    & $+$00:25:46     & S + S & \textit{H, J}     & 0.65     &  0.21    & 6.0   & 80  & 1.84   \\
Deep3a-6004    & 2.3867 & 11:25:03.8    & $-$21:45:33     & S + S & \textit{H, J}     & 0.62     &  0.20    & 5.1   & 214 & 31.6     \\
Deep3a-6397    & 1.5138 & 11:25:10.5    & $-$21:45:06     & S + S & ...               & 0.90     &  0.20    & 5.9   & 563 & 12       \\
Deep3a-15504   & 2.3826 & 11:24:15.6    & $-$21:39:31     & S + S & \textit{H, J}     & 0.50     &  0.20    & 6.0   & 150 & 10.9   \\
K20-ID6        & 2.2345 & 03:32:29.1    & $-$27:45:21     & K + S & \textit{H, J}     & 0.63     &  0.25    & 3.9   & 45  & 2.67     \\
K20-ID7        & 2.2241 & 03:32:29.1    & $-$27:46:29     & S + S & \textit{H, J}     & 0.68     &  0.19    & 8.4   & 112 & 3.95     \\
GMASS-2303     & 2.4507 & 03:32:38.9    & $-$27:43:22     & K + S & \textit{H, J}     & 0.86     &  0.20    & 1.6   & 21  & 0.72     \\
GMASS-2363     & 2.4518 & 03:32:39.4    & $-$27:42:36     & K + S & \textit{H, J}     & 0.73     &  0.22    & 2.3   & 64  & 2.16     \\
GMASS-2540     & 1.6146 & 03:32:30.3    & $-$27:42:40     & S + S & \textit{H, J}     & 0.88     &  0.29    & 8.5   & 21  & 1.89     \\
SA12-6339      & 2.2971 & 12:05:32.7    & $-$07:23:38     & S + S & ...               & 0.52     &  0.18    & 1.2   & 620 & 2.57     \\
ZC400528       & 2.3873 & 09:59:47.6    & $+$01:44:19     & S + S & \textit{H, J}     & 0.57     &  0.19    & 2.4   & 148 & 11       \\
ZC400569       & 2.2405 & 10:01:08.7    & $+$01:44:28     & S + S & \textit{H, J}     & 0.71     &  0.18    & 7.4   & 241 & 16.1     \\
ZC401925       & 2.1412 & 10:01:01.7    & $+$01:48:38     & S + S & \textit{H, J}     & 0.60     &  0.25    & 2.6   & 47  & 0.58     \\
ZC403741       & 1.4457 & 10:00:18.4    & $+$01:55:08     & S + S & ...               & 0.72     &  0.21    & 2.2   & 113 & 4.45     \\
ZC404221       & 2.2199 & 10:01:41.3    & $+$01:56:43     & S + S & \textit{H, J}     & 0.70     &  0.23    & 0.8   & 61  & 1.57     \\
ZC405226       & 2.2870 & 10:02:19.5    & $+$02:00:18     & S + S & \textit{H, J}     & 0.48     &  0.27    & 5.4   & 117 & 0.93     \\
ZC405501       & 2.1539 & 09:59:53.7    & $+$02:01:09     & S + S & \textit{H, J}     & 0.56     &  0.19    & 5.8   & 85  & 0.84     \\
ZC406690       & 2.1950 & 09:58:59.1    & $+$02:05:04     & S + S & \textit{H, J}     & 0.79     &  0.20    & 7.0   & 200 & 4.14     \\
ZC407302       & 2.1819 & 09:59:56      & $+$02:06:51     & S + S & \textit{H, J}     & 0.68     &  0.20    & 3.6   & 340 & 2.44     \\
ZC407376       & 2.1729 & 10:00:45.1    & $+$02:07:05     & S + S & \textit{H, J}     & 0.76     &  0.30    & 5.5   & 89  & 2.53     \\
ZC409985       & 2.4569 & 09:59:14.2    & $+$02:15:47     & S + S & \textit{H, J}     & 0.84     &  0.15    & 1.9   & 51  & 1.61     \\
ZC410041       & 2.4541 & 10:00:44.3    & $+$02:15:59     & K + S & \textit{H, J}     & 0.80     &  0.20    & 4.7   & 47  & 0.46     \\
ZC410123       & 2.1986 & 10:02:06.5    & $+$02:16:16     & S + S & \textit{H, J}     & 0.73     &  0.30    & 3.2   & 59  & 0.42     \\
ZC411737       & 2.4442 & 10:00:32.4    & $+$02:21:21     & S + S & \textit{H, J}     & 0.59     &  0.24    & 1.8   & 48  & 0.34     \\
ZC412369       & 2.0281 & 10:01:46.9    & $+$02:23:25     & S + S & \textit{H, J}     & 0.61     &  0.18    & 3.1   & 94  & 2.17     \\
ZC413507       & 2.4800 & 10:00:24.2    & $+$02:27:41     & S + S & \textit{H, J}     & 0.55     &  0.18    & 2.6   & 111 & 0.88     \\
ZC413597       & 2.4502 & 09:59:36.4    & $+$02:27:59     & S + S & \textit{H, J}     & 0.62     &  0.22    & 1.6   & 84  & 0.75     \\
ZC415876       & 2.4354 & 10:00:09.4    & $+$02:36:58     & S + S & \textit{H, J}     & 0.60     &  0.18    & 2.4   & 94  & 0.92     \\
COSMOS-110446  & 1.5199 & 9:59:50.82    & $+$02:04:50     & K + O & \textit{I}        & 0.84     &  0.11    & 2.35  & 49  & 3.3      \\
COSMOS-171407  & 1.5247 & 9:59:33.96    & $+$02:20:54     & K + O & \textit{H}        & 0.72     &  0.39    & 4.03  & 31  & 2.6      \\
COSMOS-130477  & 1.4651 & 10:00:0.70    & $+$02:19:47     & K + O & \textit{I}        & 0.59     &  0.38    & 4.03  & 33  & 2.6      \\
COSMOS-127977  & 1.6200 & 9:59:37.9     & $+$02:18:02     & K + O & \textit{I}        & 0.72     &  0.11    & 3.69  & 45  & 1.1      \\
UDS-78317      & 1.5247 &  02:17:34     & $-$05:10:16     & K + O & \textit{H}        & 0.69     &  0.11    & 3.02  & 45  & 3.1      \\
UDS-124101     & 1.4832 & 02:18:51      & $-$04:57:23     & K + O & ...               & 0.76     &  0.13    & 4.51  & 28.5  & 6.82   \\
COSMOS-128904  & 1.4626 & 10:00:07.6    & $+$02:18:44     & K + O & \textit{I}        & 0.60     &  0.12    & 4.94  & 9.78  & 9.05 \\
\hline
\end{tabular}
\end{table*}

\begin{figure}
	\includegraphics[width=0.47\textwidth]{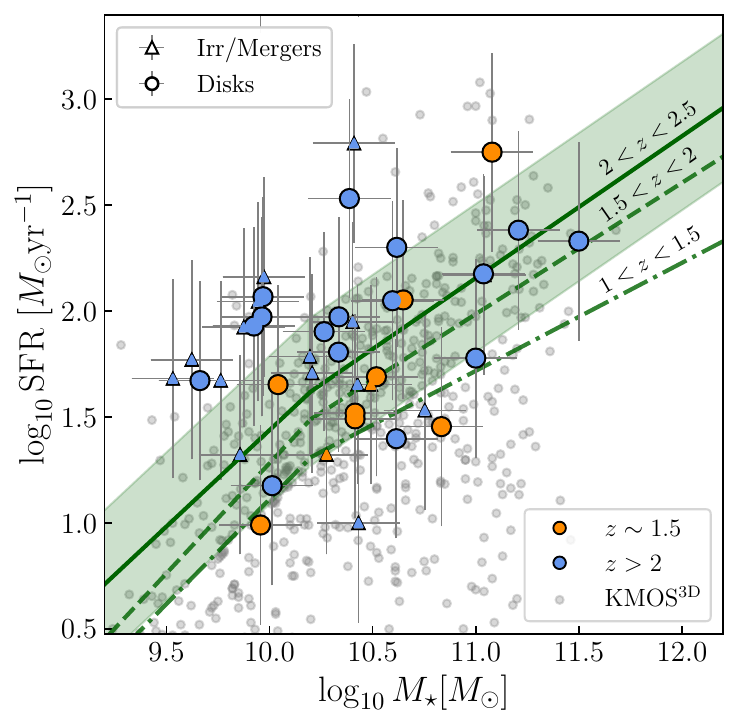}
	\caption{Star-formation rate (SFR) \textit{vs} stellar mass $M_\star$ of the full sample. Dots represent the disks and triangles represent those identified as either irregulars or mergers. Blue markers correspond to galaxies at $z>2$ and orange markers at $z\sim 1.5$, while the grey dots correspond to the KMOS$^\mathrm{3D}$ sample, which is one of the parent seeing-limited samples. The broken laws are extracted from \protect\cite{Whitaker2014}, and the scatter band for the $2<z<2.5$ of 0.35 dex represents the upper limit of the observed scatter from 25 studies compiled in \protect\cite{Speagle_2014}.}
    \label{fig: MS of sample}
\end{figure}

The sample studied in this paper consists of 41 galaxies (which we determine below to have 24 disks) in the redshift range $1.5<z<2.5$, representative of star-forming galaxies at that cosmic epoch in terms of their size, mass range, and star formation rates (see discussions in \citealp{Forster_2008_no_AO, Mancini_2011, Forster_2018}). See Figure \ref{fig: MS of sample} for a visualization of the sample in the star-formation rate (SFR) \textit{vs} stellar mass ($M_\star$) ``star-formation sequence''. This sample has deep integral field spectroscopy (IFS) observations available at both high and low spatial resolutions, with PSF FWHMs ranging from [0.1, 0.4] arcseconds and [0.5, 0.9] arcseconds, respectively, and typical on-source exposure times of the order of several hours, making it the largest sample with such a combination of datasets at $z\gtrsim1.5$. The high-spatial-resolution data were obtained with adaptive optics at SINFONI(VLT) and OSIRIS(Keck), and the low-resolution observations at natural seeing were obtained using both KMOS and SINFONI at VLT. The dataset is divided into two main subsamples, defined by the AO instrument used in each case (as well as the redshift range). The first subsample is the SINS sample at $z \sim 2.2$ (\citealp{Forster_2008_no_AO}; \citealp{Forster_2018}), obtained using SINFONI. The second subsample is the OSIRIS (Keck) sample at $z\sim1.5$. Finally, the majority of the galaxies (36) studied in this paper have been observed with the Hubble Space Telescope (\textit{HST}) in near-infrared bands. Table \ref{tab:sample_description} contains a summary of the full sample.

\begin{figure*}
	\includegraphics[width=0.98\textwidth]{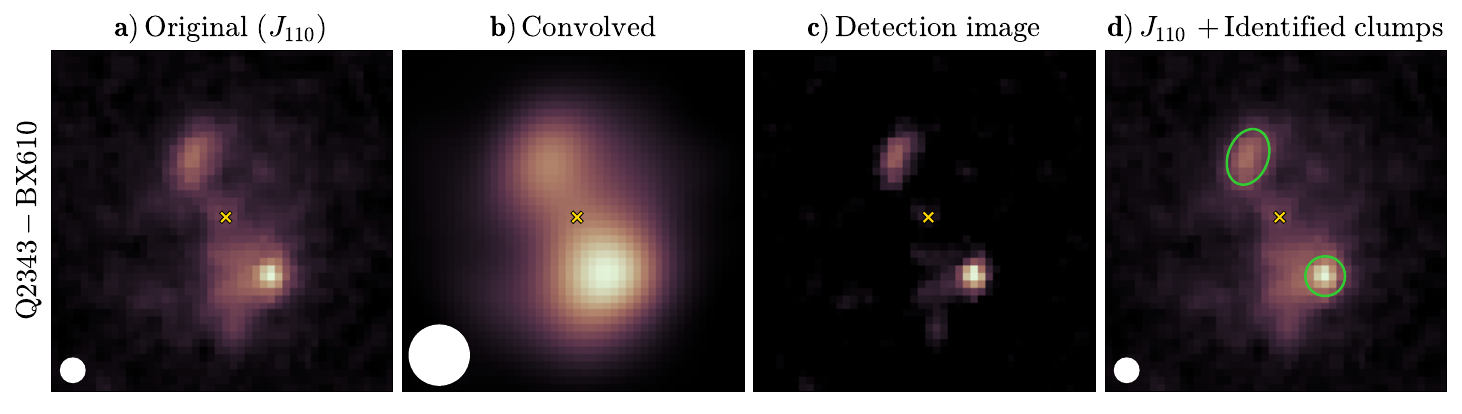}
    \caption{Clump detection method for Q2343-BX610. \textbf{a)} Original $J_{110}$ \textit{HST} image with the kinematic centre indicated with the yellow cross. \textbf{b)} Convolved image with the Gaussian kernel. \textbf{c)} ``Detection image'' where the unsharp features have been removed (Equation \ref{eq: Detection image formula}). \textbf{d)} Original image with the identified clumps in green circles/ellipses.}
    \label{fig: Method clumps}
\end{figure*}

\subsection{\texorpdfstring{$z\sim2.2$}{z2.2} sample (SINS sample - SINFONI + KMOS)}

We collected the publicly available data of the AO-assisted observations of the SINS galaxies from the SINS/zC-SINF AO survey as well as their seeing-limited counterparts from the parent sample (SINS survey). For galaxies where the SINFONI low-resolution data have a low signal-to-noise ratio (SNR), we use the KMOS data from the KMOS$^\textrm{3D}$ survey. The total number of galaxies that have seeing-limited and adaptive optics-assisted observations is 34 in the range of $1.5<z<2.5$.

\subsubsection{\textbf{SINS/zC observations (natural seeing)}}

These observations are a subset of targets from the parent sample (SINS survey \citealp{Forster_2008_no_AO})\footnote{ESO Programme IDs 070.A-0229, 070.B-0545, 073.B-9018, 074.A-9011, 075.A-0466, 076.A-0527, 077.A-0576, 078.A-0055, 078.A-0600, 079.A-0341, 080.A-0330, 080.A-0635, and 080.A-0339} that were later on followed up for AO-assisted observations. The targets in the SINS survey were originally drawn from spectroscopically confirmed targets spanning a wide redshift range of $1<z<4$, observed with the Spectrograph for Integral Field Observations in the Near Infrared \citep[SINFONI;][]{SINFONI} with typical on-source exposure times of $\sim 3.5$ hours (ranging from 20 minutes to 10 hours). The datacubes have a spatial pixel scale of 0.125 arcsec and are either in the $K$- or $H$-band where the spectral channels have a sampling of 2.45\AA$\,$ and 1.95\AA$\,$ respectively. The PSF FWHM varies in the range $\sim0.5$ to $\sim0.8$ arcsec.

There is some overlap in the SINS galaxies with those observed in the KMOS$^\textrm{3D}$ \citep{KMOS3D, Wisnioski_2019}\footnote{ESO program IDS 092A-0091, 093.A-0079, 094.A-0217, 095.A-0047, 096.A-0025,
097.A-0028, 098.A-0045, 099.A-0013, 0100.A-0039, and 0101.A-0022} survey, so in those cases, we prioritize the data with better SNR for the forward analysis since we can obtain better kinematic maps from them. Namely, the galaxies that have better quality data from KMOS$^\textrm{3D}$ are GMASS-2303, GMASS-2363, KD20-ID6, and ZC410041. This subsample has a spatial pixel scale of 0.2 arcsec and a mean PSF FWHM of 0.73 arcsec.

\subsubsection{\textbf{SINS/zC-SINF AO observations (adaptive optics)}}

We use the public release of the ``SINS/zC-SINF AO'' \citep{Forster_2018} data\footnote{\href{http://www.mpe.mpg.de/ir/SINS/SINS-zcSINF-data}{http://www.mpe.mpg.de/ir/SINS/SINS-zcSINF-data}}$^,$\footnote{ESO Programme IDs 075.A-0466, 076.A-0527, 079.A-0341, 080.A-0330, 080.A-0339, 080.A-0635, 081.B-0568, 081.A-0672, 082.A-0396, 183.A-0781, 087.A-0081, 088.A-0202, 088.A-0209, 091.A-0126}. These observations were taken as a follow-up to their low-resolution counterparts and were designed to measure emission line kinematics at spatial resolutions up to ${\sim}1.5$ kpc. They were taken with SINFONI in the natural guide star (NGS) and/or laser guide star (LGS) AO modes with on-source integration times ranging from 2 to 23 hours (median of 6 hours). As described in the parent paper, the pixel scale of the datacubes is 0.05 arcsec, the mean PSF FWHM is 0.17 arcsec, and the spectral channel sampling is 2.45$\text{\AA}$ and 1.95$\text{\AA}$ in the $K$- and $H$-band, respectively.

\subsection{\texorpdfstring{$z\sim1.5$}{z1.5} sample (OSIRIS+KMOS)}

This subsample corresponds to a total of 7 galaxies at $z\sim1.5$ observed at low-spatial resolution with \textit{K}-band multiobject spectrograph (KMOS; \citealp{KMOS}) at the VLT and our own follow-up observations at higher spatial resolution with laser guide adaptive optics using the OH-Suppressing Infra-Red Imaging Spectrograph (OSIRIS; \citealp{LARKIN2006362}) instrument at Keck.

\subsubsection{\textbf{KMOS observations (natural seeing)}}

We collected the reduced datacubes of 7 galaxies (COSMOS-110446, COSMOS-171407, COSMOS-130477, COSMOS-127977, UDS-78317, UDS-124101, and COSMOS-128904) from the KMOS Galaxy Evolution Survey (KGES; \citealp{Tiley_2021})\footnote{ESO Programme IDs: 095.A-0748, 096.A-0200, 097.A-0182, 098.A-0311,
and 0100.A-0134}, which targeted bright ($K < 22.7$) star-forming galaxies at $z\sim1.5$ in the known ECDFS, UDS and COSMOS fields. These galaxies were observed using the $K$-band multi-object spectrograph \hbox{\citep[KMOS;][]{KMOS}} where the pixel scale is 0.2 arcsec, and the mean PSF FWHM of the observations was ${\sim}0.7$ arcsec. The total exposure time (on-source) ranged between 1.5 to 11 hours, with a mean of 4.4 hours. 

\subsubsection{\textbf{OSIRIS observations (adaptive optics)}}

We selected the seven galaxies above based on the preliminary results of the KGES survey and carried out follow-up observations with LGS adaptive optics with OSIRIS (OH-Suppressing Infra-Red Imaging Spectrograph; \hbox{\citealp{OSIRIS}}) located at the Keck II telescope. We prioritized objects with kinematic maps of high quality from the low-resolution observations, well-ordered rotation, as well as proximity to stars that can be used for the tip-tilt corrections. 

We collected the data for the 7 galaxies over a total of 4 nights\footnote{Keck program IDs: W276, W146, W175 and W131.} in the span of 4 years, from which one night was lost due to an issue with the OSIRIS spectrograph and another half night was lost due to weather conditions. The observations, with typical on-source exposure times of $\sim 2-3$ hours, were taken in the $H-$band using the $Hn3$ filter (15940-16760$\text{\AA}$) and with a pixel scale of 0.1 arcsec. The mean PSF FWHM of the subsample is 0.14 arcsec. The reduction of the data, including the additional spatial smoothing, applied on COSMOS-171407 and COSMOS-130477 is explained in \cite{Espejo_Salcedo_2022} (hereafter \citetalias{Espejo_Salcedo_2022}), where an initial analysis of five of these galaxies was already carried out.

\subsection{\textit{HST} imaging}

We have collected the broad-band $HST$ images of 36 galaxies in multiple (or single) bands in the near-IR regime to infer the stellar mass profiles. The majority of the images are those used in \cite{Forster_2011a} and \cite{Tacchella_sins_sizes}, where the photometric analysis of 29 galaxies of the SINS survey was conducted. This dataset was provided directly by the PIs for our analysis. The rest of the \textit{HST} images were obtained from the different databases and are shown in the ``Photometry'' column in Table \ref{tab:sample_description}. As a summary, from the 34 galaxies in the SINS sample, 28 galaxies have both $H-$ and $J-$band images and 2 have $H-$band only (see \citealp{Tacchella_sins_sizes} for details), all with a pixel sampling of 0.05 arcsec and PSF FWHM of 0.16 and 0.17 arcsec for \textit{H}- and \textit{J}-band, respectively. 

From the $z\sim1.5$ sample, two galaxies have $H-$band images only. Namely, COSMOS-171407 from COSMOS-DASH (COSMOS-Drift And SHift; \citealp{DASH-HST}) and UDS-78317 from the Cosmic Assembly Near-Infrared Deep Extragalactic Legacy Survey \citep[CANDELS;][]{Candels}. The pixel scales are 0.1 and 0.06 arcsec, while the PSF FWHM is 0.15 and 0.18 arcsec, respectively. Additionally, four galaxies (COSMOS-110446, COSMOS-130477, COSMOS-127977, and COSMOS-128904) only have $I$-band \textit{HST} data taken with the Advanced Camera for Surveys \citep[ACS;][]{ACS} in the F814W filter with a pixel scale of 0.03 arcsec and a PSF FWHM of 0.08 arcsec. For the remaining 5 that lack \textit{HST} data, we use the H$\alpha$ intensity maps coming from the AO sample as a proxy of their mass distribution. See a discussion on the caveats of this choice in \S \ref{subsection: photometric analysis}.

\section{Substructure}
\label{section: substructure}

To investigate if the morphological complexities of star-forming galaxies at $1.5<z<2.5$ show any correlation with global quantities such as $j_\star$, we quantified the amount of substructure in the whole sample. The significant gain in spatial resolution from the AO sample and the overlap with \textit{HST} photometry allows for measuring small-scale features. More specifically, we measured star-forming clumps and central light concentrations as they can be resolved with the near-IR photometric data and thus could be measured consistently throughout the sample.

\subsection{Clumps}
\label{subsection: clumps}

In the redshift range explored in this study, spatially varying $(M/L)_\star$ ratios can arise due to the diverse extinction and stellar ages of clumps, whose prevalence has been recently confirmed in the optical and rest-frame near-IR wavelengths using \textit{JWST} imaging (\citealp{Kalita_2024}). Such variations have been quantified using spectral energy distribution (SED) modelling (e.g., \citealp{Wuyts_2012,Wuyts_2013}; \citealp{Guo_2018}). The mean PSF FWHM of the \textit{HST} near-IR data in our sample is $0.18$ arcsec which corresponds to $1.5$ kpc at $z\approx2.2$ so it allowed resolving the bright clumps from the disks, as done by \cite{Genzel_2011} who studied the properties of five of the SINS galaxies (Q1623-BX599, Q2346-BX482, Deep3a-15504, ZC407302, and ZC406690).

Clumps are less prominent (and contribute less to the integrated light) at long wavelengths so they often disappear in spatially-resolved stellar mass maps. Additionally, clumps trace largely the youngest star-forming sites (bluer) along regions of least dust obscuration. However, they are not necessarily all tracing important local enhancements in stellar mass. For these reasons, we prioritized the bluer deep infrared \textit{HST} bands for their detection.

For COSMOS-110446, COSMOS130477, COSMOS-127977, and COSMOS-128904 we used the \textit{I-}band data for the identification of clumps. For 28/41 galaxies, we used their \textit{J-}band data and for those without \textit{I}- or \textit{J}-band, we used \textit{H-}band (Q1623-BX455, Q1623-BX543, COSMOS-171407, and UDS-78317). For those galaxies that lack deep infrared imaging, we used the H$\alpha$ intensity maps from the Gaussian line fit (Q2343-BX513, Deep3a-6397, SA12-6339, ZC403741, and UDS-124101). Caveats in the choice of band are discussed at the end of this section.

The location and extent of the clumps were measured using a similar approach to the one used in \cite{Fisher_2017} and \cite{Liyu_paper_2022}, where clumps are systematically identified by detecting sharp bright regions above the galaxy mean flux. The method follows the principles of the unsharp masking technique (\citealp{Malin_unsharp_masking}). It consists of creating a ``detection image'' by convolving the original image with a Gaussian kernel with size $\sim 4 \times$ the \textit{HST} PSF FWHM. The convolved image is then subtracted from the original one, leaving only the sharper brighter regions within the disks. Finally, the detection image is normalized by dividing it by the convolved image as

\begin{equation}
    \mathrm{Detection \,\, image} = \frac{\mathrm{Original - Convolved}}{\mathrm{Convolved}}.
    \label{eq: Detection image formula}
\end{equation}

The detection image thus highlights the sharp regions in the original image that are potential clumps. However, not all of these regions correspond to clumps, given the limitations in spatial resolution and data quality, so the peaks could only be identified as clumps when they fulfilled the following set of criteria:

\textbf{1)} The clump flux must be 2$\times$ above the background scatter in the detection image.

\textbf{2)} It must have a 5$\sigma$ peak above the galaxy disk light in the original \textit{HST} image.

\textbf{3)} Criteria 1 and 2 must be fulfilled over an area equivalent to at least one resolution element.

\textbf{4)} The clump must be an independent structure with flux that declines in all directions. This criterion is checked visually.

\textbf{5)} If a clump is identified at the centre of the galaxy (within 1 kpc from the galaxy centre) by fulfilling criteria 1-4 in \textit{J-} but also in \textit{H-}band (redder) then we do not count it as a clump as it is likely to correspond to a galaxy bulge.

The construction of the detection image is shown in Figure \ref{fig: Method clumps} for galaxy Q2343-BX610 where two clumps are clearly identified.

\begin{figure}
	\includegraphics[width=0.45\textwidth]{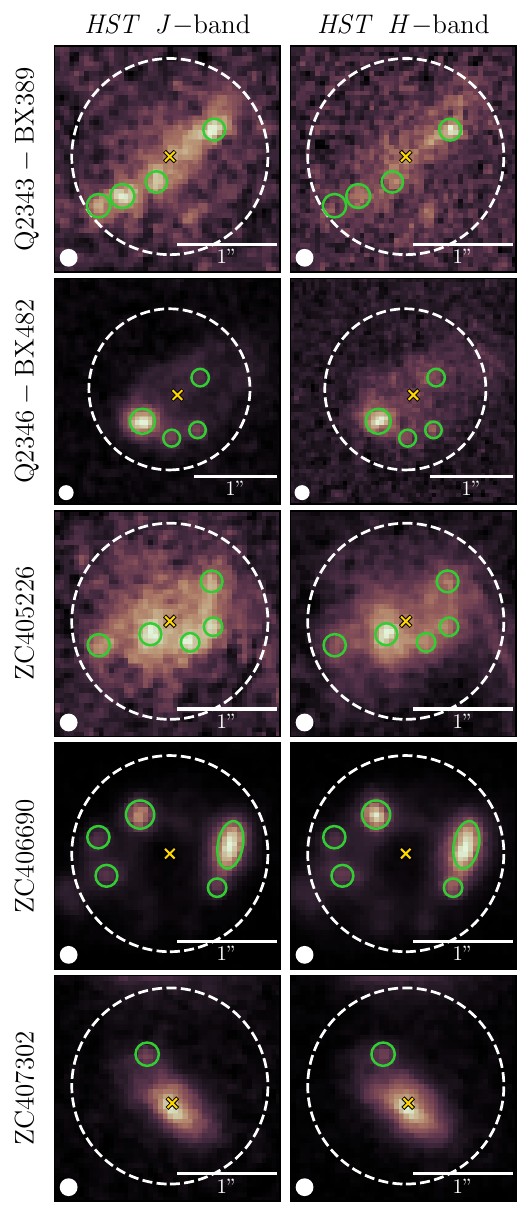}
    \caption{Visualization of clumps for 5 galaxies of the SINS sample. The left panels are in the \textit{J}-band (used for identifying the clumps), and the right panels are in \textit{H}-band.}
    \label{fig: Clumps examples}
\end{figure}

With this method, we identify a total of 102 clumps in 31 of our 41 galaxies, where galaxies with large physical sizes show a higher incidence of clumps. To show the results of the method across different galaxies, we show five examples with the detected clumps (Q2343-BX389, Q2346-BX482, ZC405226, ZC406690, and ZC407302) in Figure \ref{fig: Clumps examples} with their corresponding \textit{H}- and \textit{J}-band data.

There is some overlap between our analysis and other studies that measure the clump properties of some galaxies in the sample. In \cite{Genzel_2011}, they used the resolved H$\alpha$ line detections from the SINS observations (AO) to detect large clumps in five galaxies that overlap with this sample (Q1623-BX599, Q2346-BX482, Deep3a-15504, ZC407302, and ZC406690). They based their clump identification on two strategies: one was identifying local maxima in at least two separate velocity channel maps, and the other one was by inspecting residual kinematic maps (by removing the modelled velocity and dispersion fields). For three out of the five galaxies, they identified the same number of clumps as we did (1, 4, and 6, respectively) and at the same spatial locations. While for ZC406690, they found 4, and we found 6. However, a large discrepancy was found for galaxy ZC407302, where they identified 5 while we only identified one. In this case, the big difference is likely because the 4 additional clumps that they identify in their method are very close together and near the centre of the galaxy (see Figure 2 in \citealp{Genzel_2011}), so they do not fulfil the criteria 3 (size above the resolution element) and 5 (non-central location) in our method. In \cite{Forster_2011b}, they used \textit{HST} photometric imaging in the \textit{H}-band to identify clumps using the IRAF task \textit{daofind} (\citealp{Stetson}) which searches for local density maxima with specified size (FWHM) and above a customised background threshold. Three of the galaxies in their study overlap with our measurements (Q2343-BX389, Q2343-BX610, Q2346-BX482). For all three galaxies, we identified 4 clumps, while they identified more in all cases (6, 7, and 5, respectively), likely attributed to the choice of background threshold. Hence, our method could be tracing a lower bound as it identifies only the brightest clumps, giving a more conservative estimate of how clumpy these galaxies are.

Aside from determining the number of detected clumps, we also used this technique to quantify the ``clumpiness'' that we label $\mathcal{C}$ for the galaxies in the sample. We quantified $\mathcal{C}$ as the ratio between the light inside all the (non-central) clumps and the total galaxy light. The aperture sizes are based on the size of the region identified as belonging to a clump as long as the region is larger than the PSF size. In the few cases where two clumps overlap, we used elliptical apertures to enclose the light corresponding to both clumps as one, as is the case for ZC406690, COSMOS-110446, and Q2343-BX610. Our method is similar to that in \cite{Conselice_2003}, where ``clumpiness'' is defined as the ratio of the amount of light contained in high-frequency spatial structures -small, bright regions within the galaxy flux- to the total amount of galaxy light. In their work, the high-frequency structures (interpreted as clumps) were found using a detection method similar to ours. However, they used a different version of Equation \ref{eq: Detection image formula}, in which the denominator is the original image instead of the convolved one. Other studies have used a different metric for ``clumpiness''. 

The average number of clumps in the full sample was $N_\mathrm{clumps}=2.5$ which corresponds to $\mathcal{C}=12.7\%$, while it was $N_\mathrm{clumps}=3.3$ ($\mathcal{C} = 16.8\%$) for the galaxies where at least one clump was found. The ``clumpiness'' that we find in the sample is comparatively similar to that found by \cite{Wisnioski_2011} and \cite{Wisnioski_2012_clumps}\footnote{They do not calculate the ``clumpiness'' of the sample so the comparison is based on the spatial distribution and size of the detected clumps.} for galaxies in the WiggleZ kinematic survey (see Figure 2 in \citealp{Wisnioski_2012_clumps} for size and distribution of the detected clumps). However, it is approximately twice as high as the value found in the analysis of \cite{Wuyts_2012} for 326 galaxies in the GOODS-South field. In their study, the mass contribution of clumps was measured using the second-order moment of the brightest 20\% of the galaxy's flux (the $M_{20}$ parameter, introduced by \citealp{Lotz_2004}) from which they found that their contribution is $\lesssim 7\%$ of the integrated mass. For the 5 overlapping galaxies in \cite{Genzel_2011}, the number of detected clumps was $N_\mathrm{clumps}=4$ while for the three overlapping galaxies in \cite{Forster_2011b} it was $N_\mathrm{clumps}=6$. 

\begin{figure}
	\includegraphics[width=0.45\textwidth]{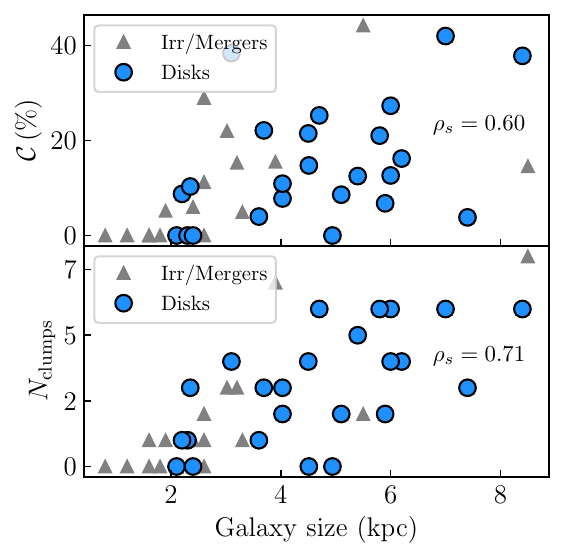}
    \caption{Size dependence of the number of clumps $N_\mathrm{clumps}$ (bottom) and galaxy ``clumpiness'' $\mathcal{C}$ (top) as a function of galaxy size $r_\mathrm{eff}$. The dots and triangles distinguish the disks from the Irregulars/mergers as discussed in \S \ref{subsection: Kinematic state of the galaxies}. The Spearman correlation coefficients are shown on the far right.}
    \label{fig:clumps_vs_size}
\end{figure}

We note that the approach used in our analysis can introduce a systematic effect related to the size of the galaxy. It is easier to identify clumps in large (extended) systems such as Q2346-BX482 and ZC406690 with $N_\mathrm{clumps}=4$ and $N_\mathrm{clumps}=6$ respectively, as opposed to small galaxies that might be clumpy despite being compact. In Figure \ref{fig:clumps_vs_size}, we show the size dependence in terms of both $N_\mathrm{clumps}$ and $\mathcal{C}$, where we find a clear correlation in both cases, as quantified by their Spearman correlation coefficients $\rho_s=0.71$ and $\rho_s=0.60$ respectively. One of the implications of these correlations is that to assess the role of clumps in the Fall relation at cosmic noon, one needs large samples in bins of fixed galaxy size.

Finally, we also investigate the effect of the choice of the photometric band in the identification of clumps. For the 28 galaxies that have both \textit{H-} and \textit{J-}band data, we find a general agreement with only a slight decrease in the total number of clumps $N_\mathrm{clumps}$ which goes from 84 in \textit{J-} band to 79 in \textit{H-}band (6\% difference). The effect is more significant when measuring the amount of light inside the clumps and quantified by the degree of clumpiness, which goes from a mean $\mathcal{C}\sim 14.7\%$ in \textit{J-} band to $\mathcal{C}\sim 12.7\%$ in \textit{H-}band (14\% change). This confirms the expectation that bluer bands are more sensitive to the presence of star-forming clumps but the difference is modest and does not affect the main results from the investigation of clumps in the sample.

\subsection{Central concentration}
\label{subsection: concentration}

\begin{figure}
	\includegraphics[width=0.48\textwidth]{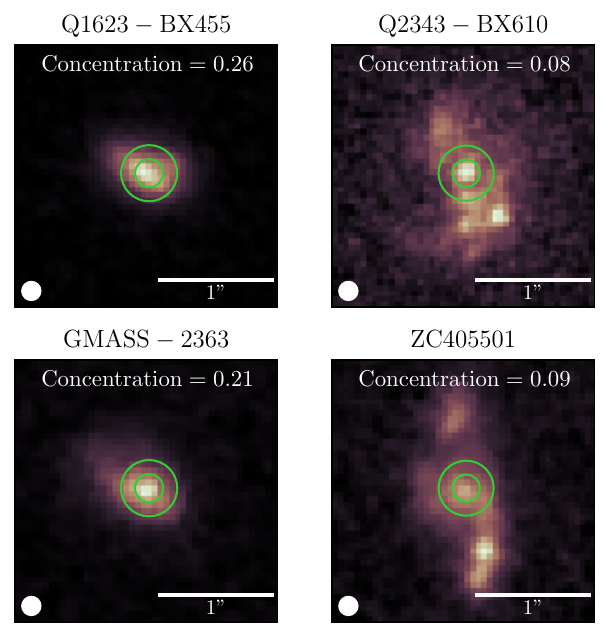}
        \caption{Examples of the central concentration estimation for four disk galaxies from their \textit{H}-band photometric images. The inner green circles indicate the size of the aperture ($r=1$ kpc) from which we extracted the central light. The outer circle indicates the outer radius ($r=2$ kpc) of the annulus used to measure the excess flux coming from the centre.}
    \label{fig: example of bulges}
\end{figure}

Another morphological feature that can affect the global angular momentum content of these systems is the presence of a massive central bulge (e.g., \citealp{Fall_Romanowsky_2018}). Noting the difficulties of measuring the bulge fractions directly in these galaxies, we measured instead the central light concentration. To quantify the concentration, we measured the flux contained inside a circular area ($A_c$) with a radius $r=1$ kpc from the galaxy centre (e.g., \citealp{Fisher_and_Drody_2016}). The centre was determined from the kinematic modelling, which will be discussed in \S\ref{subsection: kinematic modelling} as indicated by the yellow cross in the figures of \S \ref{appendix: case by case}. To determine the excess flux in the centre, we take a circular annulus of width $\Delta r =1$ kpc, defined by an inner radius $r_i=1$ kpc and an outer radius $r_o = 2$ kpc. We computed the average flux within this annulus and utilized it as the baseline flux in the centre. Then, the central flux is calculated by subtracting the baseline flux from the flux within the inner central circle. Finally, we measured the concentration by dividing the central flux by the total flux of the galaxy. This way, we ensured that the only galaxies where we measured large bulge fractions were those where the inner area $A_c$ was significantly brighter than its surroundings. In the sample, 15 galaxies (7 disks) have an off-centre clump located within the circular annulus, so this clump contribution is excluded from the measured baseline flux. See Figure \ref{fig: example of bulges} for a visualization of the areas used to measure the bulge fractions in four galaxies with their corresponding \textit{H}-band maps, where the central regions are significantly brighter than the surroundings and could indicate the presence of a real bulge.

\begin{figure}
\centering
	\includegraphics[width=0.48\textwidth]{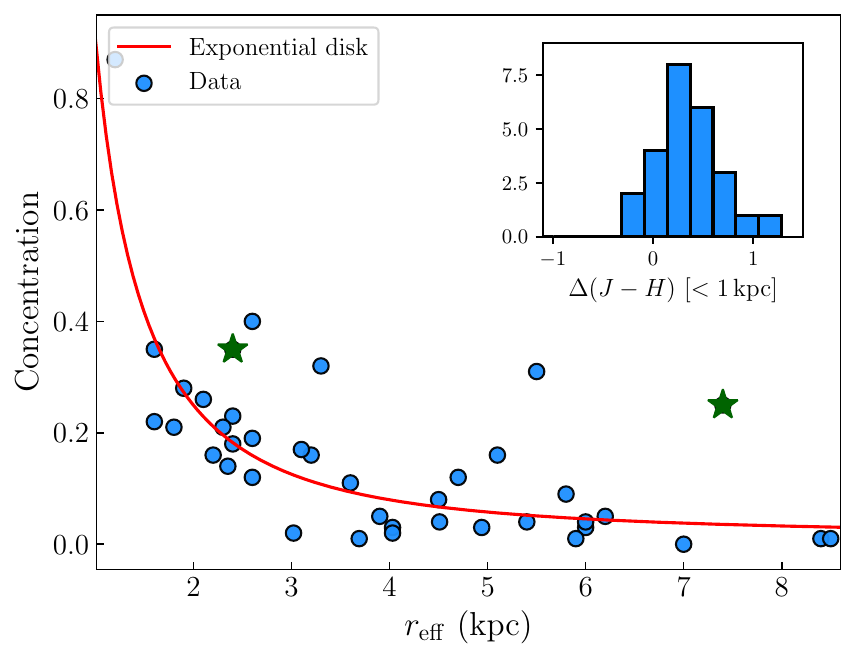}
        \caption{Comparison between the concentration expected from simulated exponential disks (red line) and from the sample of real galaxies (blue dots). The green stars correspond to galaxies ZC400528 and ZC400569, where a bulge component can be easily identified from a visual inspection (e.g. ZC400569: \ref{fig: Summary ZC400569}). The inset histogram shows the distribution of $\Delta(J-H)$ colours in the inner 1 kpc of the 28 galaxies that have the two \textit{HST} bands indicating the predominantly redder central concentrations.}
    \label{fig:concentration_vs_sims}
\end{figure}

The measured concentration can be indicative of the bulge-to-total ratios $B/T$ and is thus useful for the following discussions. However, a general trend of the surface brightness profiles is that they decrease with radius whether or not there is a bulge (except for galaxies with a ring-like structure), so it is important to point out that measuring the central light concentration does not provide a direct measurement of bulge fractions. To show this systematic effect, we show in Figure \ref{fig:concentration_vs_sims} what the measured concentration would be for perfect exponential disks (without any bulge component) using the same range in effective radii as that of the data ($1.5< r_\mathrm{eff} <8.5$ kpc). From that figure, the clear correlation between the data points and the simulated points indicates that, as expected, smaller galaxies are biased to higher concentrations while larger galaxies are biased to lower concentrations. This arises because our choice of concentration metric uses fixed physical radii apertures. However, it also highlights that some of the points that are significantly above the simulated values correspond to galaxies where a bulge component can be identified from a visual inspection, such as ZC400528 and ZC400569, both indicated by the green stars in Figure \ref{fig:concentration_vs_sims}.

In \cite{Tacchella_sins_sizes}, they measured directly the bulge-to-total ratios on the SINS sample using a double component S\'ersic fit (disk+bulge) where the disk component had a fixed S\'ersic index of $n_\mathrm{disk}=1$ and the bulge was allowed to vary with $n_\mathrm{bulge} \in [1.0,8.0]$. With this method, one can measure the contribution of a non-disk component that is not necessarily that of a classical bulge (given the vast range in S\'ersic indices), so it is a common approach in modelling the light distribution of galaxies. However, non-symmetrical features of these clumpy irregular systems can have a large effect on the modelled light distributions thus leading to an over (or under) estimation of $n$ as well as large uncertainties associated with it. In the case of the \cite{Tacchella_sins_sizes} analysis, the bulge component in five galaxies was larger than the disk component, which is improbable for this galaxy population (e.g., \citealp{Bruce_2014}). Moreover, the fit to the effective radius of the bulges for eight galaxies yields $r_{\mathrm{eff, bulge}}<0.5$ kpc, which is below the resolution limit. These limitations, addressed and discussed in detail in \cite{Tacchella_sins_sizes}, motivated our choice of addressing instead the central light concentration, where potential bulges are expected to reside.

With the expectation that bulge components are primordially constituted by old stellar populations (e.g., \citealp{Lang_2014}) and would thus translate into higher concentrations at redder wavelengths, we measured the concentrations with the photometric \textit{HST} data at the reddest wavelength available. In the sample, 32 galaxies have \textit{H}-band data, so for the rest, we used their \textit{J}-, \textit{I}-band, or H$\alpha$ if there was no near-IR photometry (e.g., Q2343-BX513, Deep3a-6397, SA12-6339, ZC403741). For the 28 galaxies that have both \textit{H}- and \textit{J}-band imaging, the concentrations are remarkably similar with an average difference of $\sim 0.02$. The concentrations were found to be the same in both bands for 8 galaxies, whereas for 10 galaxies, it was higher in the \textit{H} band (with mean $\sim 0.01$) while for 10, it was higher in \textit{J} band (with mean $\sim 0.06$). In terms of the colours, we use aperture photometry in the central 1 kpc to calculate the $J-H$ colours of the 28 galaxies that have both bands in the \textit{HST} images. We find a mean $\overline{\Delta}(J-H)=0.26$ colour with 22 galaxies being redder in the inner region, indicative of either higher obscurations of dominant older (redder) stellar populations (see inset histogram in Figure \ref{fig:concentration_vs_sims}). In \cite{Tacchella_sins_sizes}, colour profiles were measured for the sample showing that at least 10 of the galaxies show a clear negative colour gradient, likely explained by variations of stellar populations or dust content. 

In the full sample, the average concentration was $\sim 0.2$, which is similar to the estimated bulge-to-total ratios measured in \cite{Tacchella_sins_sizes} with an average value of $B/T\sim 0.19$ for the 20 galaxies with non-zero $B/T$ that overlap with our sample. On the other hand, the average $B/T$ derived from dynamical modelling (H$\alpha$) with the overlapping samples in \cite{Genzel_2020} (12 galaxies) and \cite{Shachar_2022} (22 galaxies) was $B/T\sim 0.4$ in both cases, which is a factor of 2 higher than our \textit{HST} analysis. Differences may be related to the mass contribution of components unseen in the starlight (due to extinction or in molecular gas form) but with a detectable signature in the H$\alpha$ kinematics or to the different approaches (with dynamical modelling involving a larger number of parameters compared to photometric analysis). It is worth noting that measurements of $B/T$ are challenging even in the local Universe, given the complexity of galaxies and their multiple features.

\section{Specific angular momentum \texorpdfstring{$j_\star$}{j}}
\label{section: angular momentum}

\subsection{Extraction of velocity and velocity dispersion maps}
\label{subsection: gaussian fit}

\begin{figure*}
	\includegraphics[width=1\textwidth]{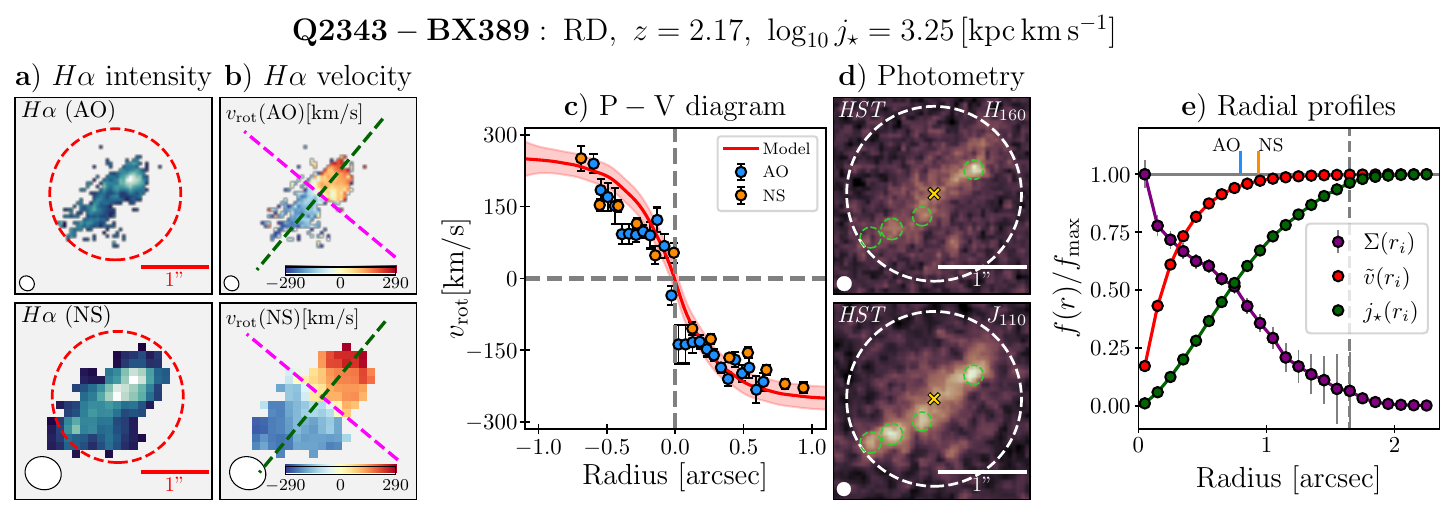}
    \caption{Summary of galaxy Q2343-BX389 which is identified as a rotating disk (RD): \textbf{a)} H$\alpha$ intensity fields at high- (top) and low-resolution (bottom) where the white circles represent the PSF FWHM, \textbf{b)} velocity fields with the main kinematic axes indicated by the dashed green lines, \textbf{c)} position-velocity (P-V) diagram along the kinematic main axis (the width of the slit is the PSF FWHM) where the red line is the model velocity curve $\tilde{v}(r_i)$ obtained with \texttt{CONDOR} and the shaded region corresponds to the uncertainty of the fit. The points correspond to those along the major kinematic axis (with a slit width given by the size of the PSF), \textbf{d)} \textit{HST} near-IR data ($H_{160}$ top and $J_{110}$ bottom) with an indication of the PSF FWHM and the location of the identified clumps, and \textbf{e)} radial normalized profiles for the mass $\Sigma(r)$ (purple), velocity $v(r)$ (red), and specific angular momentum $j_\star(r)$ (green). The vertical dashed grey line represents the extent of the photometric data, so the radial profiles are extrapolated past this boundary to reach the asymptotic value of $j_\star$. Orange and blue lines indicate the radial boundary of the kinematic datasets. The summary figures for the rest of the rotating disk galaxies are shown in the Supplementary materials \S\ref{appendix: case by case}.}
    \label{fig: Summary Q2343-BX389 main text}
\end{figure*}

To measure the kinematics of each galaxy, we first found its systematic redshift $z$ by integrating the spectra over the spaxels with H$\alpha$ detection and then fitting a Gaussian profile to that integrated spectrum using the least-squares curve fitting routine \textsc{mpfit} (\citealp{MPFIT}). The centroid of this Gaussian profile was taken as the systemic redshift of the whole system. We then trimmed each spaxel down to a length of 50 spectral channels (which corresponds to a range of $\sim1750$ km/s) around the location of the systemic redshift, and we measured the continuum in the trimmed spaxel to be used in the estimation of the line SNR. Next, we obtained the two-dimensional velocity $v(x,y)$ and velocity dispersion $\sigma(x,y)$ fields by fitting Gaussian profiles to the H$\alpha$ nebular emission line in each spaxel where the centroid, intensity, and width were free parameters. The velocity was then calculated from the position of the Gaussian profile with respect to the systemic redshift. For galaxies with a strong signal, where the [NII] doublet is clearly visible in the visual inspection of individual spectra, we incorporated the doublet into the fitting routine as additional Gaussian components. The velocity dispersion fields used in the disk classification were obtained from the width of the fitted Gaussian lines.

We took an SNR cut of $\geq3$ to the H$\alpha$ emission with respect to the baseline level in each spaxel. To do this, we measured the chi-squared $\chi^2$ associated with a fit of a straight line to the continuum ($\chi_\mathrm{cont}^2$) and the chi-squared associated with the Gaussian profile fit ($\chi_\mathrm{fit}^2$). The SNR is then calculated as $\mathrm{SNR} = \sqrt{\chi_\mathrm{cont}^2 - \chi_\mathrm{fit}^2}$ (see \cite{Stott_2016} \& \cite{Tiley_2021} for further discussions on this method). We also masked out pixels where the velocity error $v_\mathrm{err}$ calculated by \textsc{mpfit} is above a threshold of $v_\mathrm{err}> 0.5 \times v_\mathrm{max}$ where $v_\mathrm{max}$ corresponds to the pixel with the maximum velocity. This masking removes outer pixels with large uncertainties. We avoided using any spatial smoothing in the original datacubes to avoid degrading spatial resolution with only three exceptions discussed in \S \ref{appendix: spatial resolution}.


\subsection{Photometric analysis}
\label{subsection: photometric analysis}

The stellar mass profile of the galaxies in the sample was estimated from the near-infrared (near-IR) \textit{H-}band \textit{HST} photometric data, which at $z\approx1.5$ corresponds to rest-frame \textit{R}-band and at $z\approx2.2$ corresponds to rest-frame \textit{V-}band. Hence, this is an approximation that does not reflect the true stellar mass distribution since it lacks some light from old stellar populations which contribute the most to the total stellar mass (\citealp{Freeman_Bland-Hawthorn_2002}; \citealp{Kauffmann_2003}; \citealp{Conroy_2009, Conroy_2013}). However, it is expected to trace the shape of the galaxy and still provide information about bulges and clumps within these systems.

To estimate the radial mass profile, we first deprojected the surface brightness of the \textit{HST} data in the reddest wavelength available ($H$-band when available, otherwise \textit{J, I}-band, or H$\alpha$). Then, we estimated a discrete mass profile $\Sigma(r_i)$ for bins located at $r_i$ from the kinematic centre (determined in \S \ref{subsection: kinematic modelling}) by taking the azimuthal average in concentric annular rings. The width of the rings was given by the PSF FWHM of the \textit{HST} data (or the H$\alpha$ data for the galaxies that do not have \textit{HST} observations) to ensure that we were working at the resolution limit of the data. This radial approach guaranteed that the presence of a bulge and non-symmetrical features like large star-forming clumps and low brightness regions are accounted for in the measured $\Sigma(r_i)$ profile. 

The deprojection was performed using the position angle $\theta_\mathrm{PA}$ obtained from the kinematic fit (explained in \ref{subsection: kinematic modelling}) and the inclination $i$ was set using the minor-to-major axis ratios $q=b/a$ from the surface brightness profiles as $\cos ^2 i = (q^2-q_0^2)/(1-q_0^2)$ (Equation 5b in \citealp{Holmberg_inclination_axis_ratio}), with $q_0\approx0.2$ for thick disks (\citealp{Forster_2008_no_AO}; \citealp{KMOS3D}; \citealp{Wuyts_2016}). This way, we broke the well-known degeneracy of the inclination with the velocity field (e.g., \citealp{Begeman}; \citealp{Epinat2010}; \citealp{Kamphuis}; \citealp{Bekiaris2016} and \S 4.4 in \citetalias{Espejo_Salcedo_2022}). For the axis ratios, we used the reported values in \cite{Tacchella_sins_sizes} and \cite{Gillman}, both of which come from a single component \texttt{GALFIT} fit to the \textit{HST} imaging of the sample, and thus provide a reliable global estimate of the axis ratios. The only exception was galaxy ZC400569, where the kinematic fit was affected by the inclination-velocity degeneracy, so it motivated an independent measurement in both photometry and kinematics where we found an inclination $i=50.7^\circ$.

In cases where we utilized the H$\alpha$ maps, it is important to note that H$\alpha$ morphologies primarily trace the distribution of star formation rates. These can differ from the stellar continuum light observed in \textit{HST} imaging, particularly at redder wavelengths. However, visual inspection reveals that the overall H$\alpha$ light distribution (including the faint regions used in the integrated measurement of $j_\star$) displays a large level of agreement with the light continuum (in the reddest available wavelength) for the mass range that dominates this sample and only deviates significantly for galaxies in the high-mass end ($M_\star \gtrsim 10^{11} M_\odot $). This follows from the relationship between H$\alpha$ and stellar continuum disk sizes ($r_{\mathrm{H}\alpha}$ and $r_\star$) of the form $r_{\mathrm{H}\alpha} \propto r_\star (M_\star/10^{10}M_\odot)^{0.054}$ (see \citealp{Nelson_2016}). Quantitatively, this has been studied in \cite{Forster_2018} (see section 5.4 in their work) for the 29 galaxies with near-IR \textit{HST} data in the SINS sample, where they make a comparison between the sizes inferred from the \textit{HST} continuum maps (estimated in \citealp{Tacchella_sins_sizes}) as well as those inferred from the H$\alpha$ surface brightness distributions. They find that both quantities are very similar, within about 5\% in terms of the major axis and the circularized effective radius.

\subsection{Kinematic modelling}
\label{subsection: kinematic modelling}

To find the velocity profile $v(r)$ that best describes the velocity fields extracted from the IFS data (in both AO and NS resolutions), we used the multi-resolution kinematic modelling introduced in \citetalias{Espejo_Salcedo_2022} using the code \texttt{CONDOR}\footnote{\href{https://github.com/juancho9303/CONDOR}{https://github.com/juancho9303/CONDOR}}. In short, this strategy consists of creating model datacubes ($x,y,\lambda$) at both NS and AO resolutions with the kinematic model at the same pixel scale of the original datacubes. Note that the NS data reaches, on average, 1.4 times the radial extent of the AO data, allowing for better modelling of the rotation curve in the galaxy outskirts. Each model cube was then convolved with the corresponding PSF\footnote{The AO PSF is described by a combination of an Airy disk and Moffat profile while the NS PSF can be described with a single 2D Gaussian.} and subsequently with the corresponding line-spread function (LSF) given by the instrumental resolution. Velocity fields were extracted from these model cubes following the same Gaussian line fitting routine in \S \ref{subsection: gaussian fit}, and the best kinematic model or $v(r)$ was drawn from maximum likelihood estimation in 2D using the velocity fields extracted from the observed cubes. The likelihood in this calculation is a combination of the chi-squared associated with both datasets, so their individual contribution is accounted for:

\begin{equation}
\displaystyle{\mathcal{L}\propto e^{-\chi^2/2}}\quad \mathrm{with}\quad \chi^2 = \chi^2_\mathrm{NS} + \chi^2_\mathrm{AO},
\label{eq: likelihood}
\end{equation}
where $\chi^2_\mathrm{NS}$ and $\chi^2_\mathrm{AO}$ are associated with the natural seeing and adaptive optics-assisted data, respectively. See Figure 6 in \citetalias{Espejo_Salcedo_2022} for details and Appendix \S \ref{appendix: spatial resolution and PSF} for the details of the different PSF convolutions.

\begin{figure*}
\centering
	\includegraphics[width=0.96\textwidth]{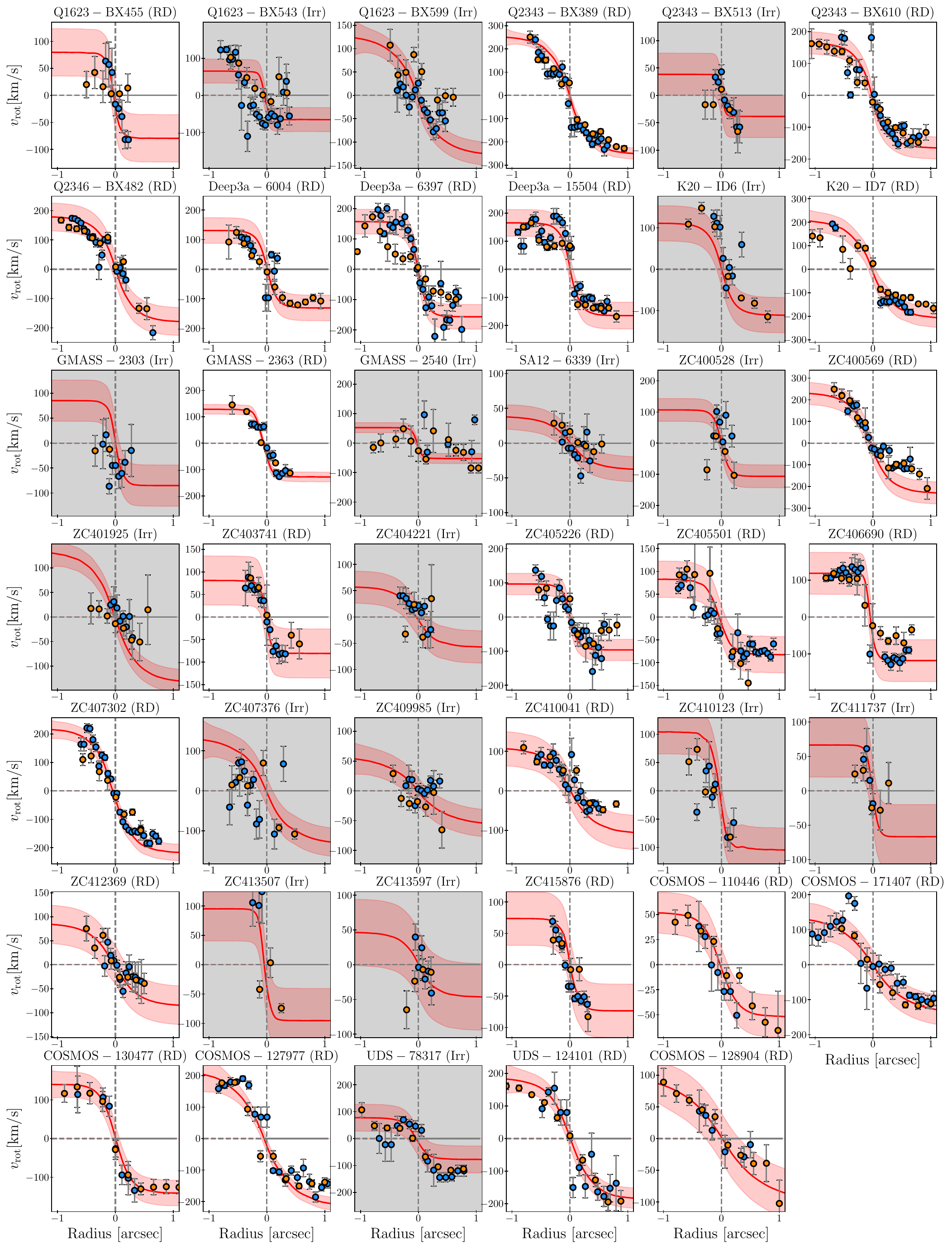}
        \caption{Position-velocity (P-V) diagrams for the full sample, where the red line is the best-fit model from our kinematic modelling and the shaded region represents the uncertainties in the fit. The blue (AO) and orange (NS) dots are extracted from a slit along the kinematic axis with a slit width corresponding to the PSF FWHM. The panels with a grey background correspond to the galaxies classified as irregulars.}
    \label{fig: all PVs}
\end{figure*}

A visual inspection of the individual position-velocity (P-V) diagrams for each disk galaxy (see Figure \ref{fig: all PVs} and Supplementary materials in \S \ref{appendix: case by case}) suggested that the rotation curves can be described with the simple functional form characterized by an asymptotic velocity $v_\textrm{flat}$ and a characteristic radius $r_\textrm{flat}$ proposed by \cite{Boissier_2003}:

\begin{equation}
   v(r) = v_{\textrm{flat}} \left( 1 - \exp \left(\frac{-r}{r_{\textrm{flat}}}\right) \right).
   \label{eq:v model}
\end{equation}

There are various important assumptions and approximations in these estimates.

First, we note that galaxies can exhibit rotation curves that cannot be modelled with a simple flat model. This difference can arise due to various factors, including limitations in data quality in the galaxies' outer regions or the presence of complex gravitational dynamics. Early work in this area by \cite{Brandt_1960} proposed an analytical expression that attempts to capture the rotational behaviour of galaxies in terms of their underlying mass distribution. This marked a significant step in understanding the complexities of galactic rotation and its relation with galaxy mass. Functional forms have emerged to enhance modelling precision. Among these, the Freeman exponential disk model \cite{Freeman_RC} considers stellar disk contributions, while the arctangent model by \cite{Couteau_1997} accommodates transitions between the inner rise and outer flattening, pertinent to extended disk or dark matter halo galaxies. Another notable example is the study of \cite{Giovanelli_2002}, who proposed a flexible model that fits well steep rising curves and allows for varying outer slopes in their so-called ``Polyex'' model. 

Recent investigations of the rotation curves of high redshift star-forming galaxies, which include the SINS sample, find that a fraction of galaxies do exhibit rotation curves that drop or keep rising at large radius (e.g., \citealp{Genzel_2017, Genzel_2020}; \citealp{Price_2021}; \citealp{Shachar_2022}). The dropping rotation curves can be attributed to a combination of effects, including high central baryonic mass concentrations, shallow dark matter halo profiles towards the galaxy centre, and elevated pressure support. For some galaxies in those studies, smoothing was applied to increase SNR, which facilitates the measurement of $v(r)$ at large radii. In our analysis of the SINS subsample, only Q2343-BX610 and ZC405226 show a mildly declining rotation curve along the kinematic axis (in agreement with their work). However, in the case of Q2343-BX610, the visual drop has a value of 30 km/s which is below the uncertainties of measured $v_\mathrm{flat}$ ($\Delta v_\mathrm{flat} = 34$ km/s). In the case of ZC405226, the dropping pattern is only visible in the blue side (negative velocities) of the NS data which corresponds to a region of large uncertainties (see figure \ref{fig:Summary ZC405226}). Since we did not apply any smoothing on the data (except for COSMOS-171407 and COSMOS-130477, which do not overlap with the SINS sample), our method is limited by the SNR and thus probes only the brightest galaxy regions, which are located at small radial extents. As seen in the individual rotation curves along the kinematic axes of the disk galaxies in Figure \ref{fig: all PVs} (as well as in the Supplementary material \S \ref{appendix: case by case}), there are no evident deviations from flat profiles in the outskirts of the majority of galaxies, which motivated the choice of Equation \ref{eq:v model}. However, if the rotation curve keeps rising at large radii, then the underestimation of $v_\mathrm{flat}$ will correspond to an underestimation of $j_\star$. A notable comparison can be done with the 3D kinematic modelling of the galaxies in the SINS sample by \cite{Forster_2018}, \cite{Price_2021} and \cite{Shachar_2022}. They used complex velocity models in their fitting which include a bulge and dark matter halo component and can continue rising or falling at large radii. In their study, they use the $v_\mathrm{rot}$ parameter to characterize the velocity profile and is thus similar to our $v_\mathrm{flat}$ parameter. The difference between the mean values of both parameters for the overlapping sample is $\overline{\Delta v} \approx 30\%$ (with a median of 18.3\%), and this difference tends to be larger in systems with irregular morphologies (non-disks) such, as ZC413597 and GMASS-2540.

Another important approximation used in our measurement of $v(r)$ is that the ionized gas kinematics (measured using H$\alpha$ emission) traces the kinematics of the stars (which are measured using absorption features). However, the rotation velocities of ionized gas and stars can differ in what is known as asymmetric drift (e.g., \citealp{Stromberg_1946}; \citealp{Shetty_2020}). Asymmetric drift quantifies the discrepancy between the rotation velocity of the gas and that of the stars at a specific radius within a galaxy. Since gas is more efficient at dissipating energy, it can cool and settle towards the circular speed associated with the galaxy's potential. On the other hand, stars are collisionless so they can retain more efficiently their non-circular motions. At low redshift, the gas component has been shown to be dynamically colder than the stellar components. A noteworthy example is \cite{Cortese_2016}, who found systematic uncertainties between the velocities inferred from the stellar and ionized gas components at the $\sim 0.1$ dex level for the SAMI survey at $z = 0$. Asymmetric drift is difficult to address at $z > 1$ since the SNR needs to be very large in the absorption features for the detection of the underlying continuum. In most cases, it is only possible to assess the stellar kinematics by measuring the velocity dispersions of the galaxy-integrated light. If we assume a similar velocity profile shape (only considering $v_\mathrm{flat}$) and take the systematic difference of 0.1 dex estimated by \cite{Cortese_2016}, we find that for this subsample (with a mean velocity of 217 km/s), the mean difference of $\pm 56$ km/s would imply small mean differences in $j_\star$ at the 0.07 dex level.

Finally, another important aspect to consider when modelling the velocity profiles is that strong outflows have been identified in some galaxies in the sample (e.g., ZC406690, Deep3a-6004, Deep3a-15504; \citealp{Genzel_2006, Genzel_2011}; \citealp{Newman_2012, Newman_2013}; \citealp{Forster_2014}, and extensive discussion in \citealp{Forster_2018}). These can induce local deviations in the extracted velocity and dispersion maps since they can affect the velocity centroid of the Gaussian fit on the H$\alpha$ line. \cite{Forster_2018} investigated the effect of an additional broad component caused by star-formation-driven outflows in the SINS galaxies and showed that it is only relevant at high SFR surface densities ($\Sigma(\mathrm{SFR}) \geq 1 M_\odot \,\mathrm{yr}^{-1} \,\mathrm{kpc}^{-2}$), which is only the case for $\leq 30\%$ of the area in the SINS sample (see Appendix C in their study). Additionally, only for the six most massive galaxies of the SINS-AO sample, an AGN-driven outflow was detected. This indicated that a single Gaussian profile fit could achieve a satisfactory representation of the observed line profiles for the individual galaxies, especially given the low signal-to-noise ratios, which limit a double-Gaussian fitting strategy. During our examination of the SINS subsample, we made a visual inspection of individual spaxels in the brightest galaxies, and we only found a notable broad component in the datacubes of galaxies ZC406690 and Deep3a-15504, which had also been previously identified in the aforementioned studies through a more comprehensive approach.

Besides $v_\textrm{flat}$ and $r_\textrm{flat}$, the other free parameters in the fit were the kinematic position angle $\theta_\mathrm{PA}$ and the kinematic centres (the inclination was fixed as explained in \S \ref{subsection: photometric analysis}). Once the optimal parameters of the velocity profile were found using \texttt{CONDOR}, we created a radial velocity profile $\tilde{v}(r_i)$ evaluated at the radial bins $r_i$ of the surface brightness profile $\Sigma(r_i)$ found in the previous subsection.

\subsection{Integrated measurement of specific angular momentum}
\label{subsection: angular momentum}

The specific angular momentum ($j_\star= {J_\star}/{M_\star}$) of a galaxy is a function of its distribution of mass $\rho(\boldsymbol{r})$, position $\boldsymbol{r}$ and velocity $\boldsymbol{v}$ as

\begin{equation}
j_\star \equiv \frac{J_\star}{M_\star} =\frac{ \big | \int_{V} \rho(\boldsymbol{r})(\boldsymbol{r} \times \boldsymbol{v}) \mathrm{d}^{3} \boldsymbol{r} \big |}{\int_{V} \rho(\boldsymbol{r}) \mathrm{d}^{3} \boldsymbol{r}}, 
\label{eq:j 3D}
\end{equation}
which under cylindrical symmetry and for $n$ equally spaced radial bins at locations $r_i$ can be reduced to a one-dimensional sum, which we incorporate as our fiducial approach for the calculation of $j_\star$:

\begin{equation}
j_\star=\dfrac{2\pi\displaystyle\sum\limits_{i=1}^{n} \, r_i^2\,\Sigma(r_i)\,\tilde{v}(r_i)}{2\pi\displaystyle\sum\limits_{i=1}^{n} \, r_i\Sigma(r_i) }, 
\label{eq:j definition}
\end{equation}
where $\Sigma(r_i)$ is the azimuthally averaged surface mass density profile\footnote{When calculating the specific angular momentum, we do not need to assume an explicit mass-to-light ratio as $\Sigma(r)$ appears both in the numerator and denominator, so it cancels out (since it is assumed to be constant).} estimated in \S \ref{subsection: photometric analysis} and $\tilde{v}(r_i)$ is the model velocity profile at the corresponding bins estimated in \S \ref{subsection: kinematic modelling}. Note that we have labelled this profile $\tilde{v}(r_i)$ instead of $v(r_i)$ to emphasize that this is a \textit{model} velocity profile obtained from the kinematic fit using the two spatial resolutions. In some cases, the extent of the \textit{HST} data was not enough for $j_\star$ to reach the asymptotic value, so we extrapolate $\Sigma(r_i)$ after the last bin with an exponential decay to guarantee that that value is reached. The difference in the measurement of $j_\star$ using the data-limited measurement and the extrapolated $\Sigma(r_i)$ is $\approx12\%$ for the sample of disks. The cumulative $j_\star(r)$ profiles are indicated in the far right of Figure \ref{fig: Summary Q2343-BX389 main text} and the Supplementary figures in \S \ref{appendix: case by case}.

This approach allows for the stellar mass profile to contain some information measured from the \textit{HST} data about asymmetrical features, substructure, bulges, and clumps in some of these systems. At the same time, it uses the rotation curve that best represents the two datasets at high- and low-spatial resolution accounting for the effects of beam smearing in the inner part of the galaxies, which is an important limitation in low-resolution studies. The use of this method to measure $j_\star$ is a key distinguishing feature of this study compared to existing studies at similar redshifts, as will be discussed in detail in \S \ref{Comparison to different ways of estimating j}.

To test the difference between our approach of calculating $j_\star$ with Equation \ref{eq:j definition} and a pixel-by-pixel measurement, we compared the values of $j_\star$ obtained using the radial approach and those obtained from the discrete sum over all the pixels of the \textit{HST} images with the corresponding velocity field (from AO) at the same pixel scale as in

\begin{equation}
j_\star= \frac{\sum\limits_{i,j} v_{i,j}\Sigma_{i,j}r_{i,j}}{\sum\limits_{i,j} \Sigma_{i,j}},
\label{eq:j pixelwise}
\end{equation}
where $i,j$ go through all the spatially matched pixels in velocity and photometry and $v_{i,j}$ is the azimuthal velocity component, orthogonal to the radii. If the velocity field is axisymmetric, both approaches are expected to yield the same results as discussed in Appendix B of \cite{Obreschkow_2014}. This is because if the velocity is just a function of $r$ (axisymmetry), then the contribution of the surface brightness profile at $r$ in the calculation of $j_\star$ is the radial average, independent of whether it is measured as a pixel or radial sum.

\begin{figure}
    \centering
	\includegraphics[width=0.48\textwidth]{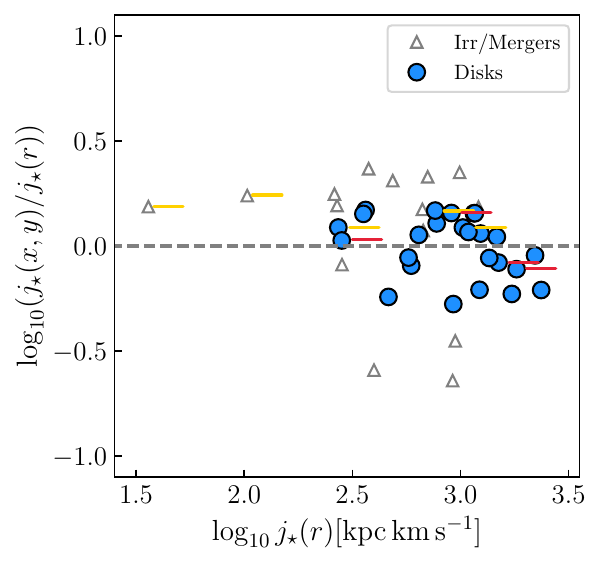}
    \caption{Difference between $j_\star$ from the the radial measurement (Equation \ref{eq:j definition}) and pixel-by-pixel measurements (Equation \ref{eq:j pixelwise}). Grey triangles represent irregular galaxies, and blue circles represent the disks. The grey dashed line represents the one-to-one correspondence. No significant trend is observed for the four galaxies where the \textit{I}-band data was used to infer the light distribution (red lines). For the galaxies with H$\alpha$ radial measurements (yellow lines), the trend is an overestimation of $j_\star$ in the pixel-by-pixel measurement.}
    \label{fig: Radial vs pixel-by-pixel}
\end{figure}

In Figure \ref{fig: Radial vs pixel-by-pixel}, we show the difference between the two methods, which we will also compare in the context of the Fall relation in \S \ref{subsection: Fall relation}. We find large discrepancies in the measurements using the two methods, with some irregular galaxies showing the largest deviations as expected from their non-symmetric morphologies and bright clumps. But even in the sample of disks, the discrepancies are large (up to 0.3 dex). The differences are likely due to the large weight of the clumpy regions in the 2D measurement (lower in the radial profile approach) and the fact that the radial approach allows for a radial extrapolation while the pixel-by-pixel approach does not allow it. If we compare the measurements using the pixel-by-pixel approach and the radial profiled-based approach with the same radial extent, we find that the differences remain high with a median dex difference of 0.2. This experiment shows how relevant the choice of method is when measuring $j_\star$ in the sample and motivated our choice to use \ref{eq:j definition} throughout this paper.

Figure \ref{fig: Summary Q2343-BX389 main text} shows the summary of the photometric and kinematic modelling for galaxy Q2343-BX389 and the supplementary materials in Appendix \S \ref{appendix: case by case} has figures with a summary of all the disk galaxies in the sample.

\begin{table*}
\centering
\caption{Measured parameters of our kinematic and photometric modelling with the corresponding uncertainties from the MCMC resampling. The inclination has been constrained from the axis ratios in the fit to the observed data and allowed to vary $\pm3^\circ$ in the resampling strategy to account for expected uncertainties in this parameter, so the errors are expected to be of that order. We show the classification (class) of each galaxy as Rotating Disk (RD) or Irregular (Irr). The last two columns correspond to the central concentration (con) as a proxy for $B/T$ and the number of detected clumps $N_\mathrm{clumps}$.}
	\label{tab: results}
\begin{tabular}{lcccccccccccc}
\toprule
Galaxy & Class & $\theta_\mathrm{PA}$ & $i$ & $r_\mathrm{eff}$ & $r_{\textrm{flat}}$ & $v_{\textrm{flat}}$ & $\sigma$(AO) & $\log_{10}j_\star$ & Con & $N_\mathrm{clumps}$\\
ID & &($^\circ$) & ($^\circ$) & (kpc) & (kpc) & (km/s) & (km/s) & (kpc km/s) &  &\\
\hline
Q1623-BX455   & RD  & $322.9 \pm17.6$ & $52.3 \pm2.5$ & 2.1 & $0.4 \pm0.5$ & $100.8 \pm43.8$ & $99.0  \pm22.1$  & $2.58 \pm 0.17$ & 0.26 & 0 \\
Q1623-BX543   & Irr & $272.6 \pm17.3$ & $52.5 \pm2.8$ & 3.3 & $0.2 \pm0.3$ & $82.7  \pm32.3$ & $154.0 \pm30.5$  & $2.60 \pm 0.18$ & 0.32 & 1 \\
Q1623-BX599   & Irr & $75.3 \pm31.4$  & $48.3 \pm1.0$ & 2.4 & $3.0 \pm2.2$ & $173.8 \pm24.5$ & $181.3 \pm21.2$  & $2.88 \pm 0.15$ & 0.23 & 1 \\
Q2343-BX389   & RD  & $50.2 \pm13.6$  & $75.5 \pm3.4$ & 6.2 & $2.2 \pm1.7$ & $263.0 \pm24$   & $129.2 \pm26.8$  & $3.25 \pm 0.17$ & 0.05 & 4 \\
Q2343-BX513   & Irr & $239.2 \pm33.4$ & $46.7 \pm2.9$ & 2.6 & $0.1 \pm1.6$ & $52.8  \pm38.6$ & $92.0  \pm19.2$  & $2.04 \pm 0.12$ & 0.40 & 0 \\
Q2343-BX610   & RD  & $85.1 \pm25.9$  & $56.2 \pm3.3$ & 4.5 & $1.9 \pm0.9$ & $200.2 \pm34.7$ & $161.1 \pm27.1$  & $3.02 \pm 0.21$ & 0.08 & 4 \\
Q2346-BX482   & RD  & $195.2 \pm22.1$ & $59.1 \pm3.5$ & 6.0 & $2.0 \pm0.9$ & $210.3 \pm47.6$ & $115.1 \pm33.7$  & $3.10 \pm 0.19$ & 0.03 & 4 \\
Deep3a-6004   & RD  & $252.8 \pm35.6$ & $33.8 \pm3.6$ & 5.1 & $0.9 \pm0.4$ & $235.4 \pm43.2$ & $86.8  \pm31.6$  & $3.10 \pm 0.26$ & 0.16 & 2 \\
Deep3a-6397   & RD  & $20.7 \pm0.2$   & $52.5 \pm2.6$ & 5.9 & $0.9 \pm0.6$ & $197.5 \pm40.5$ & $144.0 \pm27.4$  & $3.07 \pm 0.49$ & 0.01 & 2 \\
Deep3a-15504  & RD  & $232.0 \pm28.2$ & $45.4 \pm3.6$ & 6.0 & $1.0 \pm0.5$ & $231.7 \pm47.4$ & $192.5 \pm43.6$  & $3.19 \pm 0.25$ & 0.04 & 6 \\
K20-ID6       & Irr & $187.2 \pm25.6$ & $34.4 \pm1.9$ & 3.9 & $1.3 \pm0.6$ & $196.4 \pm42.5$ & $137.8 \pm35.4$  & $3.11 \pm 0.21$ & 0.05 & 7 \\
K20-ID7       & RD  & $302.9 \pm25.9$ & $62.0 \pm3.2$ & 8.4 & $2.5 \pm1.8$ & $239.7 \pm41.6$ & $95.6  \pm22.2$  & $3.35 \pm 0.24$ & 0.01 & 6 \\
GMASS-2303    & Irr & $232.1 \pm30.4$ & $48.8 \pm3.4$ & 1.6 & $0.6\pm0.6$ & $113.1 \pm41.2$ & $83.1  \pm12.9$  & $2.62 \pm 0.26$ & 0.22 & 0 \\
GMASS-2363    & RD  & $322.4 \pm24.7$ & $60.1 \pm1.0$ & 2.3 & $0.5 \pm0.2$ & $148.2 \pm18.2$  & $63.6  \pm6.8$  & $2.82 \pm 0.22$ & 0.21 & 1 \\
GMASS-2540    & Irr & $139.8 \pm2.9$  & $31.0 \pm0.5$ & 8.5 & $0.3 \pm0.3$ & $102.0 \pm17.0$  & $62.6  \pm9.3$   & $2.98 \pm 0.31$ & 0.01 & 8 \\
SA12-6339     & Irr & $227.2 \pm26.0$ & $49.9 \pm1.4$ & 1.2 & $2.5 \pm0.9$ & $50.5  \pm18.0$  & $118.8 \pm12.0$  & $1.61 \pm 0.21$ & 0.87 & 0 \\
ZC400528      & Irr  & $0.4 \pm  15.4$ & $42.3 \pm2.0$ & 2.4 & $0.6 \pm0.3$ & $158.5 \pm36.6$ & $136.9 \pm26.5$  & $3.00 \pm 0.27$ & 0.35 & 0 \\
ZC400569      & RD  & $96.7 \pm 49.8$ & $50.7 \pm22.0$& 7.4 & $2.1 \pm1.1$ & $300.8 \pm50.8$ & $99.5  \pm22.3$  & $3.39 \pm 0.26$ & 0.25 & 3 \\
ZC401925      & Irr & $166.8 \pm19.1$ & $67.9 \pm3.0$ & 2.6 & $2.6 \pm2.1$ & $145.0 \pm23.7$  & $77.9  \pm3.9$   & $2.85 \pm 0.33$ & 0.19 & 2 \\
ZC403741      & RD  & $303.2 \pm47.4$ & $46.7 \pm3.4$ & 2.2 & $0.4 \pm0.2$ & $111.4 \pm54.4$ & $80.5  \pm31.2$  & $2.46 \pm 0.18$ & 0.16 & 1 \\
ZC404221      & Irr & $149.4 \pm25$   & $64.9 \pm1.0$ & 0.8 & $2.0 \pm1.4$ & $63.7  \pm30.7$ & $108.4 \pm30.1$  & $2.52 \pm 0.19$ & 0.53 & 0 \\
ZC405226      & RD  & $237.9 \pm23.5$ & $51.8 \pm3.3$ & 5.4 & $0.8 \pm0.4$ & $122.8 \pm31.2$ & $58.4  \pm14.0$  & $2.91 \pm 0.30$ & 0.04 & 5 \\
ZC405501      & RD  & $103.1 \pm23.6$ & $73.0 \pm3.1$ & 5.8 & $1.0 \pm0.6$ & $87.0  \pm39.8$ & $79.0  \pm19.9$  & $2.79 \pm 0.14$ & 0.09 & 6 \\
ZC406690      & RD  & $25.6  \pm14.9$ & $50.2 \pm3.6$ & 7.0 & $0.2 \pm0.8$ & $153.9 \pm56.7$ & $71.8  \pm13.5$  & $3.05 \pm 0.15$ & 0.01 & 6 \\
ZC407302      & RD  & $141.8 \pm27.4$ & $62.7 \pm3.2$ & 3.6 & $2.1 \pm1.3$ & $247.0 \pm29.0$ & $115.7 \pm26.6$  & $3.18 \pm 0.28$ & 0.11 & 1 \\
ZC407376      & Irr & $292.5 \pm29.1$ & $35.7 \pm1.9$ & 5.5 & $3.0 \pm2.4$ & $229.7 \pm35.4$ & $164.8 \pm30.0$  & $3.04 \pm 0.28$ & 0.31 & 2 \\
ZC409985      & Irr & $41.8  \pm23.7$ & $46.7 \pm2.9$ & 1.9 & $3.6 \pm2.8$ & $80.8  \pm22.8$ & $80.2  \pm16.6$  & $2.50 \pm 0.17$ & 0.28 & 1 \\
ZC410041      & RD  & $209.4 \pm17.7$ & $81.6 \pm2.9$ & 4.7 & $2.5 \pm2.1$ & $111.0 \pm43.3$ & $82.1  \pm18.8$  & $2.92 \pm 0.09$ & 0.12 & 6 \\
ZC410123      & Irr & $107.3 \pm17.2$ & $73.7 \pm2.8$ & 3.2 & $0.1 \pm0.6$ & $109.4 \pm38.8$ & $65.4  \pm24.3$  & $2.84 \pm 0.12$ & 0.16 & 3 \\
ZC411737      & Irr & $17.0  \pm16.6$ & $33.1 \pm2.7$ & 1.8 & $0.4 \pm0.3$ & $122.2 \pm46.7$ & $73.4  \pm18.0$  & $2.65 \pm 0.19$ & 0.21 & 0 \\
ZC412369      & RD  & $197.9 \pm20.9$ & $67.0 \pm2.5$ & 3.1 & $2.6 \pm2.1$ & $94.2  \pm40.2$ & $77.4  \pm21.1$  & $2.80 \pm 0.17$ & 0.17 & 4 \\
ZC413507      & Irr & $41.7  \pm22.1$ & $52.7 \pm2.0$ & 2.6 & $0.4 \pm0.2$ & $119.9 \pm54.8$ & $93.9  \pm40.8$  & $2.70 \pm 0.14$ & 0.12 & 1 \\
ZC413597      & Irr & $189.8 \pm44.3$ & $67.8 \pm2.8$ & 1.6 & $1.8 \pm1.2$ & $50.5  \pm47.9$ & $97.9  \pm33.9$  & $2.47 \pm 0.15$ & 0.35 & 1 \\
ZC415876      & RD  & $233.5 \pm16.6$ & $42.6 \pm2.4$ & 2.4 & $0.8 \pm0.5$ & $108.7 \pm42.6$ & $84.3  \pm19.8$  & $2.67 \pm 0.15$ & 0.18 & 0 \\
COSMOS-110446 & RD  & $63.4  \pm23.4$ & $49.0 \pm2.7$ & 2.4 & $1.7 \pm0.9$ & $69.0  \pm20.9$ & $22.1  \pm4.9$   & $2.60 \pm 0.19$ & 0.14 & 3 \\
COSMOS-171407 & RD  & $275.8 \pm28.0$ & $51.0 \pm2.4$ & 4.0 & $3.6 \pm2.2$ & $181.6 \pm41.9$ & $81.8  \pm13.3$  & $3.10 \pm 0.22$ & 0.03 & 3 \\
COSMOS-130477 & RD  & $266.5 \pm24.2$ & $40.0 \pm2.1$ & 4.0 & $1.2 \pm0.6$ & $219.3 \pm32.9$ & $56.3  \pm11.1$  & $3.31 \pm 0.22$ & 0.02 & 2 \\
COSMOS-127977 & RD  & $137.8 \pm23.4$ & $65.1 \pm3.2$ & 3.7 & $3.2 \pm2.0$ & $242.5 \pm20.4$ & $82.0  \pm22.4$  & $3.20 \pm 0.14$ & 0.01 & 3 \\
UDS-78317     & Irr & $143.7 \pm25.7$ & $37.2 \pm2.1$ & 3.0 & $0.9 \pm0.7$ & $129   \pm50.1$ & $105.3 \pm37.7$  & $3.16 \pm 0.25$ & 0.02 & 3 \\
UDS-124101    & RD  & $185.2 \pm21.4$ & $50.3 \pm3.5$ & 4.5 & $2.2 \pm1.1$ & $242.4 \pm31.2$ & $106.0 \pm31.1$  & $2.97 \pm 0.22$ & 0.04 & 0 \\
COSMOS-128904 & RD  & $250.6 \pm15.9$ & $67.5 \pm2.1$ & 4.9 & $5.1 \pm2.9$ & $112   \pm38.7$ & $93.0  \pm30.4$  & $2.94 \pm 0.09$ & 0.03 & 0 \\
\bottomrule
\end{tabular}
\end{table*}

\section{Results}
\label{section: Results}

In this section, we describe the results of our measurements, including the morphological classification, best-fit kinematic parameters, specific angular momentum, bulge-to-total ratios, and galaxy clumps. In particular, we focus our discussion on the scaling of $j_\star$ \textit{vs} $M_\star$ (Fall relation) in our sample of disks and discuss the differences between our findings and other high-redshift studies.

The main results of the kinematic fits and other quantities relevant to the discussion are summarized in Table \ref{tab: results}. We note that our strategy to measure the velocity fields, the multi-resolution kinematic modelling, and the choice of rotation curve differ from previous studies of the SINS galaxies (e.g., \citealp{Forster_2018}; \citealp{Genzel_2020}; \citealp{Shachar_2022}) as well as the KGES $z\sim1.5$ galaxies (\citealp{Gillman}), so there are some expected differences in the results. However, there is some general agreement with those studies for the majority of the galaxies. As a noteworthy example, our measurements of the key quantity $v_\mathrm{flat}$ is on average within 30\% from the characteristic rotational velocity $v_\mathrm{rot}$ measured in \cite{Forster_2018} for the SINS galaxies along the major kinematic axis, which is within the typical uncertainties of the fit. 

To estimate the uncertainties in the individual parameters shown in Table \ref{tab: results} ($r_\mathrm{flat}$, $v_\mathrm{flat}$, $\theta_\mathrm{PA}$, $i$), we performed a Monte Carlo resampling. To do this, we resampled $10^4$ times the observational errors associated with the extracted velocity field at each observed spaxel and added them to the best-fit model. In each iteration, we used our kinematic modelling and measured $\log_{10} j_\star$ along with the corresponding best-fit parameters (see details of this strategy in \citetalias{Espejo_Salcedo_2022}). The errors in the parameters are then taken as the standard error of the mean of those resampled values. The inclination is fixed from the axis ratios in the fit to the data, but to account for some of the expected uncertainties in this parameter, we allowed it to vary $\pm 3^\circ$ in the resampling approach.


\begin{figure*}
	\includegraphics[width=0.75\textwidth]{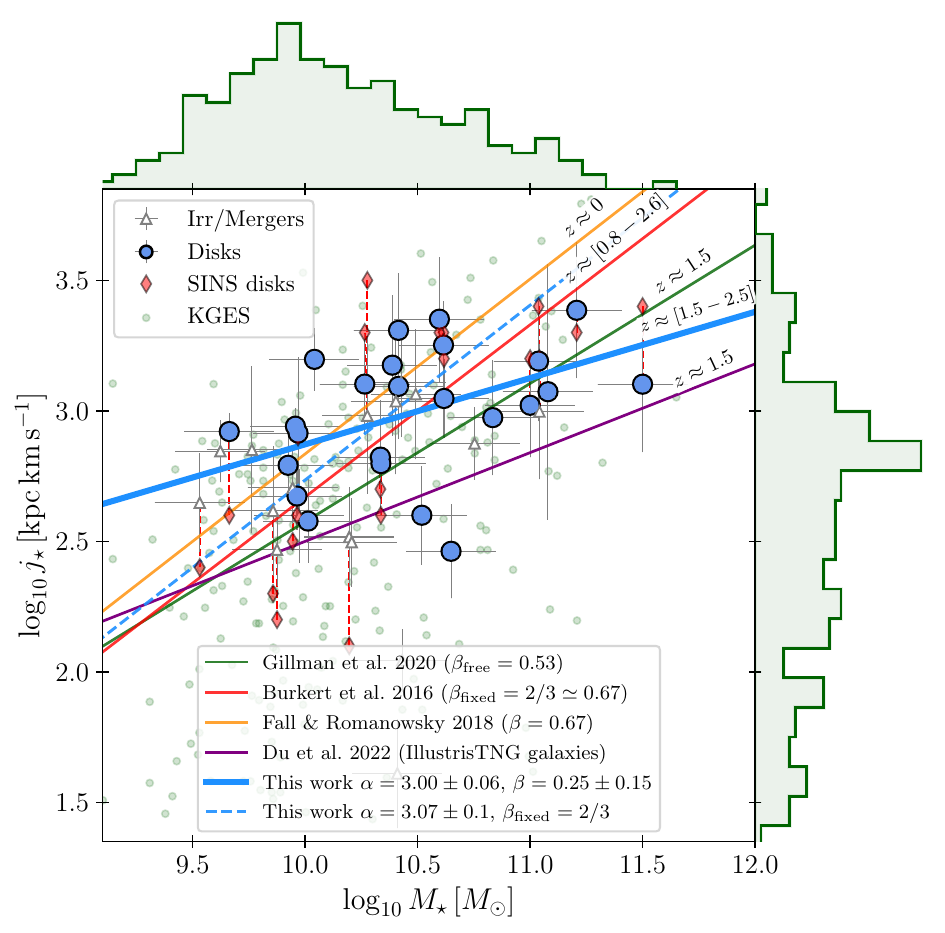}
    \caption{Specific angular momentum $j_\star$ \textit{vs} stellar mass $M_\star$ ``Fall relation'' where the measurements of $j_\star$ come from the radially integrated method using Equation \ref{eq:j definition}. The large blue dots correspond to the disk galaxies, which yield a \texttt{hyper-fit} solution with slope $\beta=0.25\pm0.15$ and indicated by the thick solid blue line. The normalization of a fit with a fixed slope $\beta=2/3$ is $\alpha=3.07\pm0.09$ and is indicated by the dashed blue line. The red line is the estimated fit by \protect\cite{Burkert} where they used 360 galaxies at $0.8<z<2.6$ with a fixed slope $\beta=2/3$. The 17 red diamonds are the \protect\cite{Burkert} $j_\star$ measurements that overlap with this sample (connected by the red vertical dashed lines) for which the \texttt{hyper-fit} free-slope solution is $\beta=0.74\pm0.16$ and for which our measurements yield a slope of $\beta=0.3\pm0.12$ (not shown in the plot to avoid overpopulating it and discussed in \S \ref{Effect of method}). To visualize the expected distribution of galaxies in $M_\star$ and $j_\star$ for typical star-forming galaxies, we show the histograms for the KGES $z\sim 1.5$ sample \protect\citep{Gillman} shown by the green dots and its corresponding fit $\beta=0.53\pm 0.1$. The \protect\cite{Burkert} and \protect\cite{Gillman} measurements were made using $\tilde{j}_\star\approx k_nv_sr_\mathrm{eff}$. We also include the scaling measured by \protect\cite{Fall_Romanowsky_2018} at $z\approx 0$ indicated by the orange line and the fit by \protect\cite{Du_2022} for IllustrisTNG galaxies at $z\approx1.5$.}
    \label{fig: Fall relation}
\end{figure*}

\begin{table*}
	\centering
	\caption{Best fit parameters of the Fall relation for different high-redshift observational studies in the form $\log(j_\star)=\alpha + \beta (\log(M_\star/M_\odot)-\gamma)$. We chose a value of $\gamma=10.5$ as it is the most commonly used in other studies and re-scaled $\gamma$ in different studies to match this value and make direct comparisons. $N$ corresponds to the number of galaxies used in each case with RD = Rotating Disk and Irr = Irregular, respectively. The bold line corresponds to our main results from the sample of 24 disks using the AO and NS data and with the integrated measurement of $j_\star$.}
	\label{tab: results_j_vs_M}
	\begin{tabular}{ccccccccc}
		\hline
		Fit type & Work & Redshift range & $N$ & Type & Method & $\beta$ & $\alpha$ & RMS\\
		\hline
		  & \textbf{This work} & \textbf{[1.45-2.45]} & \textbf{24 (RD)} & \textbf{IFU (AO+NS)}  & \textbf{Equation \ref{eq:j definition}} & $\mathbf{0.25\pm0.15}$ & $\mathbf{3.00\pm0.06}$ & $\mathbf{0.23}$\\
		  & This work          & [1.45-2.45] & 41 (RD+Irr) & IFU (AO+NS)   & Equation \ref{eq:j definition}                & $0.48\pm0.21$ & $2.93\pm0.09$ & 0.33\\

            & This work          & [1.45-2.45] & 24 (RD)     & IFU (AO)      & 2D (Eq. \ref{eq:j pixelwise})                & $0.29\pm0.23$ & $2.9\pm0.13$ & 0.38\\
            & This work          & [1.45-2.45] & 41 (RD+Irr) & IFU (AO+NS)   & 2D (Eq. \ref{eq:j pixelwise})                & $0.63\pm0.26$ & $2.65\pm0.11$ & 0.39\\
    
		  & This work          & [1.45-2.45] & 24 (RD)     & IFU (AO+NS)   & $\tilde{j}_\star\approx k_nv_sr_\mathrm{eff}$ & $0.61\pm0.21$ & $2.95\pm0.09$ & 0.33\\
		& This work          & [1.45-2.45] & 41 (RD+Irr) & IFU (AO+NS)   & $\tilde{j}_\star\approx k_nv_sr_\mathrm{eff}$ & $0.77\pm0.23$ & $2.88\pm0.10$ & 0.37\\
Free slope  & This work          & [1.45-2.45] & 24 (RD)     & IFU (NS only) & Equation \ref{eq:j definition}                & $0.3\pm0.18$ & $2.92\pm0.06$ & 0.29\\
$\beta$     & This work          & [1.45-2.45] & 24 (RD)     & IFU (AO only) & Equation \ref{eq:j definition}                & $0.24\pm0.14$ & $3.00\pm0.05$ & 0.23\\
 & \cite{Harrison_2017}          & [0.6-1.0]   & 586         & IFU (NS)      & $\tilde{j}_\star\approx k_nv_sr_\mathrm{eff}$ & $0.6\pm0.20$  & $2.83\pm0.04$ & -- \\
		& \cite{Alcorn_2018} & [2-2.5]     & 25      & Slit spectroscopy & $\tilde{j}_\star\approx k_nv_sr_\mathrm{eff}$ & $0.39\pm0.11$ & $2.99\pm0.05$ & 0.56\\
	& \cite{Gillman_Hizels}  & [0.8-3.3]   & 34          & IFU (AO)      & $\tilde{j}_\star\approx k_nv_sr_\mathrm{eff}$ & $0.56\pm0.03$ & $2.63\pm0.05$ & -- \\
	   & \cite{Tiley_2021}    & [1.25-1.75] & 288         & IFU (NS)      & $\tilde{j}_\star\approx k_nv_sr_\mathrm{eff}$ & $0.75\pm0.11$ & $2.98\pm0.04$ & -- \\
		& \cite{Gillman}     & [1.25-1.75] & 288         & IFU (NS)      & $\tilde{j}_\star\approx k_nv_sr_\mathrm{eff}$ & $0.53\pm0.10$ & $2.84\pm0.04$ & 0.56 \\
		\hline
		& This work          & [1.45-2.45] & 24          & IFU (AO+NS)   & Equation \ref{eq:j definition}                & $2/3$         & $2.97\pm0.09$ & 0.32 \\
		& \cite{Burkert}     & [0.8-2.6]   & 360         & IFU (NS/AO)   & $\tilde{j}_\star\approx k_nv_sr_\mathrm{eff}$ & $2/3$         & $3.00\pm0.10$ & -- \\
$\beta=2/3$ & \cite{Alcorn_2018} & [2-2.5]     & 25      & Slit spectroscopy & $\tilde{j}_\star\approx k_nv_sr_\mathrm{eff}$ & $2/3$         & $3.06\pm0.06$ & 0.56 \\
		& \cite{Tiley_2021}  & [1.25-1.75] & 288         & IFU (NS)      & $\tilde{j}_\star\approx k_nv_sr_\mathrm{eff}$ & $2/3$         & $2.94\pm0.03$ & --\\
		& \cite{Gillman}     & [1.25-1.75] & 288         & IFU (NS)      & $\tilde{j}_\star\approx k_nv_sr_\mathrm{eff}$ & $2/3$         & $2.86\pm0.03$ & 0.56 \\
		\hline
	\end{tabular}
\end{table*}

\subsection{Classification as rotating disks}
\label{subsection: Kinematic state of the galaxies}

Several high-redshift ($z \geq 1$) IFS studies have found that the fraction of disks ($f_\mathrm{disk}$) ranges between $\sim$40\% to $\sim$80\% between $1<z<3$ (e.g., \citealp{Epinat_2012}; \citealp{KMOS3D}; \citealp{Stott_2016}; \citealp{Forster_2018}; \citealp{Wisnioski_2019}; \citealp{Tiley_2021}). Different factors contribute to the vast range of disk fractions from the different studies. The mass dependence plays an important role since for a fixed redshift, probing different mass ranges leads to very different $f_\mathrm{disk}$ (e.g., \citealp{Kassin_disk_settling}; \citealp{Simons2019}; \citealp{Wisnioski_2019}). Additionally, the low spatial resolution in the majority of these studies leads to important uncertainties in the classification, a known limitation in high-redshift disk galaxy studies (e.g., \citealp{Rizzo_2022}). Hence, the high spatial resolution in this sample can be used for an independent test of $f_\mathrm{disk}$. We note that this has already been done for the 34 galaxies in the SINS sample in different studies (e.g., \citealp{Forster_2008_no_AO}; \citealp{Mancini_2011}; \citealp{Tacchella_sins_sizes}; \citealp{Forster_2018}; \citealp{Genzel_2020}).

Moreover, to measure $j_\star$ using the cylindrical symmetry assumption under which Equation \ref{eq:j definition} holds, we need to first distinguish the disk galaxies from those with irregular shapes using their kinematics and photometry. We used an approach based on the kinematic and photometric maps where a galaxy is classified as a disk when:

\textbf{1)} it is rotationally supported ($v/\sigma>1$, following \citealp{Genzel_2006}),

\textbf{2)} the kinematic axis (defined by a monotonic velocity gradient) is co-aligned with the morphological axis,

\textbf{3)} the centres are consistent among the photometric and kinematic maps, and

\textbf{4)} the velocity dispersion peaks at the kinematic centre and where the kinematic model yields small residuals.

This approach is suitable for the limited number of resolution elements in the different two-dimensional maps and is similar to the classification strategy employed for the KMOS$^\mathrm{3D}$ survey, which uses a set of five criteria (Section 4.1 in \hbox{\citealp{KMOS3D}}). We identified 24 systems as rotating disks or ``RD'', representing a disk fraction of $f_\mathrm{disk}\sim 58.6$\% with a binomial proportion error of 7.7\%. The remaining 17 systems (41\%) were simply labelled as ``Irregulars'' (Irr). Using only the low-resolution NS data for this classification would have identified 30 disks (73\%), reducing the quality and trustworthiness of the sample used for the measurements of $j_\star$. It is important to mention that we did not attempt to distinguish mergers from irregulars since the focus here is on the systems where we could find a reliable estimate of the angular momentum content. 

There is general agreement with previous disk classifications. \hbox{\cite{KMOS3D}} found a disk fraction of $\sim 70\%$ form the NS-only classification of over 600 KMOS$^{\mathrm{3D}}$ galaxies. For a more direct comparison, we note that the galaxies classified as rotating disks in \cite{Genzel_2014} are also classified as disks in our sample, except for ZC410123 and GMASS-2540, which we classify as irregulars. The data for these galaxies yields low SNR in the extracted velocity and dispersion fields. Thus too few pixels are useful for assessing the kinematic state and the kinematic modelling with \texttt{CONDOR}. Similarly, the galaxies from the SINS sample that we classified as rotating disks also fulfil at least the first three of the \hbox{\cite{KMOS3D}} disk criteria\footnote{\textbf{1)} A smooth monotonic velocity gradient across the galaxy defining the kinematic axis, \textbf{2)} A central peak velocity dispersion distribution with a maximum at the position of steepest velocity gradient, defining the kinematic centre and \textbf{3)} Dominant rotational support, quantified by the $v/\sigma$ ratio.} in \cite{Forster_2018}.

The galaxies for which our classification differs from the classification in \cite{Forster_2018} are Q2343-BX389, K20-ID7, ZC400569, and ZC412369, which fulfil two or less of the criteria in their classification but are consistent with our classification of disks described above. It is important to note that the velocity and dispersion fields that we obtain with our independent Gaussian fit (see figures in Supplementary materials \S \ref{appendix: case by case}) can differ from those found in \cite{Forster_2018}. Finally, we note that \cite{Forster_2018}, \cite{Genzel_2020}, and \cite{Shachar_2022} classify K20-ID6 as a rotating disk, but the SNR in our measurement is too low for a trustworthy classification given our criteria for the measurement of $j_\star$, and because the kinematic modelling is sensitive to the low number of pixels.


In Table \ref{tab: results}, we show the results of our measurements, including the best-fit parameters of our modelling, the measurements of $j_\star$, central light concentrations and the number of identified clumps per galaxy.

\subsection{The Fall relation at \texorpdfstring{$z\sim2$}{z2}}
\label{subsection: Fall relation}

It has been well established that spiral galaxies in the local Universe exhibit a consistent relationship between $j_\star$ and $M_\star$ that follows the $j_\star\propto M_\star^\beta$ Fall relation, with a slope of approximately $\beta\approx2/3$ for fixed $B/T$ (e.g., \citealp{Romanowsky}; \citealp{Fall_Romanowsky_2018}; \citealp{Posti_2018}). On the other hand, elliptical galaxies have a comparable slope, but they exhibit a significant negative vertical offset and a larger scatter (e.g., \citealp{Romanowsky}; \citealp{Obreschkow_2014}), so the relationship is not as well established as it is for spirals. It is worth noting that different galaxy types have been found to deviate from the $\beta=2/3$ slope. For instance, a study by \cite{Butler_2017} on the baryonic mass $M_b = M_\star + M_{\textrm{H}\scriptstyle\mathrm{I}}$ and specific angular momentum $j_b$ of 14 dwarf galaxies found a relation with a steeper slope than 2/3, which they argue, is likely due to lower stellar-to-halo mass ratios at decreasing mass.

Significant changes in the Fall relation, indicating a weaker or stronger correlation between $j_\star$ and stellar mass, as a function of redshift, could reveal shifts in the dominant processes that drive galaxy growth and evolution over cosmic history (e.g., \citealp{Fall_1983}). Understanding such trends provides crucial insights into the interplay between internal galaxy processes and external environmental factors (e.g., \citealp{Dekel_Burkert}) as well as the transformative stages that galaxies undergo, such as periods of rapid star formation, quenching, and morphological transformation (e.g., \citealp{Kormendy_2004}). Furthermore, it provides insights into the contributions from disk instabilities, internal angular momentum redistribution, or secular processes, which could play a more significant role in shaping the kinematics of galaxies as they evolve (e.g., \citealp{Mo_galaxy_formation}; \citealp{Dutton_van_den_Bosch}).

For instance, a steeper slope at high redshift could indicate an increasing role of gas accretion and mergers (e.g., \citealp{Dekel_2009}; \citealp{Stewart_2017}), as such merger events are more prevalent at high redshift. Their cumulative effect may manifest as a steeper slope in the Fall relation for galaxies at those cosmic epochs. Conversely, a shallower slope could suggest that low-mass galaxies are more efficient at retaining angular momentum relative to their higher-mass counterparts, potentially due to less efficient angular momentum loss through outflows and feedback processes in lower-mass systems (e.g., \citealp{Governato_2007}; \citealp{Brook_2012}). This retention of angular momentum can result in higher $j_\star$ values for low-mass galaxies, thereby flattening the slope of the Fall relation. 

Given the limitations of sample sizes at high redshift, and the difficulties of measuring $j_\star$ with good quality, many of the existing studies that investigated the Fall relation at $z>1$ used the fixed slope of $\beta=2/3\approx0.67$ and focused mainly on the normalisation and scatter around that relation (e.g., \citealp{Burkert}; \citealp{Alcorn_2018}; \citealp{Tiley}; \citealp{Gillman}). Other studies allowed the power-law slope to vary in the fit but made use of the R\&F approximation of the form $\tilde{j}_\star\approx k_n v_s r_\mathrm{eff}$ proposed by \hbox{\cite{Romanowsky}} (e.g., \citealp{Contini_2016}; \citealp{Harrison_2017}; \citealp{Tadaki_2017}; \citealp{Swinbank}; \citealp{Alcorn_2018}; \citealp{Gillman_Hizels}; \citealp{Tiley}; \citealp{Gillman}). This useful approximation allows for an overall estimate of $j_\star$ when mass and velocity profiles cannot be measured with detail (e.g., NS studies), but it relies heavily on global quantities that are subject to large uncertainties in high-redshift galaxies with morphological complexities. In particular, the factors $k_n$ (spatial weighting factor as a function of the S\'ersic index) and $r_\mathrm{eff}$ are often obtained from a single-component S\'ersic fit to the brightness profile, so the presence of clumps can heavily affect the accuracy of these factors. Additionally, the factor $v_{\textrm{s}}$ is the rotation velocity evaluated at $2r_\mathrm{eff}$ and does not contain any information about the actual shape of the rotation profile of the galaxy.

We avoid the possible bias introduced by these two assumptions (which we will also test directly in \S \ref{Effect of data type and outliers}) by doing a radially integrated measurement of $j_\star$ using Equation \ref{eq:j definition} with multi-resolution kinematic modelling, and by allowing the slope to be a free parameter in the fit to the Fall relation. In our fitting strategy, we parametrize the Fall relation with

\begin{equation}
 \log(j_\star)=\alpha + \beta (\log(M_\star/M_\odot)-\gamma),
 \label{eq:j fit equation}
\end{equation}
where $\beta$ is the power-law slope in $j_\star\propto M_\star^\beta$. The constant $\gamma$ is a scaling parameter for the stellar mass, and $\alpha$ determines the normalisation. For $\gamma$, we chose the mean value of $\log_{10}(M_\star)$ which in the sample is $\gamma=10.5$, to minimise the correlation between $\alpha$ and $\beta$. We then re-scaled the values of $\alpha$ from the literature for a direct comparison. 

We used the \texttt{hyper-fit} (\citealp{Hyperfit}) package in our fitting, which allowed us to take into account the errors in both $M_\star$ and $j_\star$ as well as the intrinsic scatter of the data. To estimate the optimal parameters and their uncertainties, we applied a Markov Chain Monte Carlo (MCMC) approach where the best-fit values $\alpha$ and $\beta$ are the mean values from the model realisations\footnote{Model realisation refers to the set of parameter values that the MCMC algorithm generates for each iteration.} and the uncertainties $\sigma_\alpha$ and $\sigma_\beta$ are calculated as the standard deviations of those realisations. In Table \ref{tab: results_j_vs_M}, we show the best-fit parameters to Equation \ref{eq:j fit equation} from our sample of disks as well as other high-redshift studies for comparison. 

\begin{figure*}
	\includegraphics[width=0.98\textwidth]{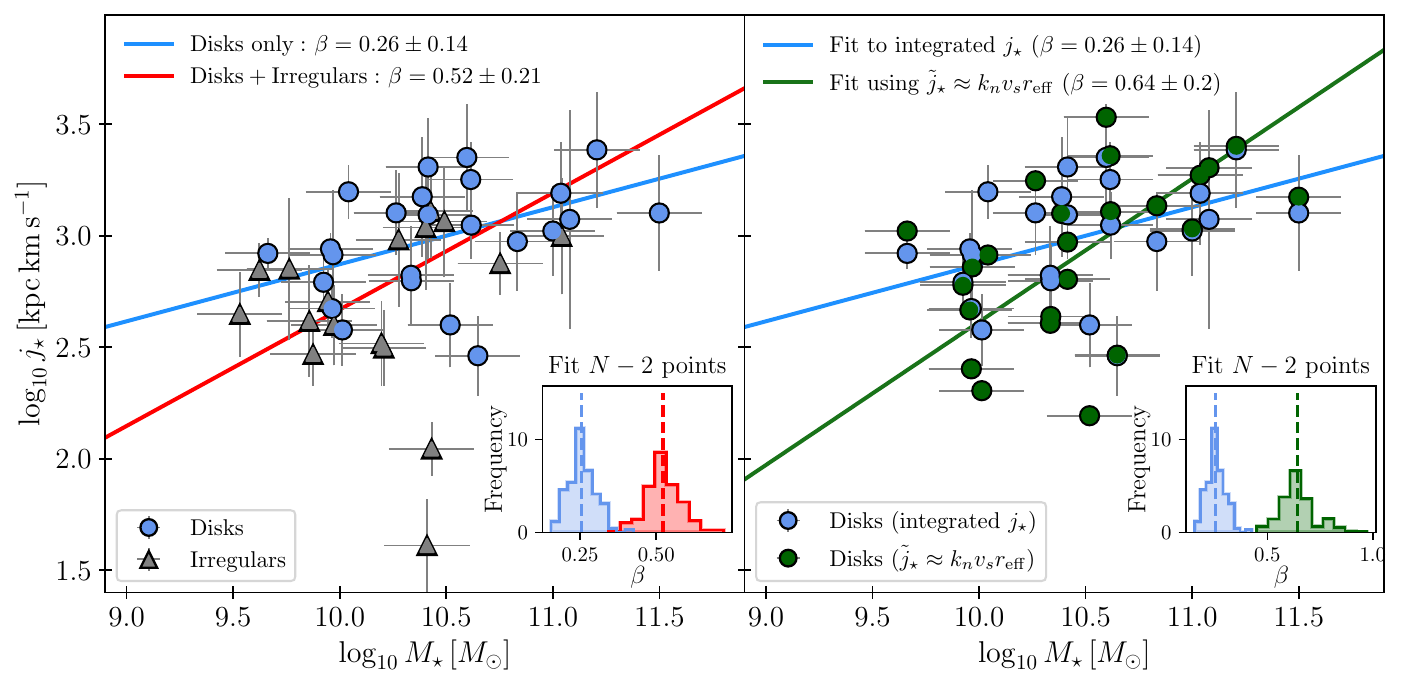}
    \caption{Comparisons between different power law slopes in the Fall relation for different analyses. The blue line on both panels indicates the best fit ($\beta=0.25\pm0.14$) using the disks with the integrated measurement of $j_\star$ (blue dots). \textbf{Left:} Red line indicates the fit using the disks+irregulars (irregulars are grey triangles) with $\beta=0.52\pm0.21$. The bottom right panel shows the distribution of slopes $\beta$ found from randomly removing two data points from the samples. \textbf{Right:} Green line indicates the fit using the values from the R\&F approximation $\tilde{j}_\star \approx k_n v_s r_\mathrm{eff}$ which yields $\beta=0.57\pm0.2$.}
    \label{fig: Fall relation comparisons}
\end{figure*}

Our focus is centred on our AO+NS combined analysis on the 24 disks (excluding irregulars/mergers) and for a free power-law slope, where we obtain a normalisation $\alpha=3.00\pm 0.06$ and a slope of $\beta=0.25\pm0.15$ with an intrinsic scatter of 0.14 dex. The total vertical root-mean-square (RMS) scatter, encompassing intrinsic, observational, and model-dependent errors, is measured at 0.23 dex, slightly larger but consistent with the RMS scatter of 0.2 dex found for local disks (e.g., \citealp{Romanowsky}; \citealp{Obreschkow_2014}; \citealp{Posti_2018}). A large scatter at high redshift compared to low redshift could point to the emergence of rotational support that assembles massive disks in the local Universe (\citealp{Kassin_disk_settling}), however the small difference inferred here is not significant enough to draw any conclusions.

More importantly, the $\beta\approx0.25$ slope is shallower than other high-redshift studies and, in particular, much shallower than the $\beta=2/3\approx0.67$ assumed by many of them (see Figure \ref{fig: Fall relation}), deviating by approximately 3 standard deviations (2.8$\sigma$). If we fit the data with a fixed power-law slope $\beta=2/3\simeq0.67$, we obtain a normalisation of $\alpha=3.07\pm0.09$ and RMS=$0.31$ dex, which is consistent with the other studies that use a fixed $\beta=2/3$. However, a visual inspection of the data in Figure \ref{fig: Fall relation} shows that the $\beta=2/3$ slope is inconsistent with the data. Table \ref{tab: results_j_vs_M} summarizes the best-fit results and a comparison to other studies that find a steeper slope $\beta$.

The reduced chi-squared for the free slope fit is $\chi^2_\nu \approx 1.86$, where $\nu = 22$ degrees of freedom (calculated as $N - k = 24 - 2$). The corresponding $p$-value for this fit is approximately 0.009, which suggests that while the fit does not perfectly represent the data, the fit is reasonable given the inherent large scatter in the data. In contrast, the fixed slope fit has a reduced chi-squared value of $\chi^2_\nu \approx 4.18$, with $\nu = 22$ degrees of freedom. The corresponding $p$-value is nearly 0 ($4.426 \times 10^{-10}$), indicating that the fixed slope model is statistically rejected. Therefore, the free slope model offers a significantly better fit for the data than the fixed slope model.

This finding aligns with the results from \cite{Du_2022}, where a shallower slope of the Fall relation at high redshift was also observed compared to the local Universe. This study used galaxies from the IllustrisTNG simulation (\citealp{Pillepich_2018}; \citealp{Marinacci}; \citealp{Nelson_2019}; \citealp{Pillepich_2019}) to study the origin and evolution of the Fall relation in disk galaxies by connecting the stellar properties to those of their parent haloes (see also \citealp{Posti_2018_2}). They find that the slope they obtain is shallow at high redshift with $\beta\approx 0.34$ at $z=1.5$ compared to $\beta\approx 0.55$ at $z=0$ (see Figure 1 in \citealp{Du_2022}), suggesting that the processes governing angular momentum acquisition and retention differ significantly between high redshift and the local Universe. They conclude that the growth of disk-like structures established the locally observed $j_\star$ \textit{vs} $M_\star$ relation with $\beta=2/3$ only at $z<1$, implying the need for a revised understanding of the origin of the Fall relation.


An upcoming complementary study of low-mass and dwarf galaxies in SHaDE (\citealp{Barat_2020}) and IllustrisTNG by Deeley et al. (in prep.) uses a spatially resolved approach (following \citealp{Sweet_2018, Sweet_2020}) to calculate the angular momentum of galaxies as small as $M_\star > 10^7 M_\odot$. They find that at $z = 0$, the overall slope is $\beta = 0.71$, with a steeper slope of $\beta = 0.84$ for dwarf galaxies. Similar to \cite{Du_2022}, they observe a shallower slope at high redshift, with an overall $\beta = 0.44$ at $z = 1.5$ and $\beta = 0.20$ for dwarfs. The evolution of the slope to the present day appears to be linked to a sudden increase in the angular momentum of galaxies settling into disks, a change more pronounced for rotationally supported dwarfs than for massive galaxies. In the next subsections, we investigate some of the potential drivers of the found slope of the Fall relation.

\subsection{Effect of data types, morphology and outliers}
\label{Effect of data type and outliers}

To understand the origin of the difference in slope from the $\beta=2/3$ value, we compare the fit using different datasets and different techniques in the measurement of $j_\star$. First, by using the $j_\star$ measurements from the AO-only and NS-only analysis, we found a shallow slope of $\beta=0.24\pm 0.14$ and $\beta=0.3\pm0.18$, respectively. Since this is consistent with the combined results, we conclude that the large difference in slope from $\beta=2/3$ is not driven by the spatial resolution used to measure $j_\star$.

Second, we calculate a fit using the full sample (i.e. disks + irregulars). In this case, we find $\beta=0.48\pm 0.21$ as indicated by the red line in the left panel of Figure \ref{fig: Fall relation comparisons} suggesting that including irregulars in the fit has the significant effect of making the slope steeper. This result is not surprising since irregular galaxies are expected to populate more in the low-mass regime where, following the Fall relation, the $j_\star$ content is low, which can affect the slope. Additionally, it shows that while the choice of AO \textit{vs} NS data does not significantly affect the Fall relation via the measurement of $j_\star$ directly, it does affect it indirectly via the disk/irregular+merger split for which AO resolution is vital.

To check this, we compare the mass and size properties of the disks and irregulars with their $j_\star$ content. Out of the 17 irregular galaxies in the sample, a significant majority (14 galaxies) have sizes below the mean, while only 3 are larger. Similarly, concerning mass, 15 of the irregular galaxies fall below the mean mass, with only 2 being above. The prevalent lower mass and size observed in these systems correspondingly result in a low content of specific angular momentum ($j_\star$). Specifically, 12 irregular galaxies are found to have values below the mean, while the remaining 5 have angular momentum above the mean value. This is not surprising, as the randomized kinematics in these irregular systems translate to $j_\star$ adding incoherently to the measured low values.

To test the influence of individual outliers in the fit, we randomly removed two points ($N-2$) from the sample $10^3$ times for both the disk-only and disks+irregular samples and re-did the fit. For the disk+irregulars sample, we found that the mean of the distribution is $\beta=0.52$ with a standard deviation of $\sigma_\beta = 0.056$. On the other hand, resampling the disks-only sample yields $\beta=0.26$ and $\sigma_\beta = 0.04$, as indicated by the histograms in the low right for each panel in Figure \ref{fig: Fall relation comparisons}. This experiment suggests that outliers are not the main driver of the large slope differences. 

Third, we make a fit to the Fall relation with the pixel-by-pixel 2D measurement ($j_\star(x,y)$) described in \S \ref{subsection: angular momentum} (Equation \ref{eq:j pixelwise}), from where we find a slope of $\beta=0.29\pm 0.29$ for the sample of disks and $\beta=0.63 \pm 0.26$ for the full sample. While the best-fit values align with the slopes found with the integrated measurement, the large uncertainties in the fit, the large RMS scatter ($0.41$ and $0.39$, respectively) and the limited spatial extent of the data used make these estimates unreliable.

Fourth, we compare the fit to the Fall relation using our radially integrated method (Equation \ref{eq:j definition}) to the fit using the approximation $\tilde{j}_\star \approx k_n v_s r_\mathrm{ef}$ using the S\'ersic indices from \cite{Tacchella_sins_sizes} and \cite{Gillman}. The slope we find using the $\tilde{j}_\star$ approximation is $\beta=0.61\pm0.2$, which is comparable to the other IFU-based high-redshift studies that also use the approximation as seen in the right panel in Figure \ref{fig: Fall relation comparisons}. By resampling the dataset (removing two data points as done in the previous experiment), we found that the distribution of fits to the $N-2$ points remains high with $\beta=0.64$ with $\sigma_\beta=0.058$, again suggesting that the slope difference is not driven only by a couple of points. Finally, if we use the measurements of $\tilde{j}_\star$ for the full sample (disks + irregulars), which would resemble the approach used in low-resolution studies that use the R\&F approximation, then the slope is $\beta=0.77\pm0.23$, which is consistent with the commonly used $2/3$, suggesting again both using approximate low-resolution methods and the disk classification can lead to significant biases in the slope of the Fall relation.

\subsection{The effect of clumps}
\label{effect of clumps}

Aside from external factors affecting the global properties of galaxies at cosmic noon, internal processes (that could be associated with the lack of angular momentum as discussed in \citealp{Obreschkow_2015}) lead to galaxy-wide instabilities and the formation of large star-forming clumps. The prominent presence of these structures in the galaxy sample motivates the question of whether these are correlated to the observed slope or scatter (above or below the mean relation) in the Fall relation when compared to the results for smooth disks in the local Universe. From the analysis of \S \ref{subsection: clumps}, we found that 31/41 galaxies have clumps with an average of 2.5 clumps per galaxy. We measured the percentage of galaxy light contained within the clumps and found an average of 12.7\% and a median of 10.6\%. 

\begin{figure*}
	\includegraphics[width=0.98\textwidth]{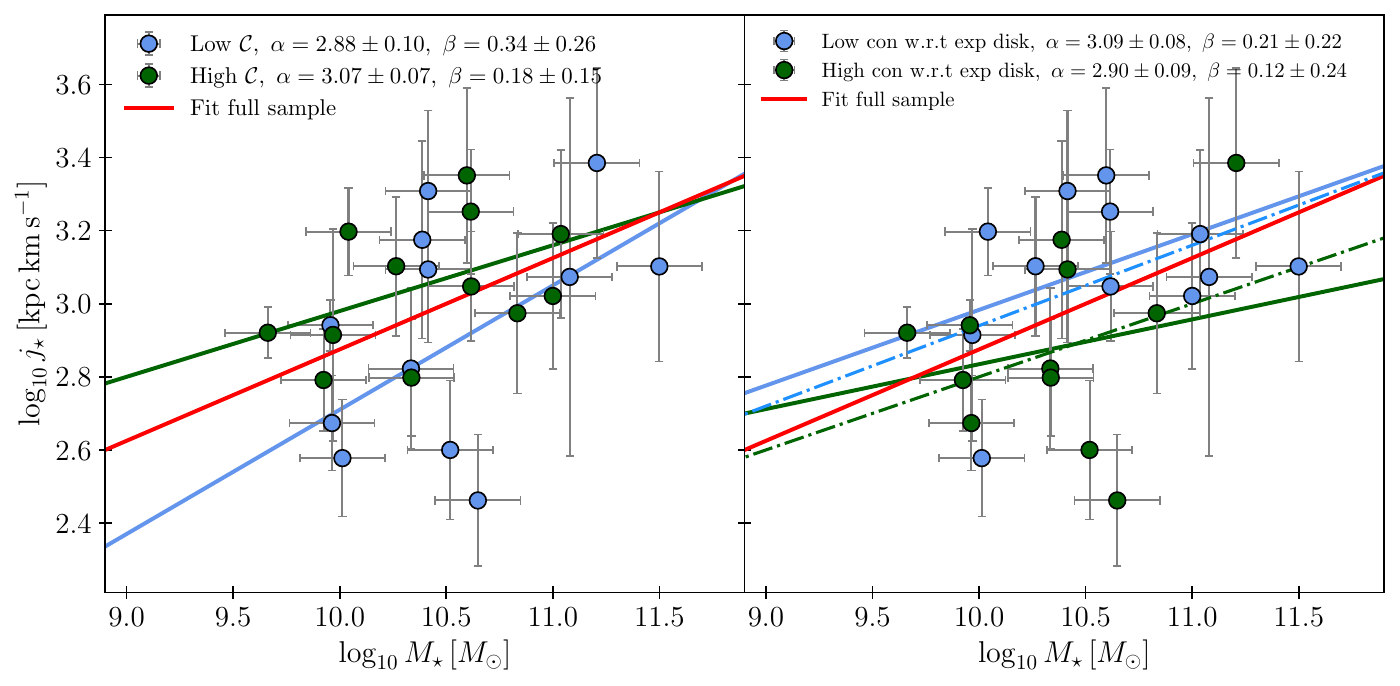}
        \caption{\textbf{Left:} Specific angular momentum $j_\star$ \textit{vs} stellar mass $M_\star$ ``Fall relation'' for the disk galaxies coloured depending on whether they are below (blue) or above (green) the median ``clumpiness'' $\mathcal{C}=11.72\%$ as defined in \S\ref{subsection: clumps}. The solid red line indicates the $j_\star\propto M_\star^\beta$ relation with $\beta\approx0.25$ found in the sample of disks. The green line indicates the fit using the galaxies above the median, while the blue line represents the fit to the sample below that value. \textbf{Right:} Similar as in the right but coloured as a function of whether the points are above or below the concentration of an exponential disk model in Figure \ref{fig:concentration_vs_sims}. The dots and fits in blue correspond to those below the exponential disk model, while the green dots and lines correspond to those above the model. The dashed-dotted lines correspond to the fit to the subsamples using the \textit{J}-band \textit{HST} imaging.}
    \label{fig: Fall relation vs clumps vs bulges}
\end{figure*}

To check if the presence of clumps has a systematic effect on the Fall relation, we broke the sample into two. One ``clumpy'' subsample corresponds to galaxies where $\mathcal{C}$ is above the median clumpiness $\mathcal{C}>11.72\%$, and the other subsample corresponds to those with $\mathcal{C}<11.72\%$. An independent fit to the Fall relation for both subsamples shows that the difference in slope is within the uncertainty of the fit, with $\beta=0.34\pm0.26$ for the low $\mathcal{C}$ subsample and $\beta=0.18\pm0.15$ for the high $\mathcal{C}$ subsample. In terms of the normalisation $\alpha$, the fit to the clumpy systems shows a negative vertical offset from the less clumpy sample of $\Delta \alpha \sim 0.2$ as indicated in Figure \ref{fig: Fall relation vs clumps vs bulges}. The small difference is barely outside the uncertainties in the fit (the errors in $\alpha$ in the low- and high-clumpiness cases are $\pm 0.1$ and $\pm 0.07$, respectively) but suggests that the high $j_\star$ galaxies in the sample are clumpier. This contradicts the expectation that a high content of $j_\star$ opposes the presence of instabilities that give rise to the formation of clumps, as discussed by \cite{Obreschkow_2015} who defined a disk-averaged Toomre parameter $\overline{Q}$ as $\overline{Q}\propto j_\star \sigma_0 M^{-1}$ where $\sigma_0$ is simply a dispersion scale. However, this could be explained by the fact that galaxies with a large amount of angular momentum are also larger ($j_\star \propto r$), where the identification of clumps is easier with the current method. Similarly, galaxies with clumps at large radii could bias the size to be large and thus translate to larger $j_\star$. This bias in the measurement could also have a systematic effect on the determination of the clumpiness for low-mass galaxies (which are small given the mass-size relation), so clumpiness in the low-mass regime is likely to be underestimated. This is discussed in more detail in \ref{subsection: clumps}.

To determine the dominant factor in driving disk instabilities and to verify the accuracy of the \cite{Obreschkow_2015} model, one could use ALMA or the Plateau de Bure interferometers to make molecular gas measurements of a large sample of galaxies that also have $j_\star$ measurements. However, given the size bias discussed in the previous paragraph, this comparison should be made for large galaxies where the measurement of clumpiness is more precise.

\begin{table}
	\centering
	\caption{Summary of the slopes found for different fits to the different samples and methods. The last two columns correspond to the results obtained from resampling each sample $10^3$ times (randomly removing two points), where $\overline\beta$ is the mean slope of the distribution and $\sigma_\beta$ is the standard deviation associated with it.}
	\label{tab:comparisons}
	\begin{tabular}{cccccc}
		\hline
		Data & Method & \multicolumn{2}{c}{Slope} & \multicolumn{2}{c}{Resampling} \\
		\cmidrule(lr){3-4}
		\cmidrule(lr){5-6}
		 &  & $\beta$ & $\Delta \beta$ & $\overline\beta$ & $\sigma_\beta$\\
		\hline
		 Disks & Equation \ref{eq:j definition}                    & $0.25$ & $0.15$ & 0.26 & 0.04\\
		 Disks + irregulars & Equation \ref{eq:j definition}       & $0.48$ & $0.21$ & 0.52 & 0.056\\
		 Disks & $\tilde{j}_\star \approx k_n v_s r_\mathrm{eff}$  & $0.61$ & $0.21$  & 0.64 & 0.058\\
		\hline
	\end{tabular}
\end{table}

\subsection{The effect of central concentration}
\label{Effect of bulge-to-total ratios}

The morphology of a galaxy is related to its baryonic mass and angular momentum. Understanding these relationships can provide insights into the merger histories and angular momentum transport (e.g., \citealp{Wang_2019}). In this theme, \cite{Obreschkow_2014} found a strong correlation between the baryonic mass, angular momentum, and bulge-to-total ratio ($B/T$) in disk galaxies. It is thus useful to address the effect of the central light concentration as a proxy for the $B/T$ in each galaxy in the sample, to investigate a possible correlation with stellar mass that could affect the slope and scatter of the Fall relation.

The effect of $B/T$ ratios in the Fall relation has been discussed in a recent study of 564 nearby galaxies in xGASS (eXtended GALEX Arecibo SDSS Survey), where they employed a similar integrated measurement of $j_\star$ through the combination of $\textrm{H}\scriptstyle\mathrm{I}$ velocity widths and stellar mass profiles (\citealp{Hardwick}). They found that for a fixed bulge-to-total ratio, the slope of the Fall relation is consistent with $\beta\approx 2/3$. However, when considering all galaxy types (varying $B/T$) the slope becomes significantly lower with $\beta\sim 0.47$. They argue that this is caused by the change in galaxy morphology as a function of mass and suggest that sample selection is critical when constraining galaxy formation models using the Fall relation.

To address the possible effect of $B/T$ in this sample, we used the measurements of central light concentration obtained in \S\ref{subsection: concentration}, with the caveat that the choice of method introduces a systematic bias due to the galaxy size as discussed in that section. For the whole sample, we obtained a low average concentration of $\sim 0.2$ (median $\sim 0.1$), which, if taken as a proxy for $B/T$, at low redshift is often used as the threshold between galaxies whose light can be modelled with a single component ($B/T<0.2$) and those that need a disk and a bulge component ($B/T>0.2$) (e.g., \citealp{Barsanti}; \citealp{Casura}).

Considering the experiment shown in Figure \ref{fig:concentration_vs_sims}, where the measured concentration is compared with the expected concentration of a pure exponential disk, we separated the sample into galaxies that lay above or below the exponential disk model, e.i., those that have large and low concentrations respectively. For these two subsamples, we made a fit to the Fall relation and found shallow slopes in both cases ($\beta = 0.12\pm0.24$ for galaxies with high concentrations and $\beta = 0.21\pm0.22$ for those with low concentrations), as shown in the right panel of Figure \ref{fig: Fall relation vs clumps vs bulges}, i.e., no notable effect on the slope. This is an example of Simpson's paradox (\citealp{Simpson_paradox}), as the slopes of both subsamples are lower than the slope of the whole sample, contrary to the expectation that they would bracket that value. These results cannot be naively interpreted as evidence of central light concentration playing a significant role in driving the slope of $j_\star \propto M_\star^\beta$ due to the large uncertainties. One visible trend in the data is that galaxies with large concentrations tend to correspond to low-$j_\star$ systems, while the opposite is true for galaxies with large concentrations. The difference in the normalization is modest with $\alpha=3.09\pm0.08$ for low concentration and $\alpha=2.9\pm0.09$ for high concentration.

\begin{figure}
	\includegraphics[width=0.48\textwidth]{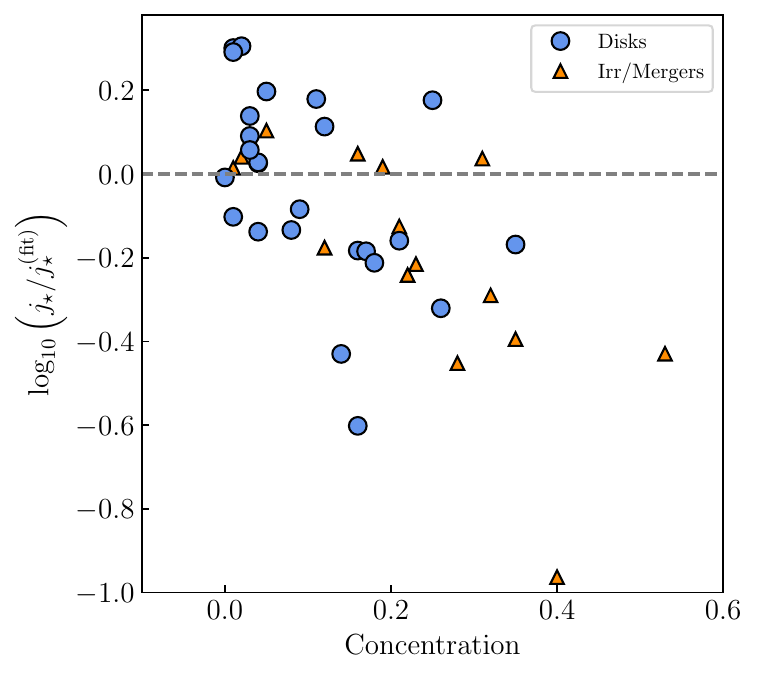}
        \caption{Residuals of the specific angular momentum $j_\star$ with respect to the fit to the Fall relation $j_\star^\mathrm{(fit)}$ \textit{vs} concentration showing the anticorrelation of both quantities with Spearman correlation $\rho_s=-0.6$ for disks and $\rho_s=-0.67$ for the full sample (disks+irregulars). To aid visualization, we removed SA12-6339 with $\log_{10}(j_\star/j_\star^\mathrm{fit})=-1.4$ since it is very compact ($r_\mathrm{eff}=1.2$ kpc) and irregular so the estimation of the central concentration ($\sim 0.78$) and the assumptions of well-ordered rotation and cylindrical symmetry used to calculate $j_\star$ do not hold for this system.}
    \label{fig: j vs bulges}
\end{figure}

For an alternative approach to assessing the effect of concentrations (without splitting the sample into those above or below the exponential disk model), we checked the residuals of $j_\star$ with respect to the fit to the Fall relation and found an anti-correlation of the measured concentration and $j_\star$ (Spearman correlation $\rho_s = -0.6$) shown in Figure \ref{fig: j vs bulges}. 

Both these results are consistent with low-redshift studies where galaxies with large $B/T$ have a negative vertical offset in the Fall relation compared to those with low $B/T$ (e.g., \citealp{Obreschkow_2014}; \citealp{Fall_Romanowsky_2018}; \citealp{Sweet_2018}). As a clear example, \cite{Romanowsky} showed that $z=0$ galaxies with $0.6< B/T <0.8$ can have 2.5-8 times less $j_\star$ than those with $B/T\approx 0$ (see Figure 2 in their study). This could be explained by different scenarios of bulge formation (such as merging and accretion events), which result in the decrease of angular momentum or its transfer from the disk to the bulge. This trend (anticorrelation between $j_\star$ and $B/T$) is unchanged when using the concentrations from the \textit{J}-band imaging instead of \textit{H}-band, with slopes $\beta = 0.11\pm0.17$ and $\beta = 0.22\pm0.18$ for the galaxies above and below the median $B/T$, respectively and normalisations $\alpha=3.13\pm0.08$ for high concentration and $\alpha=2.85\pm0.09$ and for low concentration.

\section{Discussion}
\label{section: Discussion}

\subsection{Comparison to different ways of estimating $j_\star$}
\label{Comparison to different ways of estimating j}

In this section, we discuss potential systematic effects associated with various methods for estimating $j_\star$ and their impact on measuring the slope in the $j_\star$ \textit{vs} $M_\star$ relation.

\subsubsection{The effect of the integrated radial method}
\label{Effect of method}

When looking at the direct comparison of our individual measurements of $j_\star$ for the disks that overlap with the SINS disks in \cite{Burkert} (red diamonds in Figure \ref{fig: Fall relation}), we find that the main difference is a general trend of higher $j_\star$ towards the low-mass end (and lower $j_\star$ towards high masses). Since the data used in their analysis for those galaxies was also the AO sample (when available), the resolution effects on the kinematics do not appear to be the dominant factor in driving the discrepancies. Instead, the differences in this analysis seem to be associated more closely with the disk/irregular classification as well as the method used to measure $j_\star$.

Regarding the disk classification, the low-mass galaxies with the lowest $j_\star$ in the \cite{Burkert} measurements (ZC404221, ZC413597 and GMASS-2303) are also some of the lowest $j_\star$ in our measurements ($j_\star < 10^{2.7}$ kpc km s$^{-1}$), but they are identified as irregular systems in our work and thus they are not included in the main fit which is a contributing factor to the different slope. On the other hand, the choice of method used to measure $j_\star$ could also be a contributor to the large discrepancies. In this section, we investigate the difference in the slopes that we get when adopting global properties (R\&F approximation $\tilde{j}_\star$) and when integrating the mass and velocity profiles radially using \ref{eq:j definition}.

\begin{figure}
	\includegraphics[width=0.48\textwidth]{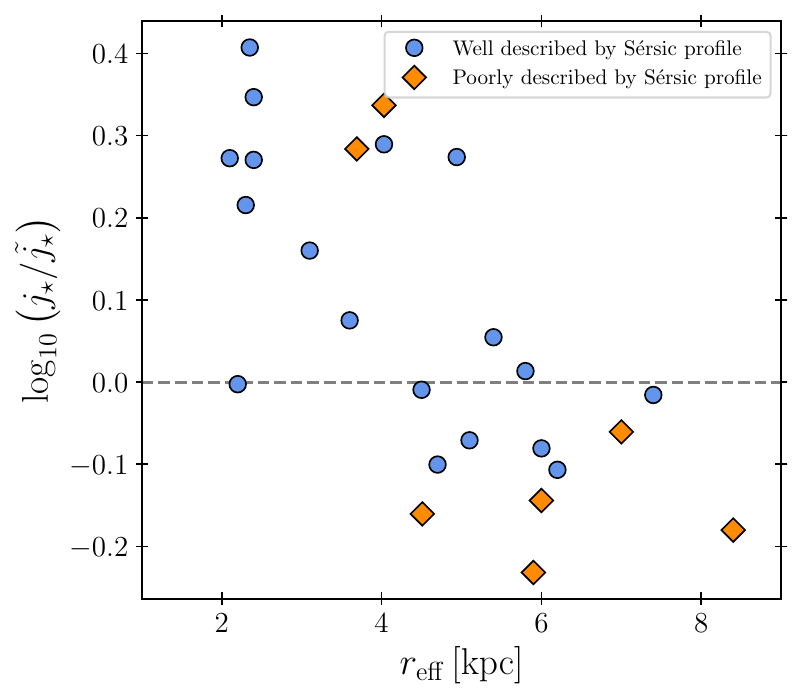}
    \caption{Comparison between the $j_\star$ measurements from the integrated method and from the R\&F approximation $\tilde{j}_\star \approx k_n v_s r_\mathrm{eff}$ as a function of the effective radius where the blue dots correspond to the galaxies where the light profile is well described by a single S\'ersic profile and the orange diamonds correspond to those where the light distribution is not well described by it.}
    \label{fig: comparison integrated vs approximation}
\end{figure}

To investigate the general effect of the choice of method, we initially categorized galaxies based on the quality of their azimuthally averaged light profiles, distinguishing between those well-described by a single S\'ersic profile and those poorly described. This was done through a visual inspection of the light profiles (see Figures in the supplementary Section \ref{appendix: case by case}). The galaxies with poorly described light profiles are Q2346-BX482, Deep3a-6397, K20-ID7, ZC406690, COSMOS-130477, COSMOS-127977 and UDS-124101. Then we compared the difference between the integrated measurement $j_\star$ and the R\&F approximation $\tilde{j}_\star$ as a function of effective radius as indicated in Figure \ref{fig: comparison integrated vs approximation} where it is evident that the galaxies with poorly described light profiles coincide with some of the galaxies with the largest discrepancies. Since the $\tilde{j}_\star \approx k_n v_s r_\mathrm{eff}$ approximation depends on the values of $k_n$ and $r_\mathrm{eff}$ from the S\'ersic profile fit, then it is clear that this can introduce a systematic uncertainty in the determination of $j_\star$ for galaxies with irregular morphologies such as those at cosmic noon.

Some notable exceptions to this trend are Q1623-BX455, ZC415876, COSMOS-110446, COSMOS-128904 and COSMOS-171407 (top left in Figure \ref{fig: comparison integrated vs approximation}) which also deviate significantly ($\log_{10}(j_\star/\tilde{j}_\star)>0.25$). However, three of these systems (Q1623-BX455, ZC415876, COSMOS-110446) are compact and have $r_\mathrm{eff}<2.5$ kpc, which is well below the average $r_\mathrm{eff}$ in the sample of $\approx 3.91$ kpc. This suggests a notable bias in the $\tilde{j}\star \approx k_n v_s r_\mathrm{eff}$ method relative to $r_\mathrm{eff}$, as discussed in \cite{Romanowsky}. In their work, they highlight the impact of the choice of radius for measuring $v_s$, noting that estimates become less accurate below 2$r_\mathrm{eff}$ (see Figure 8 and Appendix 4 in their work). Consequently, in compact galaxies where the velocity profile does not extend far, the approximation of $j_\star$ based on this method may be compromised. An examination of the full sample reveals a decreasing trend in the $j_\star/\tilde{j}_\star$ ratio with effective radius (Spearman correlation coefficient $\rho_s=-0.71$), indicating a systematic discrepancy between the two methods as a function of $r_\mathrm{eff}$.


\subsubsection{Integrated radial method \textit{vs} R\&F approximation using mock galaxies}
\label{Mock galaxies}

We performed an additional simple experiment to assess the discrepancy between methods used to calculate $j_\star$. To do this, we created a set of $10^4$ mock galaxies in the range $10^{9.5} < (M_\star/M_\odot) < 10^{11.5}$, mirroring the mass distribution of our real sample. We used widely accepted relations to calculate the effective radius and velocity to ensure that we use mock galaxies that resemble realistic systems. Specifically, the effective radius was estimated using a simple mass-size relation for local Universe late-type galaxies in the form $(r_\mathrm{eff}/\mathrm{kpc})\propto (M_\star/M_\odot)^{0.22}$ (\citealp{Van_der_Wel_2014}) and the velocity $v_\mathrm{flat}$ was set using the Tully-Fisher relation $(M_\star/M_\odot) = 50 \, (v_\mathrm{flat}/\mathrm{km\,s^{-1}}) ^4$ of \cite{McGaugh_2005}. Additionally, to allow for variations in galaxy morphology in this test, we generated a random S\'ersic index for each galaxy in the range $n\sim[1-10]$.

Overall, the mock galaxies resemble the observed sample well across key parameters. In terms of stellar mass, the 75th percentiles differ by less than 0.16 dex between the two samples. For effective radii and velocities, the 75th percentiles show differences of less than 20\%. The lower and upper bounds of the parameters are provided in Table \ref{tab:toy-model-params}.

The above parameters suffice for calculating $j_\star$ using the R\&F approximation. However, to compute $j_\star$ using the radial profiles, an additional parameter, $r_\mathrm{flat}$, is required. We note that there is no correlation between $r_\mathrm{flat}$ and $r_\mathrm{eff}$ in our sample, which is a similar conclusion found in \cite{Mocz_Glazebrook} for a sample of 25,698 late spiral-type galaxies in the SDSS survey, who found no significant correlation between the velocity turnover radius $r_t$ (analogous to the $r_\mathrm{flat}$ parameter) and the exponential scale radius which relates directly to $r_\mathrm{eff}$. An even more relevant example due to the similar redshift range ($0.6 < z < 2.6$) is the analysis by \cite{Lang_2017} where a large scatter but no discernible correlation was found between the kinematic turnover radius and the effective radius of a large sample of 101 resolved galaxies (see Figure 12 in their study for details). Therefore, for each galaxy, we generated a random $r_\mathrm{flat}$ within the distribution present in our real sample ($0.2 < r_\mathrm{flat} < 6$ kpc). Since there is no explicit dependence on $r_\mathrm{flat}$ in the R\&F approximation, by construction, this approach is likely to introduce discrepancies when comparing the two methods. In summary, we simulated galaxies that broadly cover the same parameter space as our real sample to be used to test the different methods in the measurement of $j_\star$. See the low and upper bounds of the parameters in Table \ref{tab:toy-model-params}.

\begin{table}
	\centering
	\caption{Distribution of parameters employed in resampling the mock galaxies used to test the method. $v_\mathrm{flat}$ and $r_\mathrm{eff}$ are estimated from $M_\star$ using scaling relations and $n$ and $r_\mathrm{flat}$ are generated randomly.}
	\begin{tabular}{ccc}
 \toprule
            Data & Lower bound & Upper bound \\
		\toprule
            $\log (M_\star / M_\odot)$ & 9.5 & 11.5 \\
            $v_\mathrm{flat} (\mathrm{km/s})$ & 80 & 320 \\
            $r_\mathrm{eff} (\mathrm{kpc})$ & 2 & 8 \\
            $n$ & 1 & 10 \\
            $r_\mathrm{flat}(\mathrm{kpc})$ & 0.2 & 6 \\
  \bottomrule
	\end{tabular}
	\label{tab:toy-model-params}
\end{table}

We used these $10^4$ mock galaxies to calculate $j_\star$ using the R\&F approximation $\tilde{j}_\star\approx k_n v_s r_\mathrm{eff}$ (with $v_\mathrm{flat}$ as $v_s$) as well as with the integrated method using Equation \ref{eq:j definition}. The latter is expected to result in a better estimate of $j_\star$ as the $v(r)$ and $\Sigma(r)$ have the explicit radial dependence of the light and kinematic profiles of the mock galaxies. 

\begin{figure}
	\includegraphics[width=0.48\textwidth]{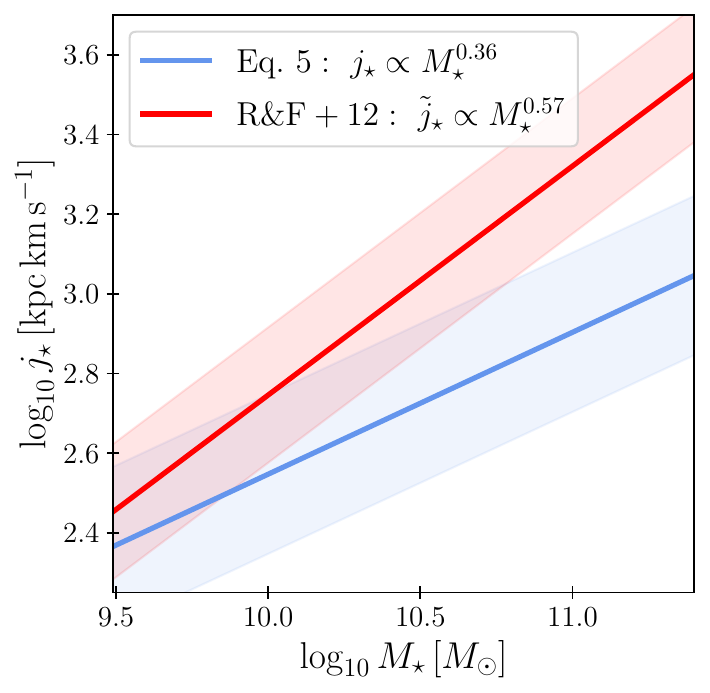}
        \caption{Fits to the $j_\star$ \textit{vs} $M_\star$ plane (Fall relation) using mock disk galaxies. The blue line corresponds to the fit from the integrated method in Equation \ref{eq:j definition} using the velocity and mass profiles, which results in a slope $\beta=0.36\pm0.05$. The red line indicates the fit using $\tilde{j}_\star \approx k_n v_s r_\mathrm{eff}$ with $\beta=0.57\pm0.05$. The shaded regions correspond to the region that encompasses 50\% of the simulated galaxies, which shows a larger scatter in the integrated method.}
    \label{fig: simulations}
\end{figure}

Finally, we used these estimates to find the best fit to the Fall relation, and we found that the slope from the integrated method is significantly lower $\beta=0.36\pm0.06$ than the slope we found using the $\tilde{j}_\star$ approximation $\beta=0.57\pm0.05$ as shown in Figure \ref{fig: simulations}. It is important to point out that these slopes may not necessarily represent the true slope that one must expect from real data due to inherent degeneracies in the parameters used to construct the mock galaxies, particularly those associated with $r_\mathrm{flat}$ or the choice of using $v_\mathrm{flat}$ as the $v_s$ in the R\&F approximation. However, it is instructive to see the systematic difference between the different methods used to measure $j_\star$. 

This experiment seems to indicate that the choice of method contributes to the difference in the slopes. In the integrated method, the velocity profile $v(r)$ is weighted by the stellar light profile, which is taken as a proxy for stellar mass and assumed to be constant in $(M/L)_\star$ radially (in contrast, the R\&F approximation relies less on explicit weighting, since there is only an implicit radial dependence of $k_n$). However, there is a well-known dependence of $(M/L)_\star$ with galaxy $M_\star$ such that stellar light profiles tend to be shallower than underlying mass profiles (e.g., \citealp{Szomoru}; \citealp{Wuyts_2012}, \citealp{Lang_2014}, \citealp{Tacchella_sins_sizes}, \citealp{Suess_2019}; \citealp{Suess_2022}; see discussion in \citealp{Forster_review_2020}). The overestimation of $M_\star(r)$ from the shallower light profiles could play a significant role in driving the high $j_\star$ for galaxies in the low-mass end which subsequently affects the slope of the Fall relation. This does not impact the experiment of the mock galaxies as we have explicitly assumed a constant $(M/L)_\star$ but could have an effect on the real data.

\subsection{Potential observational reasons for the shallow slope}
\label{Potential observational reasons}

Besides the quantified effects of data types (the choice of NS and/or AO), morphological classification and outliers in the sample as discussed in \ref{tab: results}, there are other limitations that affect the determination of the slope, limiting its trustworthy interpretation.

In particular, the sample size and its limited dynamic range are both small to fit the Fall relation with high confidence. The significant dependence on galaxy morphology (deviations from the thin axisymmetric disk approximation) and the ambiguity in the disk classification for some of the complex systems limit the interpretation of the inferred slope. The interpretation of a ``shallow'' slope comes from our comparison of the fitted slope for galaxies classified as rotating disks at $z \approx 2$ to the reference slope of $\beta\approx2/3$ from \cite{Romanowsky} at $z \approx 0$, which many high-redshift studies use as a fixed value in their parametrization of the Fall relation. However, our definition of rotating disks is very broad, whereas the 2/3 slope from \cite{Romanowsky} (or \citealp{Posti_2018}) applies to either ``pure disk'' or systems with a fixed bulge-to-total $(B/T)$ ratio. When combining all rotating systems at $z \approx 0$ that meet our RD definition, the slope is actually smaller than 2/3, with $\beta = 0.52 \pm 0.04$ (see Table 2 in \cite{Romanowsky}). This suggests that the tension between our $z \approx 2$ result and the $z \approx 0$ relation is not as large as initially suggested.

Another important limitation relates to the possible systematics in the fit. The found slope and simple statistical errors ($\beta = 0.25 \pm 0.15$ for disks) do not account for systematics. Most importantly, they do not account for systematics in our estimation of statistical uncertainties in $j_\star$. For example, if we attempt the fit to the Fall relation with our sample of disks while doubling the uncertainties in $j_\star$, then we obtain a best-fit value of the slope of $\beta = 0.53 \pm 0.38$, twice the original slope.

Additionally, potential correlations between $M_\star$ and $j_\star$ or non-normally distributed uncertainties could affect the robustness of the results. This suggests that the true uncertainties might be larger than the 1-$\sigma$ range indicates.

\subsection{Possible physical interpretation of the shallow slope}
\label{Possible physical interpretation of the shallow slope}

In this section, we qualitatively discuss potential interpretations of the slope of the Fall relation. To do this, we examine the relationships between the properties of the host haloes and their corresponding stellar counterparts. We base this discussion on a simple analytical prescription where baryons reside inside an isothermal spherical cold dark matter (CDM) halo (e.g., \citealp{Peebles}; \citealp{Mo_galaxy_formation}) characterized by the dimensionless spin parameter $\lambda$ (\citealp{Steinmetz_1995}). This parameter can be conveniently expressed as a function of some of the global properties of the halo:

\begin{equation}
    \lambda \equiv \frac{J_h E_h^{1/2}}{G M_h^{5/2}},
\end{equation}
where $G$ is the gravitational constant and $E_h$, $J_h$, and $M_h$ are the energy, angular momentum and mass of the halo, respectively. In scale-free gravity, the expected relation between the specific angular momentum of the halo $j_h$ and $M_h$ is $j_h \propto M_h^{2/3}$. 

We estimated the halo mass using the abundance matching framework, which establishes a statistical connection between observable galaxy properties like stellar mass and the properties of their dark matter haloes, such as halo mass. To compute the halo mass using the stellar mass and redshift, we used the redshift-dependent abundance matching relations from \cite{Moster_2013} (see Table 1 in their study), considering the average redshift of the sample at $z\approx 2.2$.

The abundance matching framework has some important limitations. One such limitation, as highlighted by \cite{Posti_2019}, is a tendency to overpredict halo masses. This tendency is particularly pronounced at low values of $v_\mathrm{flat}$ but also appears (less strongly) at high $v_\mathrm{flat}$. In their study, \cite{Posti_2019} applied the \cite{Moster_2013} model to a large sample of disks at $z\approx 0$ and found that the velocity fraction $f_V$ (the ratio between the circular velocity at the edge of the galactic disk and the velocity at the virial radius), expected to be close to unity, deviates significantly at low $v_\mathrm{flat}$ values (see their Figure 3). This discrepancy leads to an overprediction of halo masses, following $f_V \propto f_M^{1/3}$ (see Section 4.3 in their work).

These disparities raise questions about the consistency of the abundance matching framework's predictions with observational data, particularly in cases where disk galaxies are expected to reside in much more massive dark matter haloes than suggested by their \( \textrm{H}\scriptstyle\mathrm{I} \) rotation curves. This incongruity has been acknowledged in the literature and is often referred to as the ``too big to fail'' problem, typically associated with dwarf galaxies  (\citealp{Papastergis_2015}). This effect is more pronounced for galaxies with \( v_\mathrm{flat} < 40 \) km/s as seen in Figure 5 in \cite{Posti_2019} and Figure 6 in \cite{Papastergis_2015}. Given that all galaxies in our sample have velocities of \( v_\mathrm{flat} > 50 \) km/s, the potential overprediction in $M_h$ is expected to be marginal.

Having obtained an estimate of $M_h$, one can combine the scaling equations of an isothermal halo (see discussion in Section 4 in \citealp{Obreschkow_2014}) to obtain the specific angular momentum of the halo $j_h$ as:

\begin{equation}
    j_h = \frac{\sqrt{2}\lambda G^{2/3}}{(10H(z))^{1/3}}M_h^{2/3}.
    \label{eq: specific angular momentum of halo}
\end{equation}

Multiple $N$-body simulations that focus on the formation of haloes and the build-up of angular momentum find a distribution of spin parameters that peaks at an average value of $\langle \lambda \rangle = 0.035$ with a dispersion of 0.2 dex (\citealp{Bullock_2001}; \citealp{Hetznecker_Burkert_2006}; \citealp{Maccio_2007}) and with little dependence on redshift (e.g., \citealp{Munoz_Cuartas}). Thus, we used this value in the calculation, taking into account that $j_h$ represents the expectation for a given halo mass under the assumption that $\lambda=\langle \lambda \rangle$.

\begin{figure*}
	\includegraphics[width=0.85\textwidth]{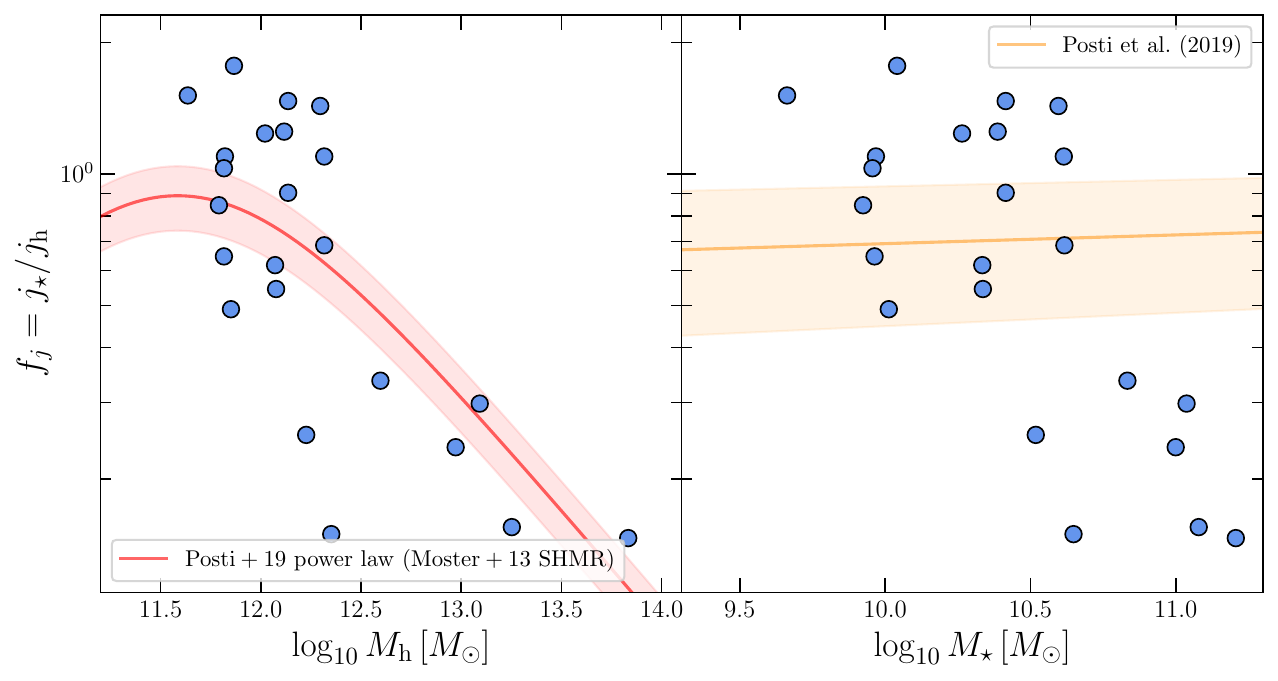}
        \caption{Angular momentum retention factor $f_j$ as a function of halo mass $M_h$ (left) and stellar mass $M_\star$ (right) for the 24 rotating disks in the sample. Blue dots correspond to the estimations based on the \protect\cite{Moster_2013} abundance matching relations. The red line corresponds to the best fit of the double power-law model from \protect\citealp{Posti_2018_2} (see Equation 18 in their work) using the \protect\cite{Moster_2013} SHMR relation. The orange line in the right panel corresponds to the linear relation found in \protect\cite{Posti_2019}. The declining trend in the retention factors seems to indicate that low-mass haloes can retain more angular momentum and thus affect the slope of the Fall relation.}
    \label{fig: fj and fM}
\end{figure*}

Using Equation \ref{eq: specific angular momentum of halo} and our measurements of $j_\star$, we now calculate a quantity that measures the proportion of the halo's angular momentum that is retained\footnote{The word ``retention'' in this context does not imply only the conservation of angular momentum, since stars can both gain or lose $j_\star$ relative to both all baryons and dark matter.} by the stars, referred to as the angular momentum retention factor $f_j=j_\star/j_h$ (\citealp{Romanowsky}). There are various physical processes involved in the retention or loss of angular momentum, including dynamical friction, interplay between inflows and outflows, hydrodynamical viscosity, and galactic winds (e.g., \citealp{Governato_2007}; \citealp{Brook_2012}; \citealp{DeFelippis_2017}). When there are no mergers, disk-like galaxies can retain a higher amount of angular momentum from the parent halo, resulting in high values of $f_j$. 

The angular momentum retention factors have been shown to depend strongly on galaxy morphology. In a comprehensive examination of the $j_\star$ \textit{vs} $M_\star$ relation, \cite{Romanowsky} explore a theoretical framework based on the hierarchical assembly of galaxy haloes within a $\Lambda$CDM cosmology and connect it to observational findings from a sample of approximately 100 nearby galaxies. To do this, they generate mock systems by drawing $M_h$ from uniform logarithmic distributions and use simplified analytical expressions for $f_j$ and $f_\star=M_\star/M_h$ to investigate their behaviour across different galaxy morphologies. Assuming spherically symmetric halo profiles with a spin parameter $\lambda=0.035$ and a power-law relation of $j_\star\propto M_\star^\beta$ with $\beta=2/3$, they find a clear dependence of $f_j$ on galaxy morphology. They report mean values of $\langle f_j \rangle \approx 0.6$ for spirals and $\langle f_j \rangle \approx 0.1$ for ellipticals. Notably, they observe that this dependence on morphology remains relatively consistent across varying mass ranges (see also \citealp{Harrison_2017} and \citealp{Posti_2019} for related studies). These results align with prior research in the local Universe, which has shown a similar range of $\langle f_j \rangle$ values, typically around 0.5 to 0.6, for spirals under similar assumptions (e.g., \citealp{Navarro_Steinmetz}; \citealp{Dutton_van_den_Bosch}; \citealp{Kassin_disk_settling}).

We present inferred angular momentum retention factors in our sample of disks as a function of $M_h$ and $M_\star$ in Figure \ref{fig: fj and fM}, followed by a discussion on the potential interpretation in the resulting trend. It is important to note that the various assumptions and approximations used to calculate $M_h$ and $j_h$ may introduce systematic biases and artificial trends. In particular, the explicit assumption that the angular momentum of the halo scales with halo mass as $j_h \propto M^{2/3}$ impacts the inferred $f_j$ by construction. Therefore, caution is advised in interpreting these tentative outcomes, which are primarily qualitative in nature.

\subsubsection{Interpretation}
\label{subsection: theory discussion}

Firstly, we found high angular momentum retention factors in galaxies with $\log_{10}M_h[M_\odot] < 12.5$, in some cases $f_j>1$, i.e the specific angular momentum of the disk is larger than that of the parent halo. Values of $f_j$ above unity contrast with findings in the local Universe reported by e.g., \cite{Fall_Romanowsky_2018} and \cite{Posti_2018} who found $f_j \leq 1$ at $z\approx0$. However, the predicted wide distribution of halo spins in CDM haloes (\citealp{Bullock_2001}), combined with complex baryonic processes such as feedback, gas accretion, and differential angular momentum transfer, can result in scenarios where the stellar component $j_\star$ exceeds the specific angular momentum of the halo, thus making $f_j>1$ values feasible. 

These would imply that gas in the halo has acquired angular momentum through mergers, accretion, inflows, outflows, or interactions with neighbouring galaxies. In simulations, the prescriptions used to reproduce such processes significantly influence disk formation (e.g., \citealp{Dekel_2009}; \citealp{Danovich}; \citealp{Genel_2015}; \citealp{Stevens_2016}). A notable example is \cite{Ubler_2014}, who conducted cosmological zoom-in simulations using different feedback models and found that strong feedback favours disk formation, sometimes resulting in galaxies with comparable or even larger specific angular momenta than their parent haloes.

Observationally, angular momentum retention factors above unity have also been measured. At $z\sim0.9$, \cite{Harrison_2017} used a similar methodology as the one we have used in this work to measure $f_j$ and found a significant scatter around $f_j\sim1$, with some $f_j$ values that exceed unity. Moreover, in the $1<z<3$ study of \cite{Burkert}, a close correspondence between $j_\star$ and $j_h$ was found. They introduced the angular momentum parameter $\lambda \times f_j$, which exhibited an inferred log-normal distribution with a mean of $0.037\pm0.015$, assuming $\lambda=0.035$. This distribution accounted for the scatter around $f_j\sim1$, as illustrated in Figure 3 of their work.

Considering the full samples shown in Figure \ref{fig: fj and fM}, we observed a discernible trend that could (at least qualitatively) point to a physical interpretation of the flattening of the Fall relation. This trend involves both the large $f_j$ at low mass and the decrease in $f_j$ with increasing $M_h$ and $M_\star$, supported by Spearman correlations of $\rho_s=-0.65$ in both cases. This suggests that at high redshifts, low-mass haloes could possess a higher capacity to retain angular momentum, leading to elevated $j_\star$ values at lower $M_\star$. This observation potentially contributes to the shallower slope of the Fall relation.

Understanding the apparent large retention of angular momentum in low-mass galaxies at high redshift requires consideration of various physical processes. At $z\sim2$, galaxies are characterized by an abundance of cold gas (see review by \citealp{Tacconi_review_2020}), facilitating efficient angular momentum transport primarily within the gas component via viscous torques (see \citealp{Lesch_1990} for a detailed description of viscous torques' role in angular momentum transport). This gas-rich environment fosters the transfer of angular momentum from the dark matter halo to the central galactic disk. Moreover, during the cosmic noon epoch, significant gas accretion occurred (e.g., \citealp{Dekel_2009}), where gas with lower angular momentum collapses earlier, carrying intrinsic angular momentum. This process amplifies the overall angular momentum of galaxies and is aided by the dissipative effects within infalling gas (e.g., see Section 6.3.2 in \citealp{Romanowsky}). This is particularly important in low-mass systems, which are even more predominantly gas-rich than the high-mass systems at cosmic noon, as discussed in \cite{Tacconi_2018, Tacconi_review_2020}. Therefore, this richness of gas in these low-mass systems could contribute to their large $f_j$.

Another potential mechanism contributing to increased $j_\star$ in the disks is strong feedback-driven outflows, which can preferentially remove low angular momentum material (\citealp{Sharma_2012}). For systems in the low-mass regime, this material can be completely ejected from the galaxy, while for more massive systems, this can lead to a redistribution of the angular momentum within the disk as some of the gas is accreted back towards the disk (\citealp{Brook_2012})). In this scenario, the feedback-driven outflows in low-mass galaxies could completely remove low-$j_\star$ material, leading to large $f_j$ values in that mass range. However, further investigations are needed to understand how the various physical processes mentioned above depend on the mass and how this could amplify the angular momentum retention factor in low-mass galaxies.

\section{Conclusions}
\label{section: Conclusions}

We have collected a sample of 41 galaxies in the range $1.5<z<2.5$ with IFS observations at both high- and low-spatial resolution (with their corresponding \textit{HST} near-IR imaging) to make a detailed measurement of their 
specific angular momentum content $j_\star$ combining both data types. Our integrated calculation of $j_\star$ using radial profiles and combining the different resolutions provides an independent measurement from the low-resolution studies and from the commonly used approach $\tilde{j}_\star\approx k_n v_s r_\mathrm{eff}$ (R\&F approximation), which we show can be biased for galaxies with complex morphologies, as is the case for some of the galaxies in this sample.
    
\begin{itemize}    
    \item \textbf{Morphological classification} We identified 24 galaxies as disks with well-ordered rotation ($f_\mathrm{disk}\approx 58.6\pm 7.7\%$) and the remaining 17 galaxies as Irregular/Merger systems. The main purpose of this classification was to identify systems where we could achieve reliable measurements of $j_\star$ under assumptions of cylindrical symmetry.

    \item \textbf{The Fall relation:} For the disk galaxies, we found that the power law relationship in the $j_\star$ \textit{vs} $M_\star$ relation (Fall relation) is of the form $j_\star \propto M_\star^\beta$ with $\beta=0.25\pm 0.15$, which is significantly shallower than the commonly adopted $\beta=2/3\approx0.67$ in studies at similar redshift and well established at $z=0$ for fixed galaxy types. While the disk sample size is modest, and there is a significant scatter in the Fall relation, the measured slope could point to a different scaling at high redshift, likely associated with the complexities of high-redshift galaxies. We ran different experiments to address the significance of this finding and point out the existing systematic uncertainties in previous studies that need to be considered with caution in future studies (see next bullet points).
    
    \item \textbf{Data dependence:} A fit to the 41 galaxies from the full sample (disks+irregulars) yields a slope of $\beta=0.48\pm0.21$, a factor of $\sim 2$ higher than the slope found for the disks. While this is formally consistent with the fit to the disks within the uncertainties, the notable difference in the best-fit values can be plausibly explained by the fact that the angular momentum measurements for irregular systems are unreliable since the approximation of cylindrical symmetry does not work for them. Additionally, some of those galaxies are likely to be mergers (or be disrupted by mergers) and thus have a low content of $j_\star$, making the slope of the Fall relation steeper in low-resolution studies that are not able to determine disk morphologies with certainty. It is important to emphasize that the galaxies identified as disks in the sample are, on average, more massive systems with higher $j_\star$ content. This fact has a notable impact on the determination of the slope and would benefit from a larger sample of disk galaxies in the low mass regime. Specifically, the majority of galaxies categorized as irregulars fall below the mean mass of the sample (14 out of 17), and similarly, a majority of the irregulars have specific angular momentum values below the mean (12 out of 17).
    
    \item \textbf{Method to measure $j_\star$:} The adopted method to measure $j_\star$ has a significant effect on the Fall relation. A fit using the measurements of $j_\star$ using the R\&F approximation $\tilde{j}_\star\approx k_n v_s r_\mathrm{eff}$ for the disk galaxies yields a slope of $\beta=0.61\pm0.21$, significantly steeper than $\beta=0.25$. We quantified the systematic difference to be expected from the choice of method by creating $10^4$ mock galaxies and measuring the different slopes. This experiment shows that the significant difference remains, with $\beta \approx 0.26\pm 0.14$ for the integrated method and $\beta \approx 0.64\pm 0.2$ for the approximation, likely attributed to the stellar mass weighting in the integrated method. Furthermore, if we use both the R\&F approximation and the full sample (disks+irregulars), we obtain an even steeper slope of $\beta=0.77\pm0.23$, which points to the strong bias of these approximations.

    \item \textbf{Clumps:} We used a systematic approach to identify and measure clumps in the sample. We found that 31 galaxies have clumps, with an average of 2.5 clumps per system. By dividing the sample into galaxies above and below the median ``clumpiness'' $\mathcal{C}\sim 11.72\%$, we found no significant trend that shows an effect of clumps in the slope of the Fall relation and only a minor difference in the normalization where clumpier galaxies have a higher content of $j_\star$. This could be associated with the detection method where it is easier to find clumps in larger galaxies (which have higher $j_\star$).

    \item \textbf{Central concentrations:} We measured the central light concentration in the sample from the \textit{HST} near-IR imaging with an average of $\approx 0.2$. By separating the sample into those below and above the median $\approx 0.1$, we found that galaxies with higher concentrations have a lower content of $j_\star$. Under the assumption that the concentration can serve as a proxy for the bulge-to-total ratios, this trend is consistent with the results that predict negative vertical offsets in the Fall relation for galaxies with large $B/T$ (e.g., \citealp{Obreschkow_2014}; \citealp{Fall_Romanowsky_2018}; \citealp{Sweet_2018}). However, central light concentrations in the sample do not seem to drive the slope of the Fall relation, in agreement with results at $z=0$.

    \item \textbf{Potential observational reasons for shallow slope}: The shallow slope in the Fall relation is influenced by small sample size and limited dynamic range, dependence on galaxy morphology, and ambiguity in the disk classification. The comparison between broad rotating disk classifications at $z \approx 2$ and narrower definitions at $z \approx 0$ suggests that the observed tension may be less significant. Additionally, systematic uncertainties in fitting and estimating statistical errors in $j_\star$, along with potential correlations between $M_\star$ and $j_\star$, impact the robustness of the results. For instance, doubling the uncertainties in $j_\star$ leads to a best-fit slope of $\beta = 0.53 \pm 0.38$, emphasizing the role of systematic errors.

    \item \textbf{Physical interpretation:} Finally, we quantified the angular momentum retention factor $f_j$ based on a set of simple assumptions for CDM haloes to search for a potential explanation or reframing of the shallow slope based on physical principles. We used the abundance matching framework to find the halo mass of the galaxies and found a trend that motivated a qualitative discussion. We found large values of $f_j$ in the low-mass regime that show only a monotonic decrease as a function of both stellar and halo mass. This suggests that low-mass haloes seem to retain more angular momentum, which populates the high-$j_\star$ \textit{vs} low-$M_\star$ region and thus contributes to the shallow slope. In our qualitative discussion, we highlight that gas-rich environments at $z\sim2$ facilitate efficient angular momentum transport via viscous torques and gas accretion, amplifying the overall angular momentum of galaxies. Moreover, feedback-driven outflows could completely remove low-$j_\star$ gas from the galaxy in low-mass systems, thus leading to the observed shallow slope in the Fall relation.
    
\end{itemize}

Further investigations are required to confirm the shallow slope of the Fall relation in the cosmic noon period ($1<z<3$) and should aim to address several key aspects. Firstly, expanding the sample size of disk galaxies, especially in the low-mass regime, would provide a more robust understanding of the relationship between angular momentum and galaxy mass. Second, refining the methods for measuring angular momentum and accounting for systematic uncertainties are essential steps towards obtaining more accurate results. The ongoing near-IR IFU surveys conducted with \textit{JWST} NIRSpec and ERIS/VLT offer unprecedented opportunities for refining kinematic measurements. In future work, we will combine this high-resolution IFU data and use estimates of stellar mass distributions from \textit{JWST} photometry for a more precise estimation of $j_\star$ and the exploration of the role of clumps, non-circular motions and central light concentrations in shaping the Fall relation. Moreover, incorporating insights from theoretical frameworks such as the abundance matching framework and refining our understanding of the physical mechanisms governing angular momentum retention in galaxies will be crucial for advancing our comprehension of galaxy formation and evolution.

\section*{Acknowledgements}

We sincerely thank the anonymous referee for their valuable comments and suggestions, which have greatly enhanced the paper's quality and clarity. We thank Mark Swinbank and Martin Bureau for the useful feedback and suggestions. We thank Sandro Tacchella for providing the photometric data and for the useful comments. JE is funded by the Swinburne University Postgraduate Research Award (SUPRA) and ASTRO3D, as well as the GALPHYS Project: 101055023 (ERC-2021-ADG) at MPE. SMS acknowledges funding from the Australian Research Council (DE220100003). Parts of this research were conducted through the Australian Research Council Centre of Excellence for All Sky Astrophysics in 3 Dimensions (ASTRO 3D) through project number CE170100013. JE, DBF, KG, DO, and SMS acknowledge support from ARC DP grant DP160102235. DO is a recipient of an Australian Research Council Future Fellowship (FT190100083) funded by the Australian Government. Some of the data presented herein were obtained at the W. M. Keck Observatory, which is operated as a scientific partnership among the California Institute of Technology, the University of California, and NASA. The Observatory was made possible by the generous financial support of the W. M. Keck Foundation. The KMOS data were obtained at the Very Large Telescope of the European Southern Observatory, Paranal, Chile, and provided by the KGES survey team and the public release of KMOS$^\mathrm{3D}$. The SINFONI data were obtained at the same facility and provided in the public release of the SINS/zC-AO survey. \textit{HST} data were obtained from the data archive at the Space Telescope Science Institute. The kinematic modelling in this work was performed on the OzStar national facility at Swinburne University of Technology and the National Collaborative Research Infrastructure Strategy (NCRIS). We acknowledge the open-source software packages used throughout this work, including \textsc{astropy} (\citealp{astropy}), \textsc{scipy} (\citealp{scipy}), \textsc{numpy} (\citealp{numpy}), CM\textsc{asher} (\citealp{cmasher}), \textsc{matplotlib} (\citealp{matplotlib}).

\section*{Data Availability}

The reduced datacubes of the adaptive-optics assisted data from the SINS sample are available at \href{http://www.mpe.mpg.de/ir/SINS/SINS-zcSINF-data}{http://www.mpe.mpg.de/ir/SINS/SINS-zcSINF-data}. The datacubes of the KMOS$^\mathrm{3D}$ survey are available at \href{https://www.mpe.mpg.de/ir/KMOS3D/data}{https://www.mpe.mpg.de/ir/KMOS3D/data}. The rest of the data underlying this article will be shared upon reasonable request to the corresponding author.



\bibliographystyle{mnras}
\bibliography{references} 




\appendix

\section{Spatial resolution and PSF}
\label{appendix: spatial resolution and PSF}

\subsection{Spatial resolution}
\label{appendix: spatial resolution}

Most near-IR kinematic samples are observed with the natural seeing (NS) of the atmosphere, which under good seeing conditions (PSF FWHM $\sim 0.5$ arcsec) correspond to $\sim4$ kpc at a redshift range of $z\sim[1.5-2.5]$. To distinguish small-scale structures such as star-forming clumps and bulges, determine the kinematic state and measure more accurate velocity and dispersion maps, one needs spatial resolutions that trace kpc or sub-kpc scales, only possible from the ground with adaptive optics (AO) which in ideal conditions can reach PSF FWHM of $\sim0.1$ arcsec or $\sim1$ kpc at $z\sim[1.5-2.5]$. 

For our AO sub-sample, the mean spatial resolution, achieved as a combination of the seeing conditions and the AO performance, is $\sim1.75$ kpc ($\sim0.21$ arcsec) with variations that ranged from 0.11 arcsec in the best cases to 0.29 arcsec in the worst case. Variations in the Strehl were also significant, with some objects having a poor performance ($\mathcal{S}<10\%$) and some showing a better performance ($\mathcal{S}>30\%$). For the natural seeing (NS) sub-sample, the mean spatial resolution in the observations is a factor of $\sim 3$ higher than for AO with $\sim5.6$ kpc ($\sim0.67$ arcsec). The variations in the seeing for the NS observations were large with $0.52 < \mathrm{FWHM}\,\,\mathrm{(arcsec)} < 0.88$. See Table 8 in \cite{Forster_2018} for details on the observing conditions of the SINS sample, including the average optical seeing, coherence time $\tau_0$ and airmass over the individual exposures of the star used for PSF calibration.

The clear gain in spatial resolution provided by the AO data is contrasted by the very long exposure times necessary to achieve similar levels in signal-to-noise as that of the seeing-limited samples. This is one of the main drawbacks of these types of observations. Some of the galaxies in the SINS sample had long exposure times (7 hours in some cases), so the SNR is enough for the kinematic modelling, but for the majority of galaxies, the AO data had lower surface brightness sensitivity so the pixels with useful information is limited to the brightest regions in the galaxy, often located solely at the centre (see Figures \ref{fig:Summary Q1623-BX455} - \ref{fig:Summary COSMOS-128904} for individual summaries and a comparison of the spatial extent from the AO and NS data). To compensate for that, we use the deeper NS data that allows us to put constraints on the shape of the rotation curves at large radii from the galaxy centre. Furthermore, we do not perform any spatial binning or smoothing on the original AO datacubes to avoid degrading the spatial resolution, as the NS data probes the outskirts of the galaxies (with an average of 1.4 times the radial extent of the AO data). The only exceptions were galaxies with low surface brightness, such as GMASS-2540, which is a large disk with a face-on orientation, as well as galaxies with low SNR due to bad weather conditions during the observations (COSMOS 171407 and COSMOS 130477). For these galaxies, we applied a median filter smoothing to increase SNR, but we made sure the degraded resolution was still better than the NS resolution. It is worth noting that the extra level of smoothing introduces correlations among adjacent pixels, so the significance of the $\chi^2$ in Equation \ref{eq: likelihood} formally changes. However, the errors in the parameters are calculated from the MC resampling of the model, in which the model cubes are smoothed with the extra median filter, so this extra smoothing does not affect our modelling strategy.



\subsection{PSF modelling}
\label{appendix: PSF modelling}

\begin{figure*}
\begin{subfigure}{15.5cm}
    \centering\includegraphics[width=0.67\textwidth]{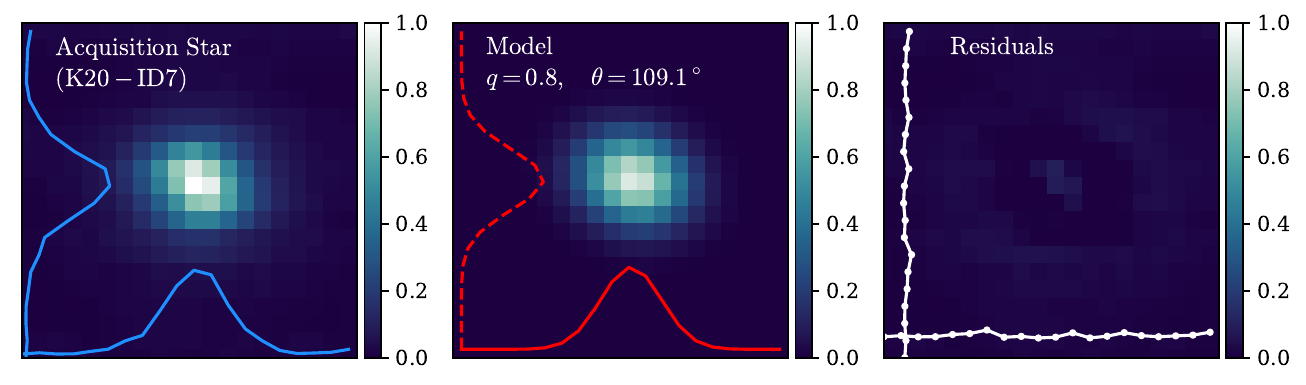}
  \end{subfigure}
\begin{subfigure}{15.5cm}
    \centering\includegraphics[width=0.67\textwidth]{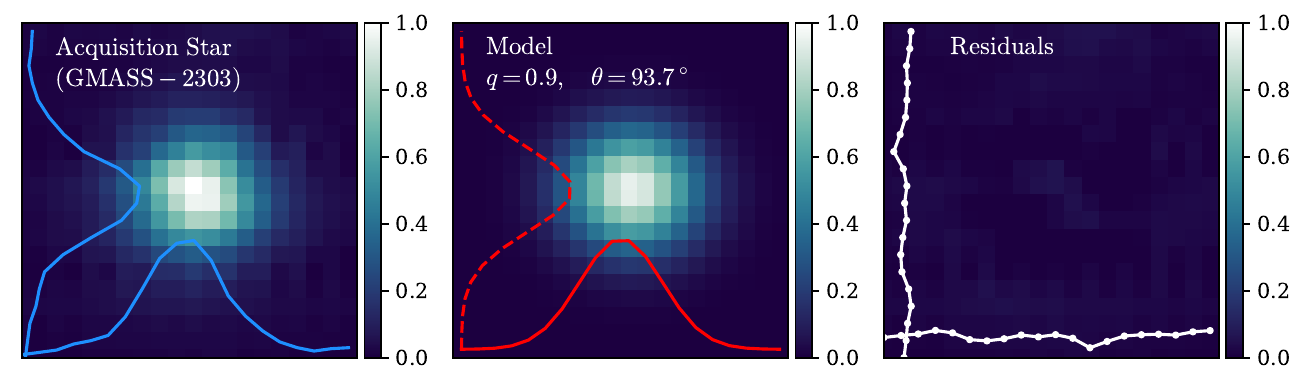}
\end{subfigure}
    \caption{PSF modelling associated with the acquisition stars of K20-ID7 (top) and GMASS-2303 (bottom). From left to right, the columns represent the original acquisition star, the model using a Gaussian kernel, and the corresponding residuals.}
    \label{fig:NS psf modelling}
\end{figure*}

\begin{figure*}
\begin{subfigure}{17.9cm}
    \centering\includegraphics[width=0.93\textwidth]{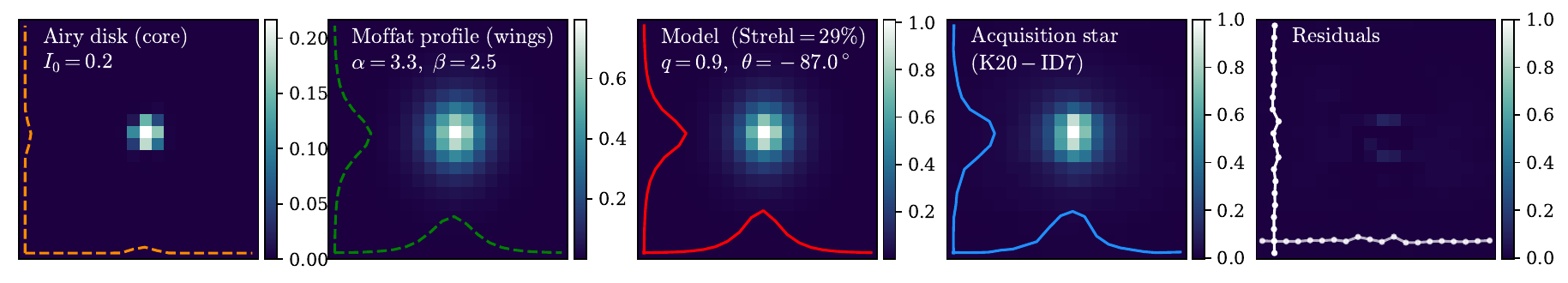}
  \end{subfigure}
\begin{subfigure}{17.9cm}
    \centering\includegraphics[width=0.93\textwidth]{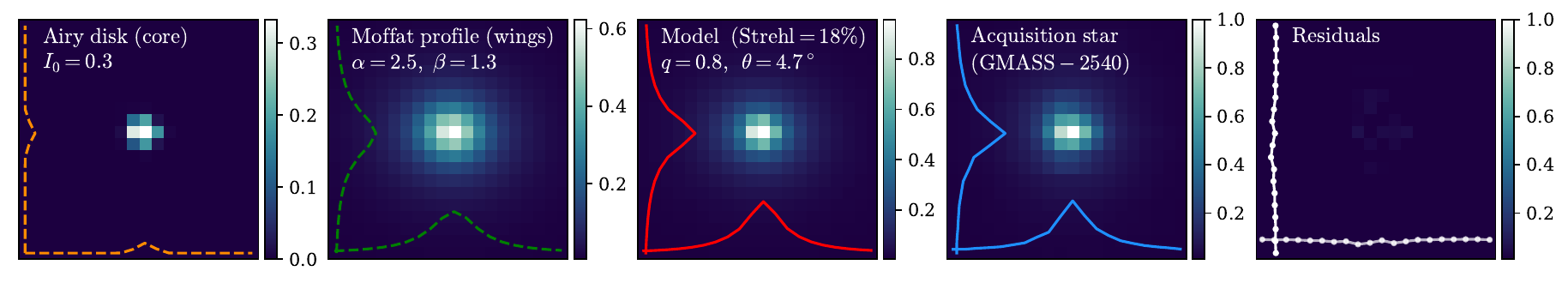}
\end{subfigure}
    \caption{PSF modelling for the acquisition stars of K20-ID7 (top) and GMASS-2540 (bottom). From left to right, the columns represent the core of the PSF modelled with an Airy disk profile, the wings of the PSF modelled with a Moffat profile, the full model (Airy disk + Moffat), the original acquisition star, and the corresponding residuals.}
    \label{fig:AO psf modelling}
\end{figure*}


A critical step in building realistic velocity models from mock datacubes is the convolution with the associated PSF of the observations. The PSF modelling in this work is the same as that used in \citetalias{Espejo_Salcedo_2022}. In summary, for the case of seeing limited observations, we model the PSF from the acquisition stars with a Gaussian kernel as indicated for three examples (SA12-6339, K20-ID7, and GMASS-2303) in Figure \ref{fig:NS psf modelling}.

In the case of the AO-assisted observations, the shape of the PSF is more complex and has two major components. The component at the core can be modelled using an Airy disk model (\citealp{Davies_AO}), while the wings of the PSF, which correspond to the residual blurring of the atmosphere can be modelled using a Moffat (\citealp{Moffat}) profile. We show two examples (stars of K20-ID7 and GMASS-2540) of this modelling in Figure \ref{fig:AO psf modelling}. The efficiency of the AO correction, quantified by Strehl, and the basic parameters of the PSF associated with the acquisition stars are indicated in Table \ref{tab:PSF_info}. The Strehl is calculated by taking the ratio of the peak intensity from the diffraction-limited model with the same throughput as the model (Airy disk plus Moffat) with respect to the peak intensity of the model.

\begin{table*}
\centering
\caption{Summary of the information of the PSF for each galaxy associated with both spatial resolutions. In the ``Quality'' column, 1 indicates good quality, and 0 indicates bad quality, based on the visual inspection of the PSF data and the goodness of the fit. $q$ is the projected minor-to-major axis ratio of the PSF model and $\theta$ is the angle. Rows in bold font indicate that the PSF associated with that specific observation was not available or had bad quality, so we used an average of all the other PSF files in the dataset.}
	\label{tab:PSF_info}
\begin{tabular}{c|ccccc|cccc}
\toprule
Acquisition Star & \multicolumn{5}{c}{AO PSF} & \multicolumn{4}{c}{NS PSF}\\
\cmidrule(lr){1-1}
\cmidrule(lr){2-6}
\cmidrule(lr){7-10}
(Galaxy ID) & Quality & $q$ & $\theta$ & FWHM & Strehl & Quality & $q$ & $\theta$ & FWHM \\
 & (0,1) & $(a/b)$ & ($^\circ$) & (arcsec) & (\%) & (0,1) & $(a/b)$ & ($^\circ$) & (arcsec) \\
\hline
Q1623-BX455     & 1     & 0.9 & 7   & 0.13  & 36    & 1 & 0.9   & 106    & 0.58 \\
Q1623-BX543     & 1     & 0.9 & 35  & 0.22   & 6     & 0 & 1.0   & NaN  & 0.78 \\
Q1623-BX599     & 1     & 1.0 & NaN & 0.29  & 10    & \textbf{--} & \textbf{0.57}    & \textbf{105}    & \textbf{0.57} \\
Q2343-BX389     & 1     & 0.9 & 179 & 0.24  & 17    & 1 & 0.9   & 26     & 0.6 \\
Q2343-BX513     & 1     & 0.8 & -91 & 0.21  & 24    & \textbf{--} & \textbf{0.57}    & \textbf{105}    & \textbf{0.57} \\
Q2343-BX610     & 1     & 1.0 & NaN  & 0.31  & 13    & 1 & 0.8   & 100    & 0.6 \\
Q2346-BX482     & 1     & 1.0 & NaN  & 0.21  & 21    & 1 & 0.7   & 15     & 0.65 \\
Deep3a-6004     & 1     & 1.0 & NaN  & 0.2   & 23    & 1 & 0.8   & 111    & 0.62 \\
Deep3a-6397     & 1     & 0.8 & 5   & 0.2   & 16    & 1 & 0.9   & 179    & 0.9 \\
Deep3a-15504    & \textbf{--}     & \textbf{0.9} & \textbf{8}  & \textbf{0.5}  & \textbf{21}     & 1 & 0.6   & 6     & 0.52 \\
K20-ID6         & 0     & 0.9 & 24  & 0.25  & 11    & 1 & 0.9   & 179    & 0.63 \\
K20-ID7         & 1     & 0.9 & -86 & 0.19  & 29    & 1 & 0.8   & 109    & 0.68 \\
GMASS-2303      & 1     & 0.8 & -16 & 0.2   & 17    & 1 & 0.9   & 93     & 0.86 \\
GMASS-2363      & 0     & 1.0 & NaN & 0.22  & 22    & 1 & 0.8   & 1      & 0.73 \\
GMASS-2540      & 1     & 0.8 & 4   & 0.29  & 18    & 1 & 0.9   & -179   & 0.88 \\
SA12-6339       & 1     & 0.9 & 15  & 0.18  & 26    & 1 & 0.7   & -168   & 0.52 \\
ZC400528        & 1     & 0.9 & 67  & 0.19  & 29    & 1 & 0.7   & 107    & 0.57 \\
ZC400569        & 1     & 0.9 & 13  & 0.18  & 23    & 1 & 1.0   & NaN    & 0.71 \\
ZC401925        & 0     & 0.9 & 20  & 0.25  & 15    & 1 & 0.8   & 107    & 0.6 \\
ZC403741        & 0     & 0.9 & 51  & 0.21  & 24    & 1 & 0.9   & 179    & 0.72 \\
ZC404221        & 1     & 1.0 & NaN   & 0.23  & 17    & 1 & 0.6   & -1     & 0.7 \\
ZC405226        & 1     & 1.0 & NaN & 0.27  & 16    & 0 & 1.0   & NaN    & 0.48 \\
ZC405501        & 1     & 0.6 & 12  & 0.19  & 15    & 1 & 0.7   & -161   & 0.56 \\
ZC406690        & 1     & 0.9 & 25  & 0.2   & 22    & 1 & 0.7   & 16     & 0.79 \\
ZC407302        & 1     & 0.9 & 18  & 0.2   & 21    & 1 & 0.9   & -142   & 0.68 \\
ZC407376        & 1     & 0.9 & 8   & 0.3   & 11    & 1 & 0.7   & 97     & 0.76 \\
ZC409985        & 1     & 0.9 & 8   & 0.15  & 33    & 1 & 0.8   & -141   & 0.84 \\
ZC410041        & 1     & 1.0 & NaN  & 0.2   & 24    & 1 & 0.9   & 179    & 0.8 \\
ZC410123        & 1     & 1.0 & NaN & 0.3  & 8     & 1 & 0.9   & -84     & 0.73 \\
ZC411737        & 1     & 0.9 & -11 & 0.24  & 18    & 1 & 0.8   & 112    & 0.59 \\
ZC412369        & 1     & 0.9 & 39  & 0.18  & 24    & 1 & 0.9   & -91    & 0.61 \\
ZC413507        & 1     & 1.0 & NaN  & 0.18  & 30    & 1 & 0.8   & -172   & 0.55 \\
ZC413597        & 1     & 0.9 & -15 & 0.22  & 18    & 1 & 0.7   & 18     & 0.62 \\
ZC415876        & 1     & 0.9 & 90  & 0.18  & 32    & 1 & 0.8   & 16     & 0.6 \\
COSMOS-110446   & 1     & 1.0 & NaN   & 0.11  & 29    & 1 & 1.0   & NaN      & 0.84 \\
COSMOS-171407   & 1     & 1.0 & NaN   & 0.39  & 17    & 1 & 1.0   & NaN      & 0.72 \\
COSMOS-130477   & 1     & 1.0 & NaN   & 0.38  & 14    & 1 & 1.0   & NaN      & 0.59 \\
COSMOS-127977   & 1     & 1.0 & NaN   & 0.11  & 29    & 1 & 1.0   & NaN      & 0.72 \\
UDS-78317       & 1     & 1.0 & NaN  & 0.11  & 29    & 1 & 1.0   & NaN      & 0.69 \\
UDS-124101      & 1     & 0.9 & 40  & 0.13  & 32    & 1 & 1.0   & NaN      & 0.76 \\
COSMOS-128904   & 1     & 0.9 & 22  & 0.12  & 26    & 1 & 1.0   & NaN      & 0.6 \\
\bottomrule
\end{tabular}
\end{table*}

\clearpage

\pagebreak
\onecolumn
\begin{center}
\textbf{\large Supplementary Materials:}
\end{center}
\setcounter{equation}{0}
\setcounter{figure}{0}
\setcounter{table}{0}
\setcounter{page}{1}
\makeatletter
\renewcommand{\theequation}{S\arabic{equation}}
\renewcommand{\thefigure}{S\arabic{figure}}
\renewcommand{\bibnumfmt}[1]{[S#1]}
\renewcommand{\citenumfont}[1]{S#1}

\begin{center}
\section{Individual case figures}
\label{appendix: case by case}
\end{center}

In this supplementary section, we show the figures with a summary containing the photometric and kinematic maps of each galaxy and the radial profiles inferred from them. We show the figures for the galaxies classified as rotating disks (RD), as they are the focus of this work. 
\hspace{1cm}
\hspace{1cm}

\captionsetup{hypcap=false}

\begin{center}
  \includegraphics[width=1\textwidth]{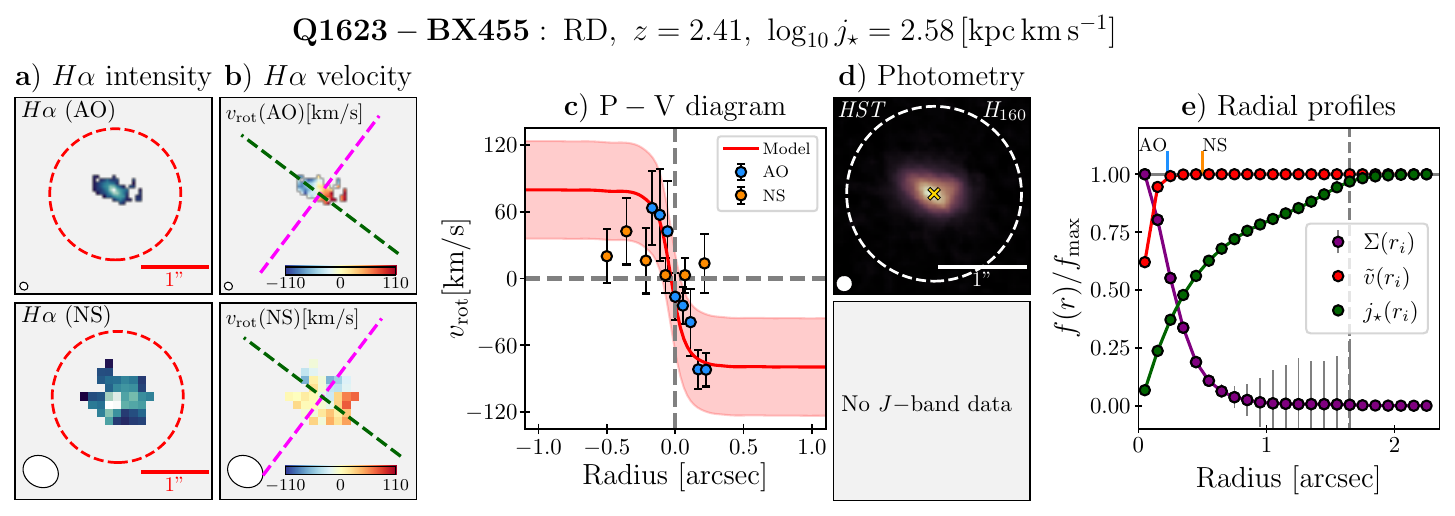}
\captionof{figure}{Summary of galaxy Q1623-BX455 (and same for the rest of the figures below): \textbf{a)} H$\alpha$ intensity fields at high- (top) and low-resolution (bottom) where the white circles represent the PSF FWHM and the red dashed line shows a boundary of radius 1 arcsec, \textbf{b)} velocity fields with the main kinematic axes indicated by the dashed green lines (with the corresponding perpendicular axis in magenta), \textbf{c)} position-velocity (P-V) diagram along the kinematic main axis where the red line is the model velocity curve $\tilde{v}(r_i)$ obtained with \texttt{CONDOR} and the shaded region corresponds to the uncertainty of the fit. The points correspond to those along the major kinematic axis, \textbf{d)} \textit{HST} near-IR data ($H_{160}$ top and $J_{110}$ bottom) with an indication of the PSF FWHM and the location of the identified clumps in green circles, and \textbf{e)} radial normalized profiles for the mass $\Sigma(r)$ (purple), velocity $v(r)$ (red), and specific angular momentum $j_\star(r)$ (green). The vertical dashed grey line represents the extent of the photometric data, so the radial profiles are extrapolated past this boundary to reach the asymptotic value of $j_\star$. Orange and blue lines indicate the radial boundary of the kinematic datasets}.\label{fig:Summary Q1623-BX455}
\end{center}

\begin{center}
  \includegraphics[width=1\textwidth]{images/summary_Q2343-BX389.pdf}
\captionof{figure}{Summary Q2343-BX389.}\label{fig:Summary Q2343-BX389}
\end{center}

\begin{figure*}
	\includegraphics[width=1\textwidth]{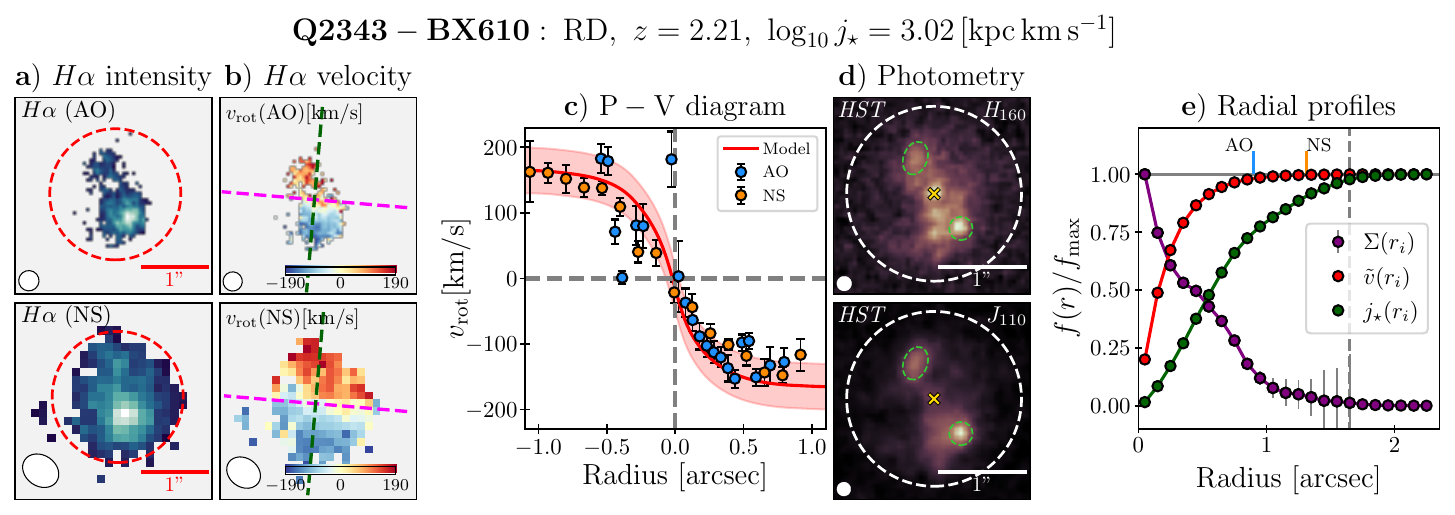}
    \caption{Summary Q2343-BX610.}
    \label{fig:Summary Q2343-BX610}
    \label{fig:Summary Q2343-BX610}
\end{figure*}

\begin{figure*}
	\includegraphics[width=1\textwidth]{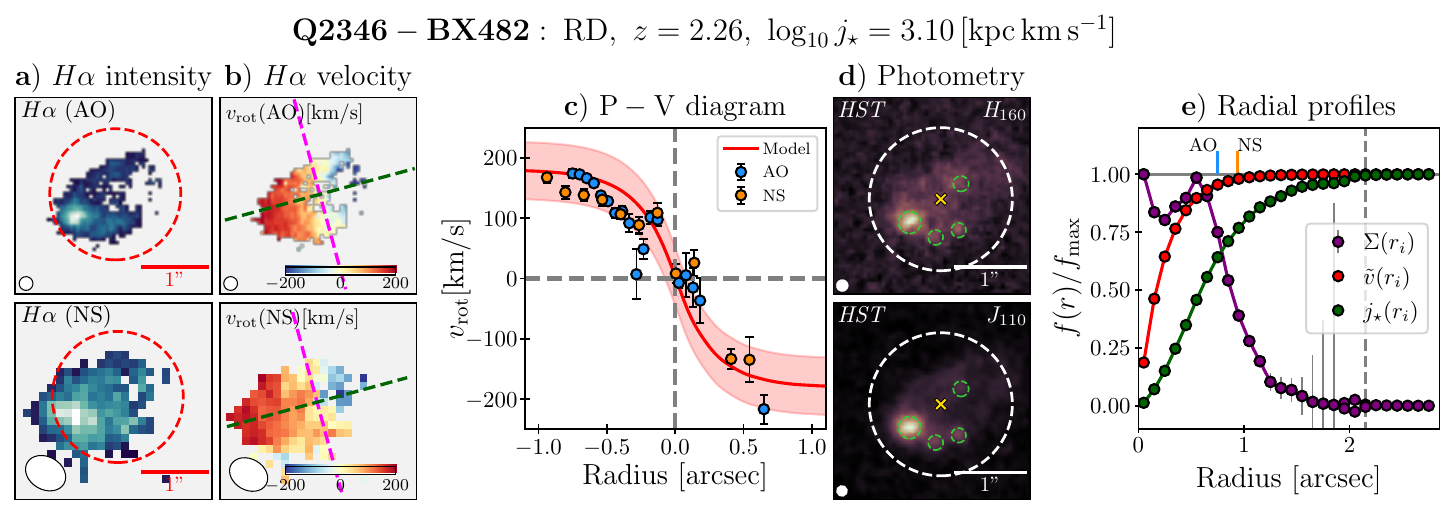}
    \caption{Summary Q2346-BX482.}
    \label{fig:Summary Q2346-BX482}
\end{figure*}

\begin{figure*}
	\includegraphics[width=1\textwidth]{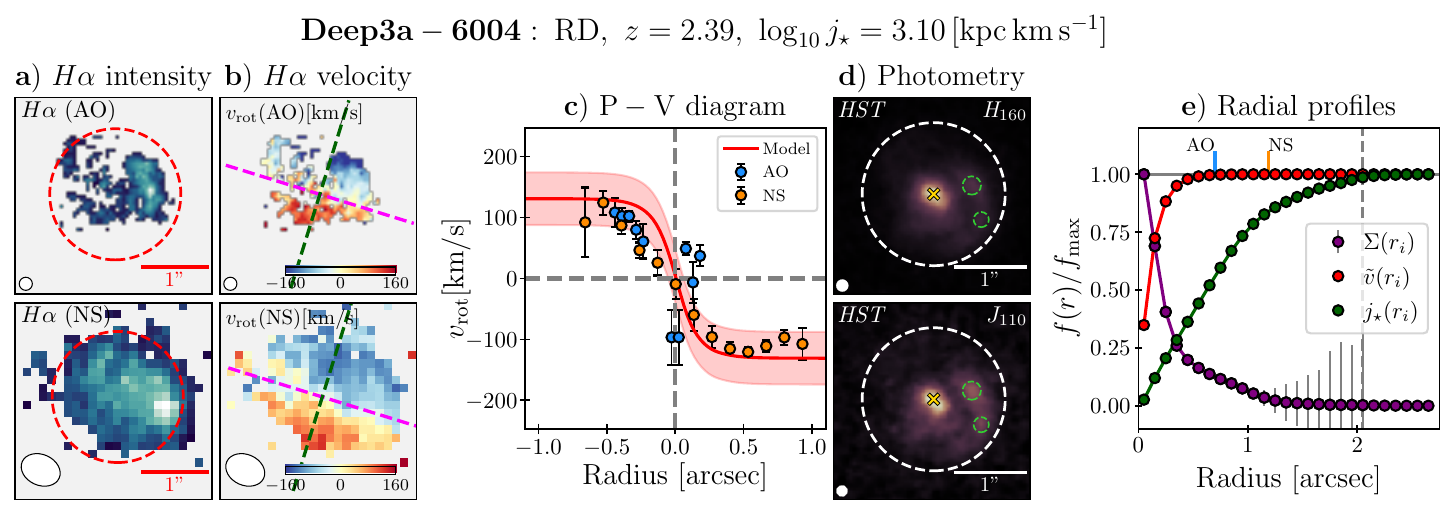}
    \caption{Summary Deep3a-6004.}
    \label{fig:Summary Deep3a-6004}
\end{figure*}

\begin{figure*}
	\includegraphics[width=1\textwidth]{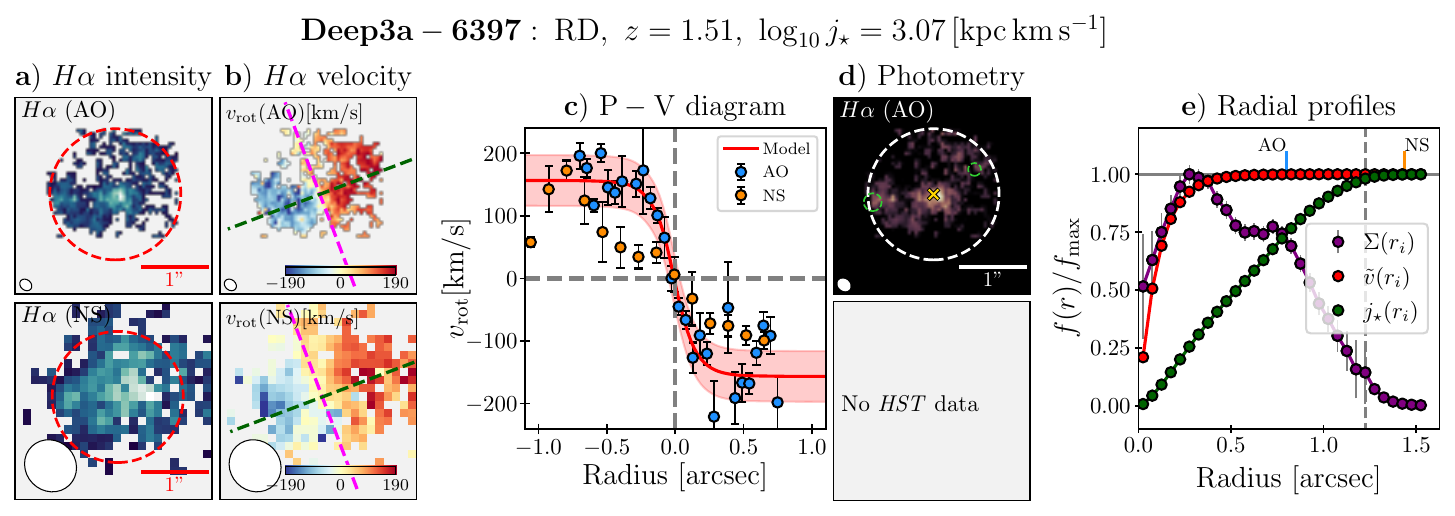}
    \caption{Summary Deep3a-6397.}
    \label{fig:Summary Deep3a-6397}
\end{figure*}

\begin{figure*}
	\includegraphics[width=1\textwidth]{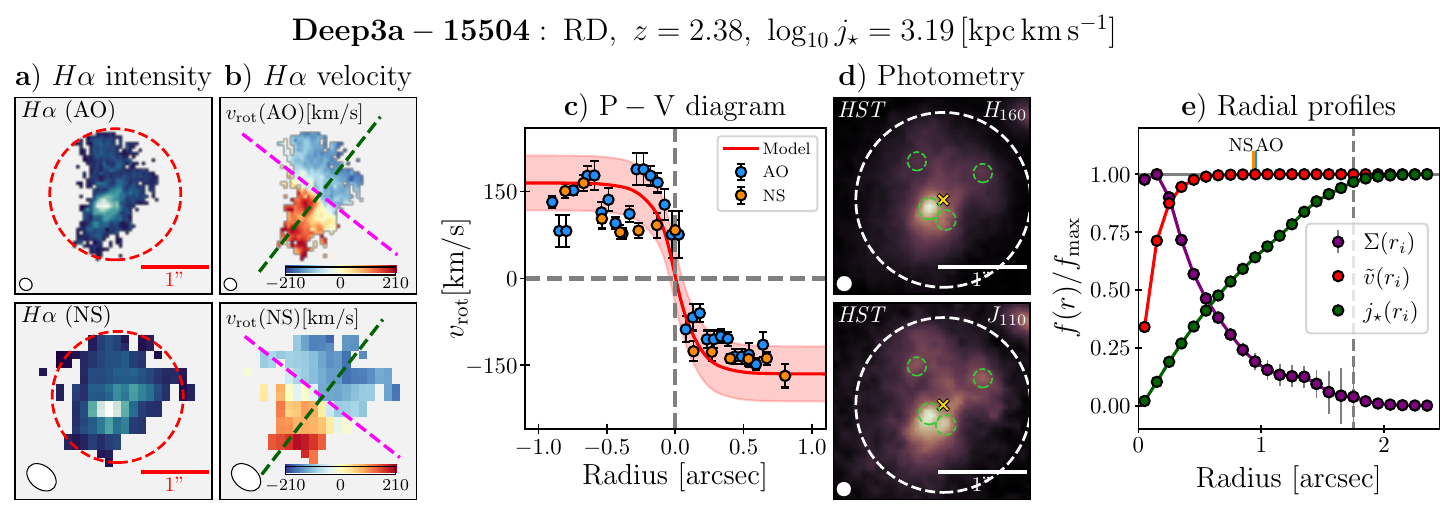}
    \caption{Summary Deep3a-15504.}
    \label{fig:Summary Deep3a-15504}
\end{figure*}

\begin{figure*}
	\includegraphics[width=1\textwidth]{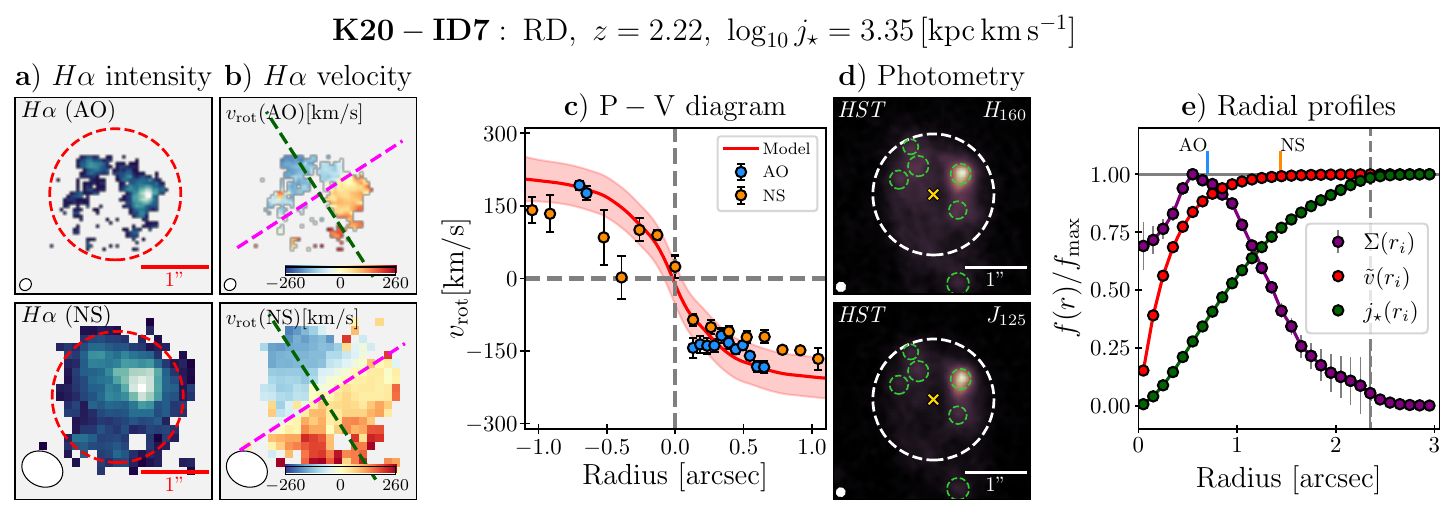}
    \caption{Summary K20-ID7.}
    \label{fig:Summary K20-ID7}
\end{figure*}

\begin{figure*}
	\includegraphics[width=1\textwidth]{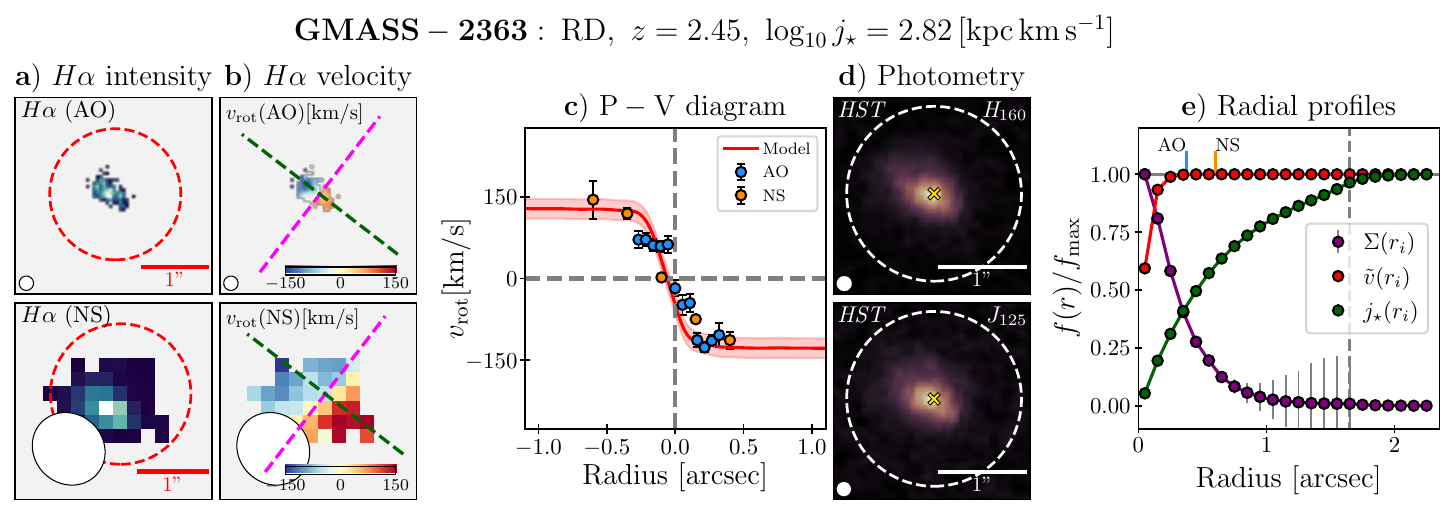}
    \caption{Summary GMASS-2363.}
    \label{fig:Summary GMASS-2363}
\end{figure*}


\begin{figure*}
	\includegraphics[width=1\textwidth]{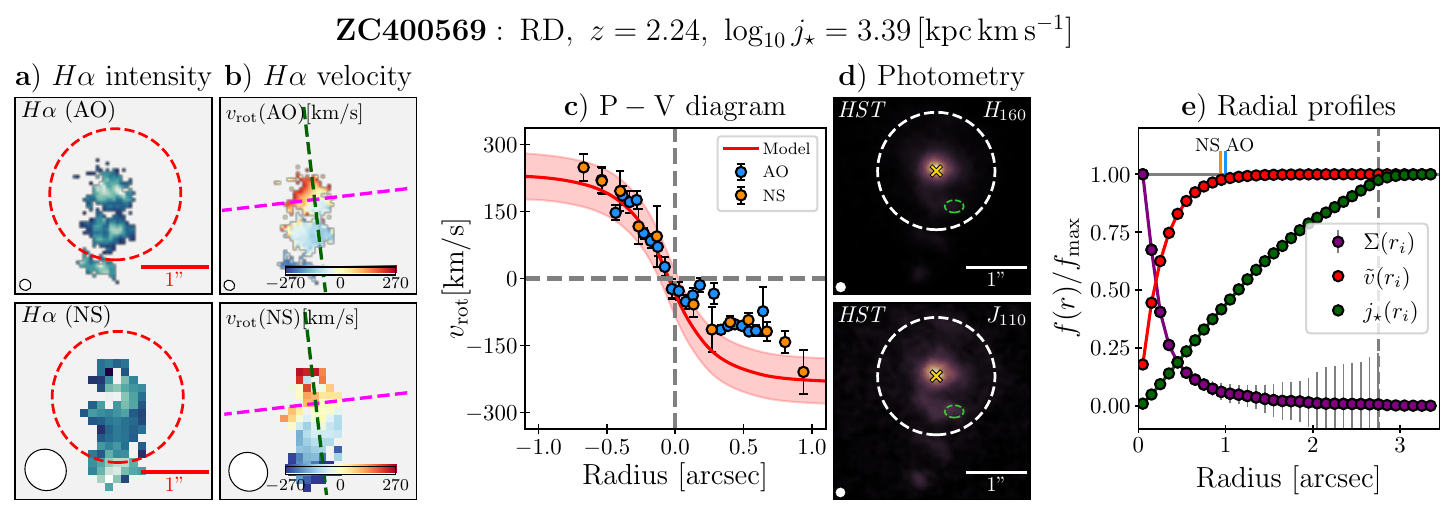}
    \caption{Summary ZC400569.}
    \label{fig: Summary ZC400569}
\end{figure*}

\begin{figure*}
	\includegraphics[width=1\textwidth]{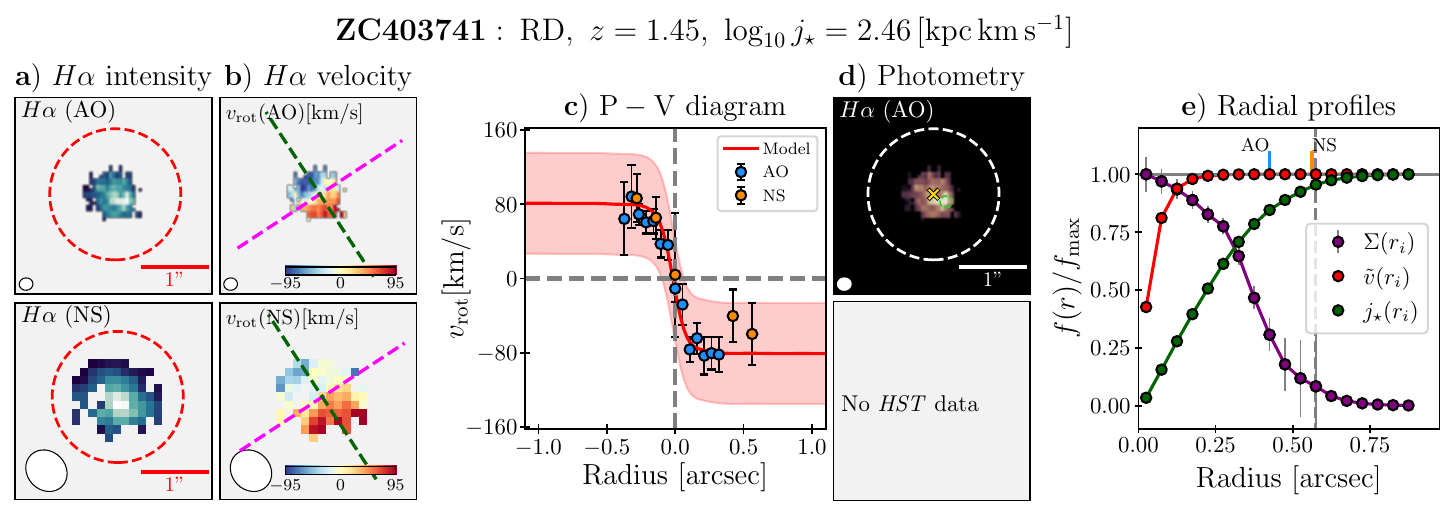}
    \caption{Summary ZC403741.}
    \label{fig:Summary ZC403741}
\end{figure*}

\begin{figure*}
	\includegraphics[width=1\textwidth]{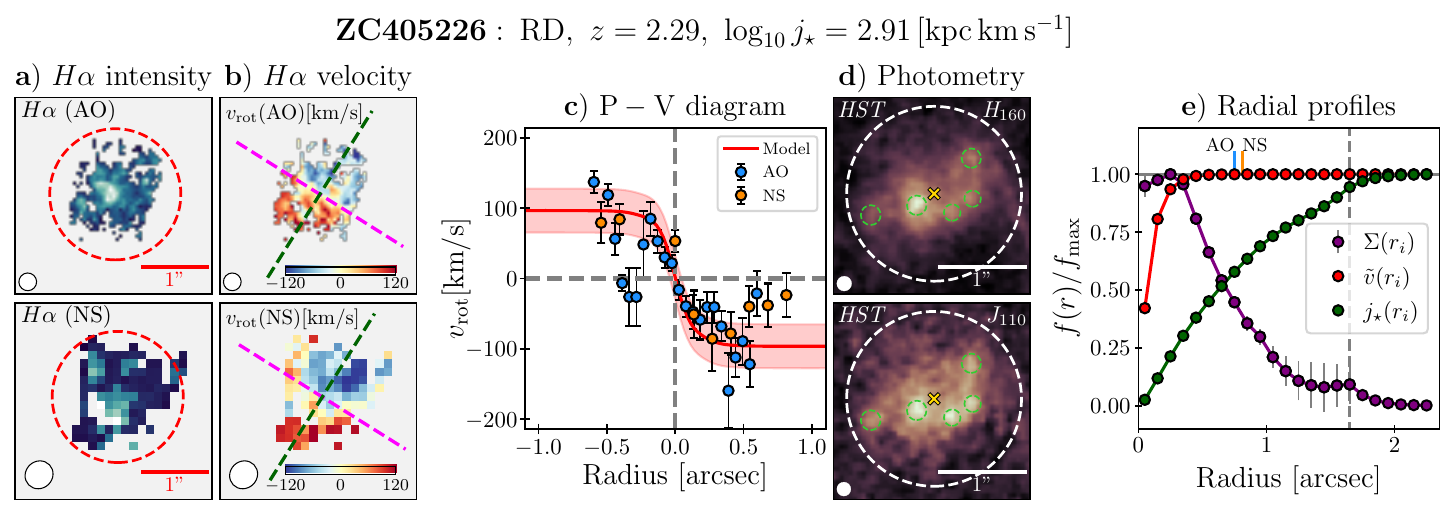}
    \caption{Summary ZC405226.}
    \label{fig:Summary ZC405226}
\end{figure*}

\begin{figure*}
	\includegraphics[width=1\textwidth]{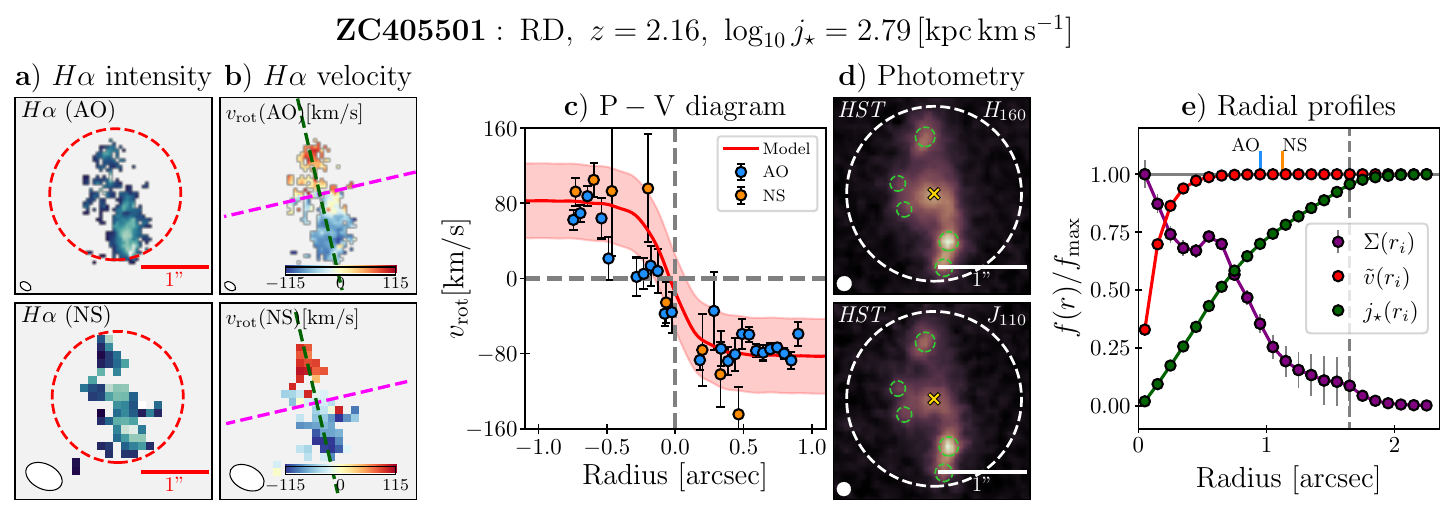}
    \caption{Summary ZC405501.}
    \label{fig:Summary ZC405501}
\end{figure*}

\begin{figure*}
	\includegraphics[width=1\textwidth]{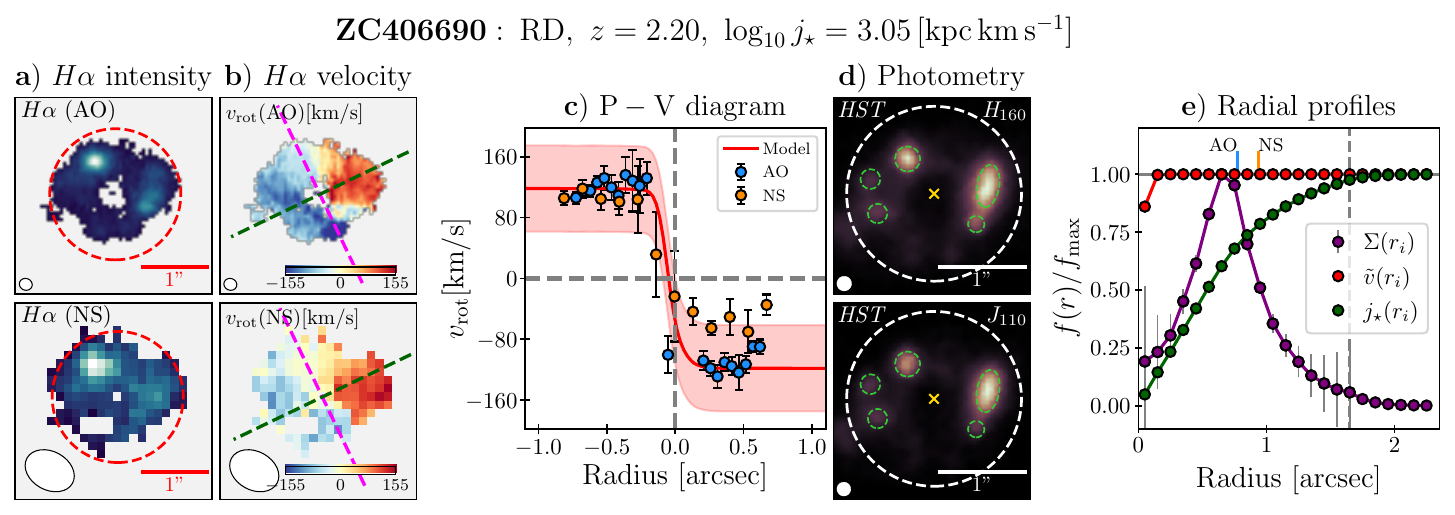}
    \caption{Summary ZC406690.}
    \label{fig:Summary ZC406690}
\end{figure*}

\begin{figure*}
	\includegraphics[width=1\textwidth]{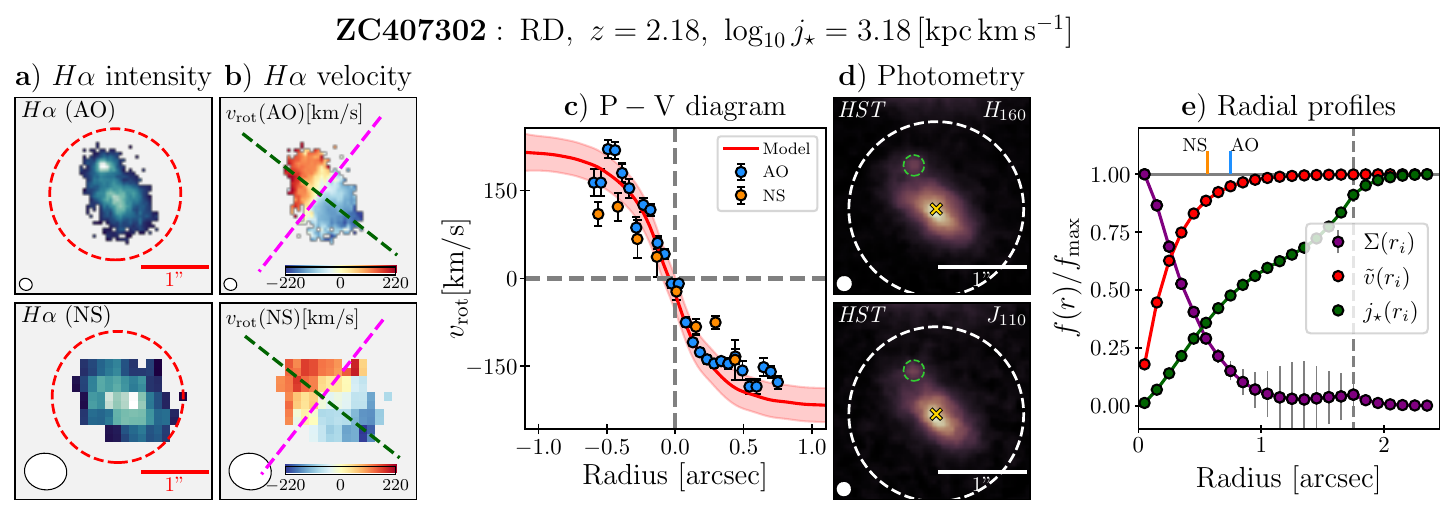}
    \caption{Summary ZC407302.}
    \label{fig:Summary ZC407302}
\end{figure*}

\begin{figure*}
	\includegraphics[width=1\textwidth]{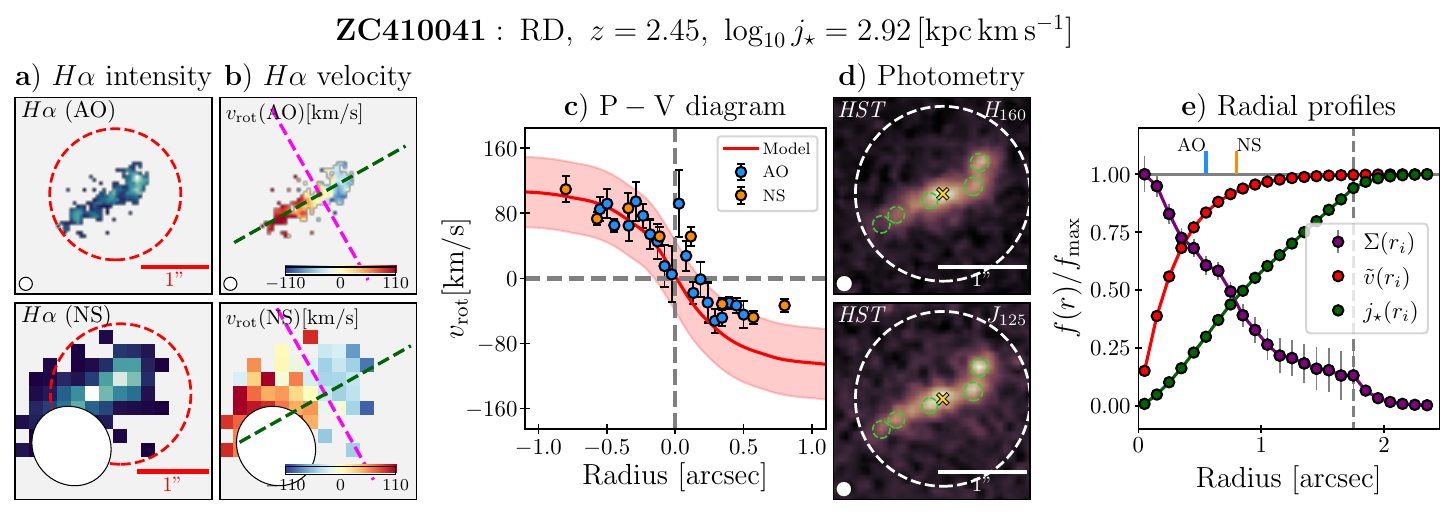}
    \caption{Summary ZC410041.}
    \label{fig:Summary ZC410041}
\end{figure*}


\begin{figure*}
	\includegraphics[width=1\textwidth]{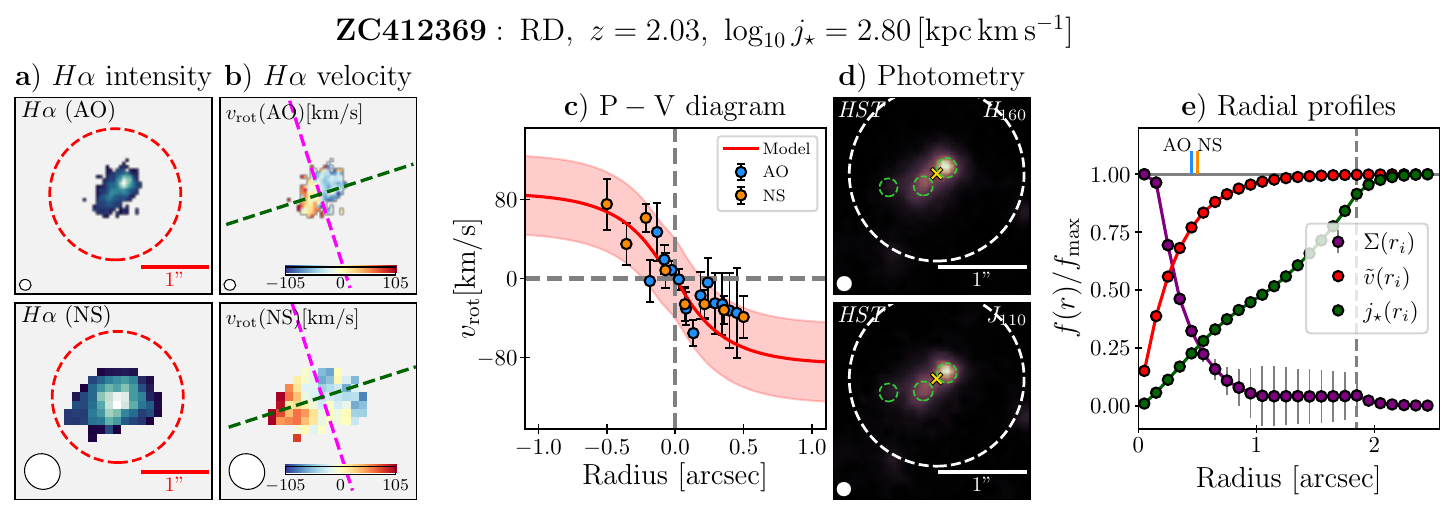}
    \caption{Summary ZC412369.}
    \label{fig:Summary ZC412369}
\end{figure*}

\begin{figure*}
	\includegraphics[width=1\textwidth]{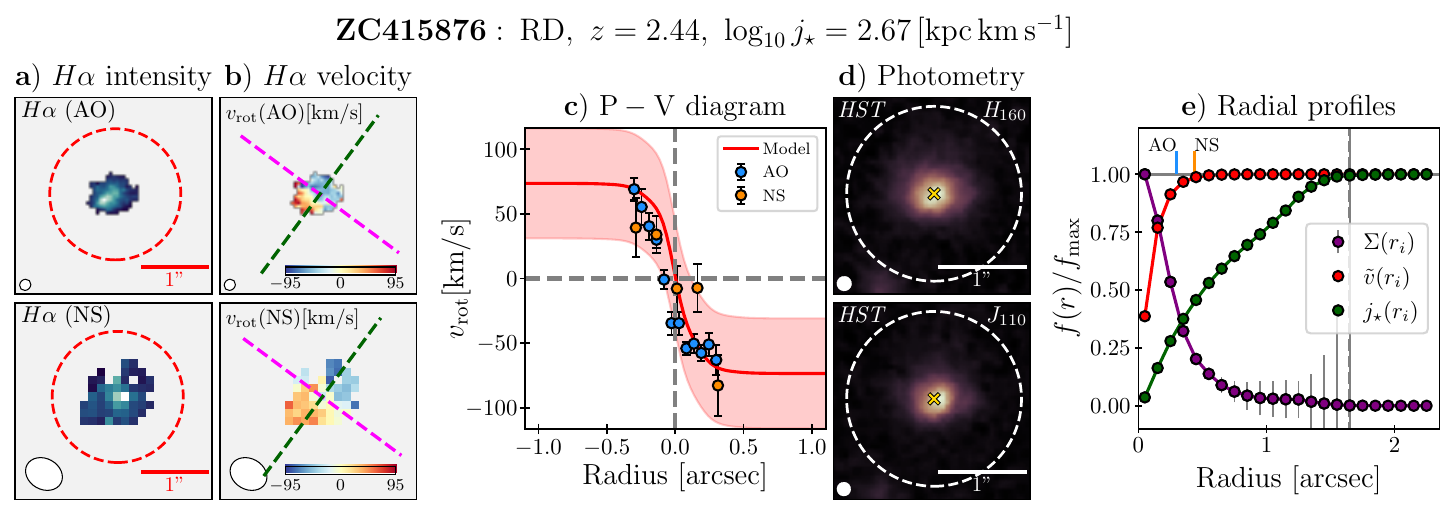}
    \caption{Summary ZC415876.}
    \label{fig:Summary ZC415876}
\end{figure*}

\begin{figure*}
	\includegraphics[width=1\textwidth]{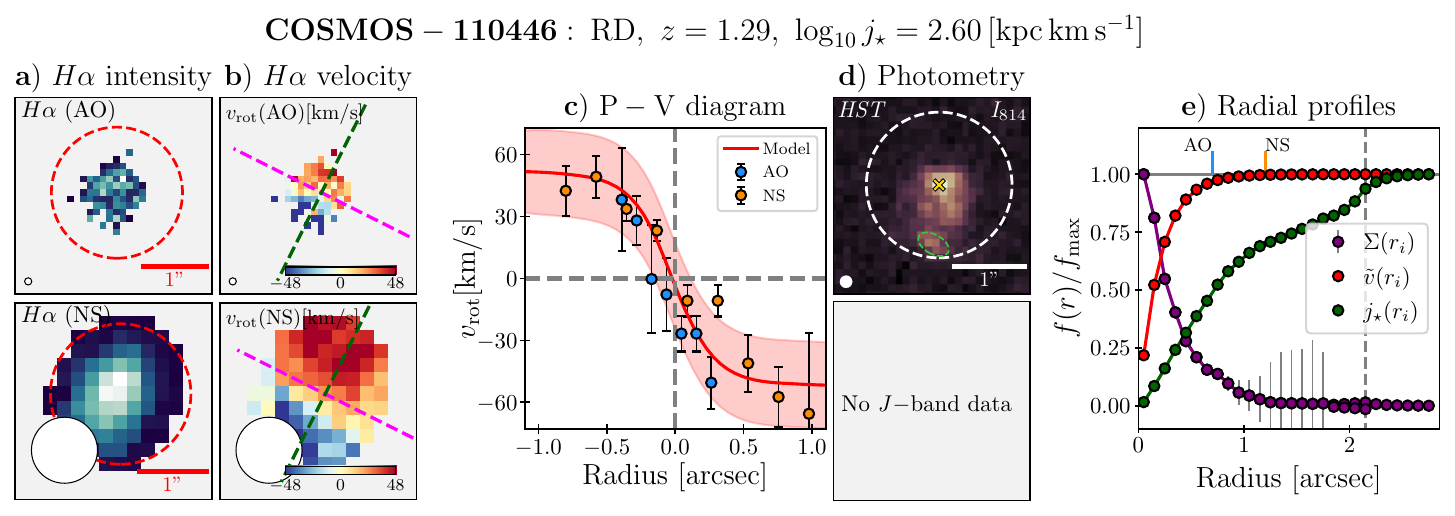}
    \caption{Summary COSMOS-110446.}
    \label{fig:Summary COSMOS-110446}
\end{figure*}

\begin{figure*}
	\includegraphics[width=1\textwidth]{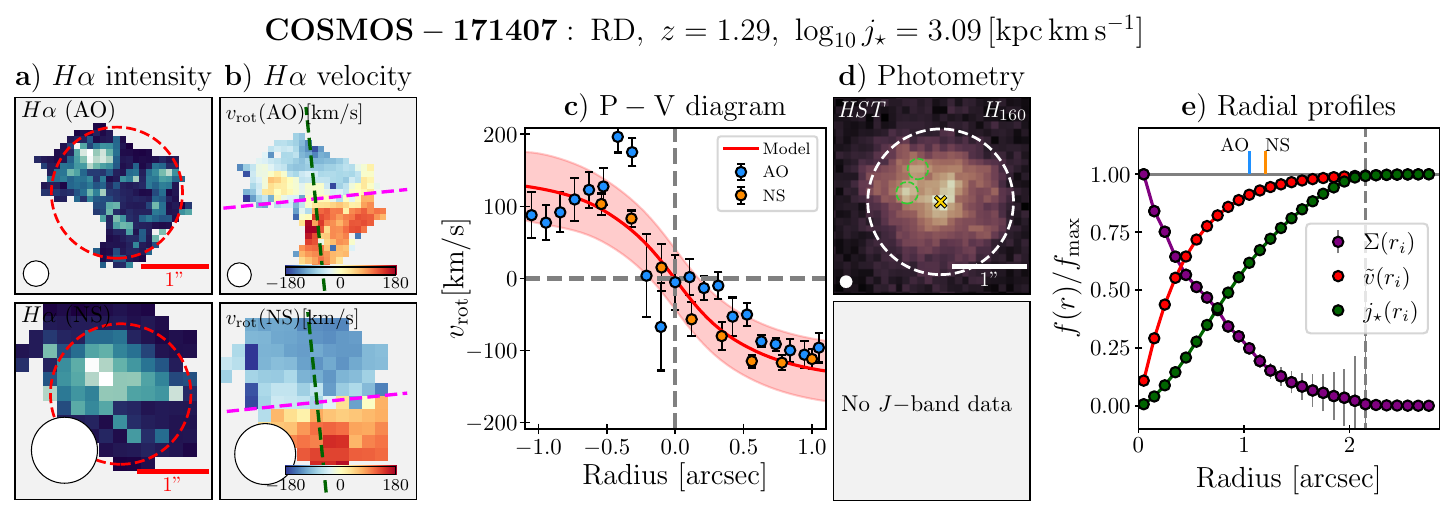}
    \caption{Summary COSMOS-171407.}
    \label{fig:Summary COSMOS-171407}
\end{figure*}

\begin{figure*}
	\includegraphics[width=1\textwidth]{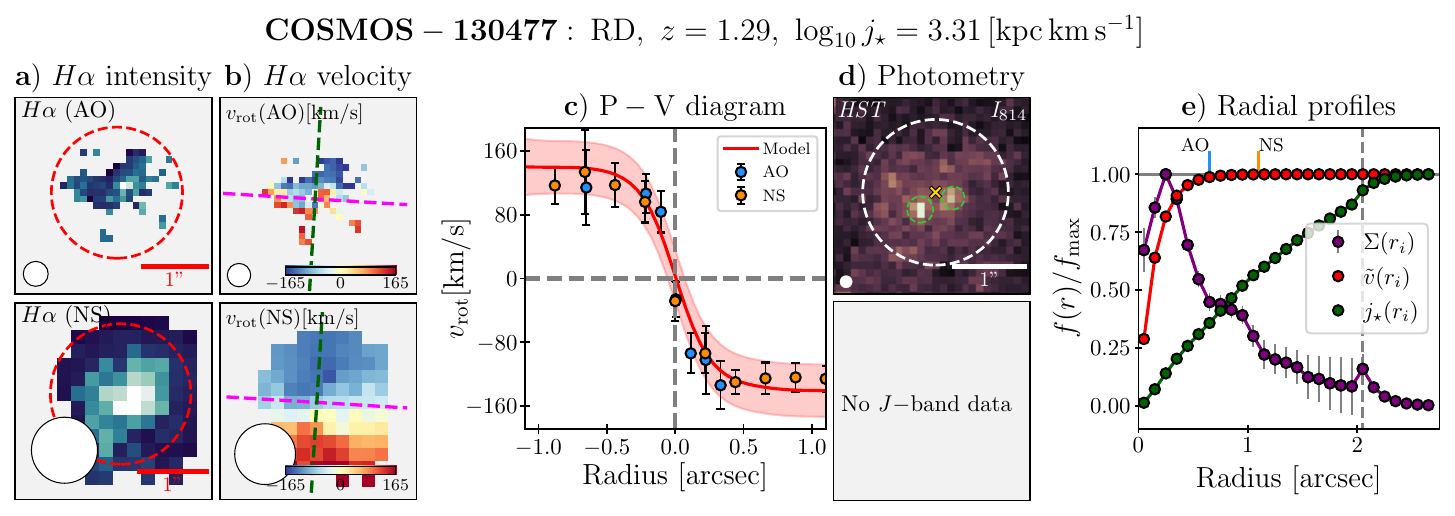}
    \caption{Summary COSMOS-130477.}
    \label{fig:Summary COSMOS-130477}
\end{figure*}

\begin{figure*}
	\includegraphics[width=1\textwidth]{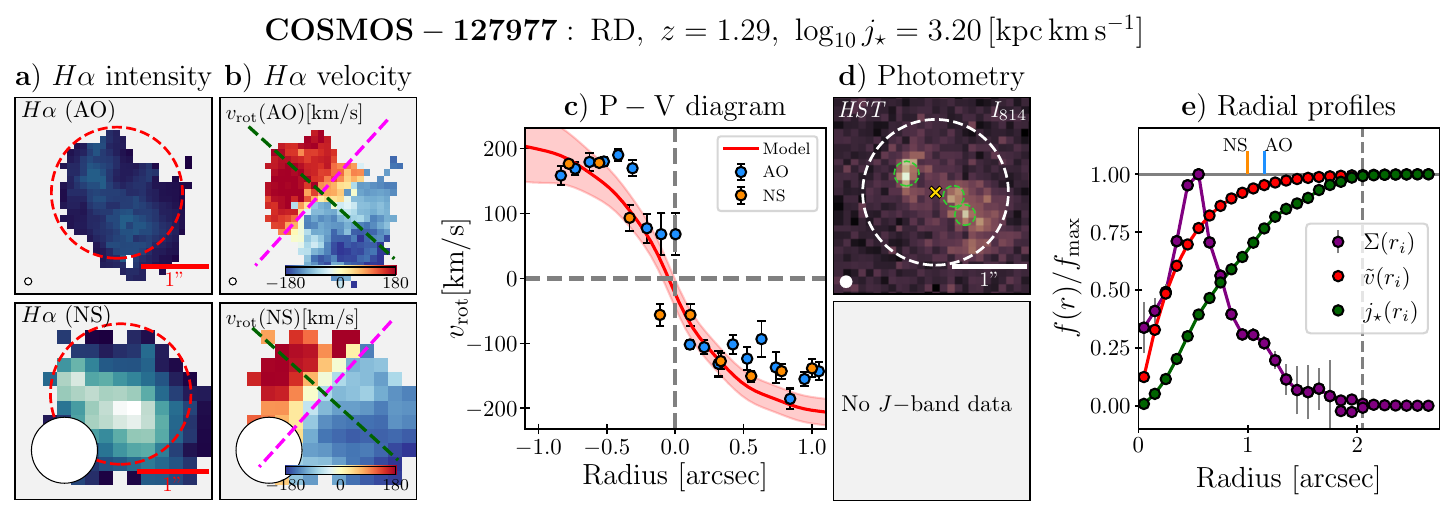}
    \caption{Summary COSMOS-127977.}
    \label{fig:Summary COSMOS-127977}
\end{figure*}

\begin{figure*}
	\includegraphics[width=1\textwidth]{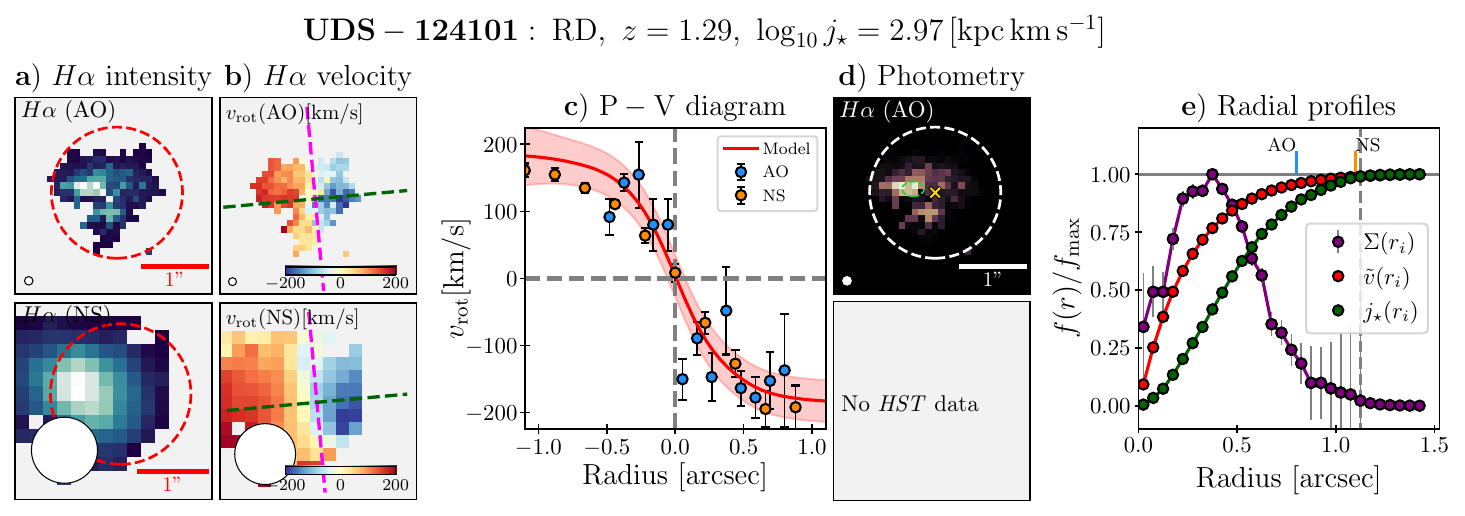}
    \caption{Summary UDS-124101.}
    \label{fig:Summary UDS-124101}
\end{figure*}

\begin{figure*}
	\includegraphics[width=1\textwidth]{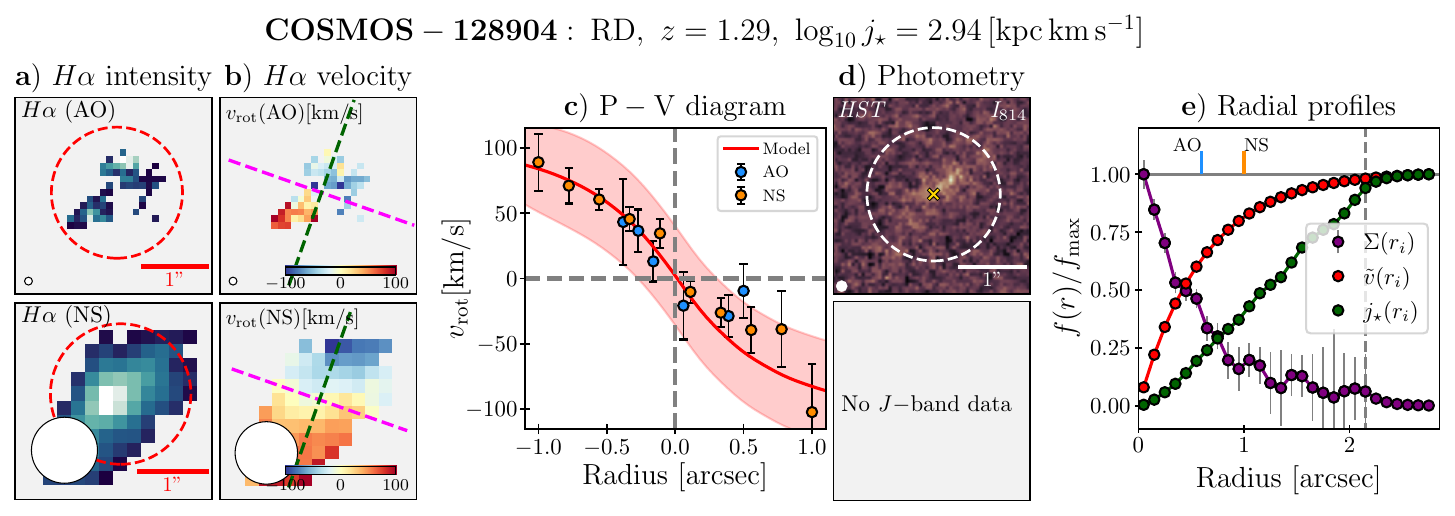}
    \caption{Summary COSMOS-128904.}
    \label{fig:Summary COSMOS-128904}
\end{figure*}


\bsp	
\label{lastpage}
\end{document}